\documentclass[review]{elsarticle}

\usepackage[english]{babel} 
\usepackage[T1]{fontenc}
\usepackage{lmodern} 
\usepackage[ansinew]{inputenc}
\usepackage{amsmath}
\usepackage{amssymb}
\usepackage{xfrac}
\usepackage{tikz}
\usepackage{tikz-dimline}
\usepackage{lineno}
\usepackage{algorithm2e}

\pgfplotsset{
  compat=1.5,
  legend image code/.code={
    \draw[mark repeat=2,mark phase=2]
    plot coordinates {
      (0cm,0cm)
      (0.15cm,0cm)        
      (0.3cm,0cm)         
    };
  }
}
\usepackage{geometry}
\usepackage{blindtext}
\usepackage{float}
\usepackage{graphicx}
\usepackage{xcolor}
\usepackage{caption}
\usepackage{hyperref}
\usepackage{pgfplots}
\usepackage{array}
\usepackage{siunitx}
\usepackage{multirow}
\usepackage{adjustbox} 
\usepackage{subcaption}
\usepackage{bm}

\usepackage{mathtools}
\usepackage[]{algorithm2e}
\usepackage{algpseudocode}
\usepackage{pgfplotstable}

\usepackage[frozencache,cachedir="minted_cache"]{minted}

\usepackage{setspace}
\DeclareMathOperator{\sign_math}{sign}

\usepgfplotslibrary{groupplots,dateplot}
\usetikzlibrary{patterns,shapes.arrows,calc,external,decorations,shapes,positioning,arrows.meta}
\usepgfplotslibrary{colormaps}
\usetikzlibrary{pgfplots.colormaps}

\usepgfplotslibrary{fillbetween}
\tikzset{>=latex}
\pgfplotsset{compat=newest}

\begin{document}

\title{JAX-Fluids 2.0: Towards HPC for Differentiable CFD of Compressible Two-phase Flows}

\author[a]{Deniz A. Bezgin\corref{cor}\fnref{fn}}
\ead{deniz.bezgin@tum.de}

\author[a]{Aaron B. Buhendwa\corref{cor}\fnref{fn}}
\ead{aaron.buhendwa@tum.de}

\cortext[cor]{Corresponding author}
\fntext[fn]{Both authors contributed equally.}

\author[a,b]{Nikolaus A. Adams\corref{}}
\ead{nikolaus.adams@tum.de}

\address[a]{Technical University of Munich, School of Engineering and Design, Chair of Aerodynamics and Fluid Mechanics, Boltzmannstra{\ss}e 15, 85748 Garching bei M\"unchen, Germany}
\address[b]{Technical University of Munich, Munich Institute of Integrated Materials, Energy and Process Engineering, Lichtenbergstra{\ss}e 4a, 85748 Garching bei M\"unchen, Germany}

\begin{frontmatter}
    \begin{abstract}
      In our effort to facilitate machine learning-assisted computational fluid dynamics (CFD),
we introduce the second iteration of JAX-Fluids. JAX-Fluids is
a Python-based fully-differentiable CFD solver designed
for compressible single- and two-phase flows.
In this work, the first version is extended to
incorporate high-performance computing (HPC) capabilities.
We introduce a parallelization strategy utilizing JAX primitive operations
that scales efficiently on GPU (up to 512 NVIDIA A100 graphics cards) and TPU (up to 1024 TPU v3 cores) HPC systems.
We further demonstrate the stable parallel computation of automatic differentiation gradients 
across extended integration trajectories.
The new code version offers enhanced two-phase flow modeling capabilities.
In particular, a five-equation diffuse-interface model is incorporated 
which complements the level-set sharp-interface model.
Additional algorithmic improvements include positivity-preserving limiters for increased robustness,
support for stretched Cartesian meshes,
refactored I/O handling, comprehensive post-processing routines,
and an updated list of state-of-the-art high-order numerical discretization schemes.
We verify newly added numerical models by 
showcasing simulation results for single- and two-phase flows, 
including turbulent boundary layer and channel flows,
air-helium shock bubble interactions, and air-water shock drop interactions.
JAX-Fluids 2.0 is released under the GNU GPLv3 license 
and made available on GitHub at \url{https://github.com/tumaer/JAXFLUIDS}.      
    \end{abstract}

\end{frontmatter}
\journal{Computer Physics Communications}

{\bf PROGRAM SUMMARY}

\begin{small}
\noindent
{\em Program Title:} JAX-Fluids \\
{\em Developer's repository link:} (\url{https://github.com/tumaer/JAXFLUIDS}) \\
{\em Licensing provisions:} GPLv3 \\
{\em Programming language:} Python, JAX \\
{\em Supplementary material:} Source code; Example scripts; Videos. \\
{\em Journal reference of previous version:} 
\begin{itemize}
  \item D. A. Bezgin, A. B. Buhendwa, N. A. Adams, JAX-Fluids: A fully-differentiable high-order computational fluid dynamics solver for compressible two-phase flows, 
  Computer Physics Communications 282 (2022) 108527.
\end{itemize}
{\em Does the new version supersede the previous version?:} Yes \\
{\em Reasons for the new version:} New features and updates of the CFD solver \\
{\em Summary of revisions:} 
\begin{itemize}
  \item JAX primitives-based parallelization for GPU and TPU clusters
  \item Automatic differentiation through distributed simulations
  \item Diffuse-interface model for two-phase flows
  \item Positivity-preserving interpolation and flux limiters
  \item Support for stretched Cartesian meshes
  \item Extended list of numerical discretization schemes
  \item Performance improvements
  \item Revised I/O handling
\end{itemize}
{\em Nature of problem:} 
The compressible Navier-Stokes equations describe continuum-scale fluid flows
which may exhibit complex phenomena such as shock waves, material interfaces, and turbulence.
The accurate numerical solution of fluid flows is computationally expensive 
and, therefore, requires high performance computing (HPC) architectures.
To this end, machine learning (ML), in particular differentiable programming,
is continuously being explored as a tool to accelerate conventional computational fluid dynamics (CFD).
With the second iteration of JAX-Fluids, 
we provide a comprehensive differentiable CFD code that
scales efficiently on HPC systems and seamlessly integrates
ML models. Furthermore, JAX-Fluids is capable of simulating
highly complex flow physics such as supersonic boundary layer flow
or interactions of shock waves with material interfaces. \\
{\em Solution method:} 
JAX-Fluids is a finite-volume solver which uses low-dissipative high-order shock capturing schemes
in combination with approximate Riemann solvers.
Two-phase flows can be simulated using the sharp-interface level-set method
or the diffuse-interface five-equation model.
The code is written in Python and builds on the JAX library.
The JAX backend allows the computation of automatic differentiation gradients.
We use a homogenous domain decomposition ansatz to implement the parallelization.
An object-oriented programming style and a modular design philosophy 
allow exchanging numerical schemes and integrating custom subroutines. \\
{\em Additional comments including Restrictions and Unusual features:}
JAX-Fluids runs on CPUs, GPUs, and TPUs in single- and multi-device settings.
JAX-Fluids requires open-source third-party Python libraries which are automatically installed.
The solver has been tested on Linux and macOS operating systems.\\\\


\end{small}

\section{Introduction}
\label{sec:introduction}
The numerical solution of partial differential equations (PDEs) is key to simulating and understanding
physical systems.
The solution of high-dimensional complex PDEs requires vast computational resources provided by
high-performance computing (HPC).
Algorithms aimed at solving PDEs on HPC systems have traditionally been written in low-level
programming languages like Fortran or C/C++.
In recent years, machine learning (ML)-assisted numerical schemes have been developed for the solution of PDEs.
The field of fluid mechanics in particular benefits
from ML-assisted modeling due the nonlinearity and
multiscale nature of the underlying physics as well as
its data-rich character \cite{Brunton2020a,Vinuesa2022}.
As ML libraries are predominantly used in Python, however, the integration
of ML into conventional PDE solvers and the joint
application on HPC systems requires novel simulation frameworks.

Differentiable solvers have gained increased attention in engineering 
and in the physical sciences due to their promise of bridging the aforementioned gap between 
computational physics and machine learning.
In particular, differentiable solvers ensure automatic differentiation (AD) \cite{Baydin2017}
of the entire algorithmic representation of the chosen numerical approximation of the PDE evolution law.
The training of data-driven models with such gradients is referred to as \textit{end-to-end optimization}.
End-to-end optimized models generally allow stable inference over long prediction horizons.
The interest in differentiable simulators has also been fueled 
by the development of powerful general purpose AD libraries,
such as TensorFlow \cite{Abadi}, PyTorch \cite{Paszke2019}, and JAX \cite{jax2018github}.
Recent applications of differentiable solvers include computational fluid dynamics (e.g., 
for molecular dynamics simulations \cite{Schoenholz2020}, incompressible single-phase flows \cite{Kochkov2021,Wang2022}, 
and Lattice Boltzmann methods \cite{Ataei2023}), structural mechanics (e.g., \cite{Xue2022}),
etc.

We reiterate some of the requirements we deem essential for next-generation physics solvers:
\begin{enumerate}
    \item Rapid prototyping capability in high-level programming languages like Python,
    \item algorithms which scale efficiently on modern HPC architectures (e.g., CPU, GPU, and TPU clusters),
    \item seamless integration of machine learning models into physics solver frameworks,
    \item differentiable algorithms which allow end-to-end optimization of data-driven models.
\end{enumerate}

In light of the aforementioned points,
we have proposed JAX-Fluids \cite{Bezgin2022} as a fully-differentiable computational fluid dynamics (CFD) solver
for compressible two-phase flows.
JAX-Fluids is build upon the JAX-based \cite{jax2018github,Frostig2018} implementation of
the popular NumPy package \cite{Harris2020},
making it easy and intuitive to use.
The JAX backend \cite{jax2018github,Frostig2018} allows automatic differentiation through every subroutine of our solver, and
it provides capabilities to run on CPU, GPU, and TPU accelerators.

In this work, we introduce the second iteration of the JAX-Fluids CFD framework.
While the first release of JAX-Fluids was restricted to a single device, 
JAX-Fluids 2.0 has been developed for use on distributed HPC architectures.
A parallelization strategy using only JAX primitives allows obtaining 
automatic differentiation gradients in distributed settings.
In particular, JAX-Fluids 2.0 supports single-host and multi-host settings.
Additionally, the two-phase modeling capabilities have been enhanced by a diffuse-interface module.
For two-phase simulations, users can now choose between a level-set-based sharp-interface model \cite{Osher1988,Sussman1994b,Hoppe2022}
and a five-equation diffuse-interface model \cite{Allaire2002,Perigaud2005,Coralic2014,Wong2021}.
Single- and two-phase modules are enhanced by positivity-preserving techniques for increased robustness.
JAX-Fluids 2.0 is complemented by performance improvements throughout the entire source code,
support for stretched Cartesian meshes,
an updated list of discretization schemes, improved I/O-handling,
and comprehensive post-processing routines. JAX-Fluids 2.0 is capable
of performing direct numerical simulations of compressible, wall-bounded
turbulence and highly resolved simulations of complex shock-interface
interactions. The computation of gradients obtained by automatic differentiation
shows stable behavior through extended integration trajectories across
multiple devices.

Differentiable CFD solvers like JAX-Fluids may contribute to a plethora of data-driven
applications in the context of fluid mechanics.
We highlight three fields which we deem especially relevant.
\begin{enumerate}
    \item \textbf{End-to-end Optimization of Surrogate Models:} 
    Up to now, many machine learning models for CFD have been optimized \textit{offline}, 
    i.e., outside the CFD solver. Offline trained models may become unstable during long
    rollout trajectories due to error accumulation (e.g., \cite{Beck2019}).
    Online training may alleviate this issue as models observe PDE dynamics during training
    and have to account for their own prediction errors \cite{Sirignano2020,Kochkov2021,Bezgin2021a}.
    
    \item \textbf{Inverse Problems:}
    An important inverse problem in fluid mechanics
    is flow reconstruction from incomplete information (e.g., sparse measurements).
    Physics-informed neural networks (PINNs) have been used extensively in this context \cite{Raissi2019,Buhendwa2021c,Jagtap2022}.
    However, recently the optimization of a discrete loss (ODIL) framework was introduced
    which utilizes differentiable solvers \cite{Karnakov2022}.
    
    \item \textbf{Uncertainty Quantification:} 
    Differentiable physics simulators can be used in probabilistic frameworks to 
    model and quantify uncertainty.
\end{enumerate}

The manuscript is structured as follows.
In section \ref{sec:physical_model}, we introduce the single-phase Navier-Stokes equations, the level-set formalism,
and the five-equation diffuse-interface model.
Section \ref{sec:numerical_models} discusses the numerical discretization of the aforementioned equations.
We introduce the JAX-based parallelization strategy of JAX-Fluids in Section \ref{sec:parallelization_strategy}.
Verification of the numerical models and the automatic differentiation gradients 
is presented in Sections \ref{sec:verification_numerical_models} and \ref{sec:verification_gradients}, respectively.
In Section \ref{sec:parallel_performance}, we analyze the parallel performance of JAX-Fluids on GPU and TPU clusters.
Section \ref{sec:conclusion} provides a summary of the achievements and gives concluding remarks. 
\section{Physical model}
\label{sec:physical_model}
JAX-Fluids solves compressible single-phase and two-phase flows.
In this section, we introduce the underlying governing equations.
Two-phase flows in JAX-Fluids can be solved by a level-set-based
sharp-interface model or
the five-equation diffuse-interface model. 
We detail both two-phase modeling approaches.

The state of the fluid at any position $\mathbf{x} = \left[ x, y, z \right]^T = \left[ x_1, x_2, x_3 \right]^T$
in the flow field at time $t$
can be described either by the vector of primitive variables $\mathbf{W}$ or by the vector of conservative variables $\mathbf{U}$.
Both representations of the fluid state are equivalent,
i.e., $\mathbf{W}=\mathcal{L}_{\mathbf{U}\rightarrow\mathbf{W}}(\mathbf{U})$,
and $\mathbf{U}=\mathcal{L}_{\mathbf{W}\rightarrow\mathbf{U}}(\mathbf{W})$.
The compressible Navier-Stokes equations in symbolic form for the vector
of conservative variables are
\begin{align}
    \frac{\partial \mathbf{U}}{\partial t}
    + \frac{\partial \mathbf{F}^c(\mathbf{U})}{\partial x}
    + \frac{\partial \mathbf{G}^c(\mathbf{U})}{\partial y}
    + \frac{\partial \mathbf{H}^c(\mathbf{U})}{\partial z} = 
    \frac{\partial \mathbf{F}^{d}(\mathbf{U})}{\partial x}
    + \frac{\partial \mathbf{G}^{d}(\mathbf{U})}{\partial y}
    + \frac{\partial \mathbf{H}^{d}(\mathbf{U})}{\partial z}
    + \sum_i \mathbf{S}_i(\mathbf{U}).
    \label{eq:DiffConsLaw1}
\end{align}
$\mathbf{F}^c, \mathbf{G}^c,$ and $\mathbf{H}^c$ denote the convective fluxes in $x$-, $y$- and $z$-direction.
Analogously, $\mathbf{F}^{d}, \mathbf{G}^{d},$ and $\mathbf{H}^{d}$ denote the dissipative fluxes in the three spatial dimensions.
The right-hand side is complemented by the sum of all source terms $\sum_i \mathbf{S}_i(\mathbf{U})$.

%
\subsection{Single-phase model}
\label{subsec:single_phase_model_phys}
For single-phase flows, the primitive variables are the fluid density $\rho$, 
the velocity components $u$, $v$, and $w$ (in $x$-,$y$-, and $z$-direction, respectively),
and the pressure $p$.
$\mathbf{u} = \left[ u, v, w \right]^T = \left[ u_1, u_2, u_3 \right]^T$ is the velocity vector.
$E = \rho e + \frac{1}{2} \rho \mathbf{u} \cdot \mathbf{u} $ denotes the total energy of the fluid.
The vectors of primitive and conservative variables are given as
\begin{align}
    \mathbf{W} = \begin{bmatrix}
        \rho \\ u \\ v \\ w \\ p
    \end{bmatrix}, \quad
    \mathbf{U} = \begin{bmatrix}
        \rho \\ \rho u \\ \rho v \\ \rho w \\ E
    \end{bmatrix},
    \label{eq:prime_cons_vector}
\end{align}
and the convective fluxes are
\begin{align}
    \mathbf{F}^c(\mathbf{U}) = \begin{bmatrix}
        \rho u \\
        \rho u^2 + p \\
        \rho u v \\
        \rho u w \\
        u (E + p)
    \end{bmatrix}, \quad
    \mathbf{G}^c(\mathbf{U}) = \begin{bmatrix}
        \rho v \\
        \rho v u\\
        \rho v^2 + p \\
        \rho v w \\
        v (E + p)
    \end{bmatrix}, \quad
    \mathbf{H}^c(\mathbf{U}) = \begin{bmatrix}
        \rho w\\
        \rho w u\\
        \rho w v \\
        \rho w^2 + p \\
        w (E + p)
    \end{bmatrix}.
    \label{eq:convective_fluxes}
\end{align}
The dissipative fluxes describe viscous effects and heat conduction.
\begin{align}
    \mathbf{F}^{d}(\mathbf{U}) = \begin{bmatrix}
        0 \\
        \tau^{11} \\
        \tau^{12} \\
        \tau^{13} \\
        \sum_i u_i \tau^{1i} - q_1
    \end{bmatrix}, \quad
    \mathbf{G}^{d}(\mathbf{U}) = \begin{bmatrix}
        0 \\
        \tau^{21} \\
        \tau^{22} \\
        \tau^{23} \\
        \sum_i u_i \tau^{2i} - q_2
    \end{bmatrix}, \quad
    \mathbf{H}^{d}(\mathbf{U}) = \begin{bmatrix}
        0 \\
        \tau^{31} \\
        \tau^{32} \\
        \tau^{33} \\
        \sum_i u_i \tau^{3i} - q_3
    \end{bmatrix}.
    \label{eq:dissipative_flux}
\end{align}
The stresses $\tau^{ij}$ are given by
\begin{align}
    \tau^{ij} = \mu \left(\frac{\partial u_i}{\partial x_j} + \frac{\partial u_j}{\partial x_i}\right) - \frac{2}{3} \mu \delta_{ij} \frac{\partial u_k}{\partial x_k},
    \label{eq:newtonian_shear_stress}
\end{align}
where $\mu$ is the dynamic viscosity.
The energy flux vector $\mathbf{q} = \left[ q_1, q_2, q_3 \right]^T$ is expressed via Fourier's heat conduction law, $\mathbf{q} = -\lambda \nabla T$,
where $\lambda$ is the heat conductivity.
\subsection{Level-set model}
\label{subsec:levelset_model_phys}
The level-set model \cite{Osher1988, Sussman1994b} (LSM) is a sharp-interface method 
and uses a signed distance function $\phi$
to track the interface. The interface location
is implicitly given by its zero level-set. The two 
phases are distinguished by its sign. The system of equations described in the previous subsection
\ref{subsec:single_phase_model_num}, i.e., Eqs.
\eqref{eq:prime_cons_vector},\eqref{eq:convective_fluxes}, and
\eqref{eq:dissipative_flux}
hold for both phases separately. In addition, the level-set function is evolved
according to
\begin{equation}
    \frac{\partial \phi}{\partial t} = - \mathbf{u}_\Gamma \cdot \nabla \phi = \mathcal{R}_{LSA},
    \label{eq:levelset_advection}
\end{equation}
where $\mathbf{u}_\Gamma$ is the interface velocity.

\subsection{Diffuse-interface model}
\label{subsec:diffuse_interface_model_phys}
In the five-equation diffuse-interface model (DIM) \cite{Allaire2002,Perigaud2005}, 
the interface is artificially thickened over a finite region.
We solve a separate continuity equation for each phase 
($\alpha_1 \rho_1$ and $\alpha_2 \rho_2$ denote the phase masses for each fluid),
a momentum equation, an energy equation,
and a volume fraction equation for either one of the two fluids.
The primitive and conservative variables are
%
%
\begin{align}
    \mathbf{W} = \begin{bmatrix}
        \alpha_1 \rho_1 \\ \alpha_2 \rho_2 \\ u \\ v \\ w \\ p \\ \alpha_1
    \end{bmatrix}, \quad
    \mathbf{U} = \begin{bmatrix}
        \alpha_1 \rho_1 \\ \alpha_2 \rho_2 \\ \rho u \\ \rho v \\ \rho w \\ E \\ \alpha_1
    \end{bmatrix}.
\end{align}
The density of the mixture $\rho = \alpha_1 \rho_1 + \alpha_2 \rho_2$,
and the volume fraction of the second fluid $\alpha_2 = 1 - \alpha_1$.
The convective fluxes in the three spatial dimensions are given as
\begin{align}
    \mathbf{F}^c(\mathbf{U}) = \begin{bmatrix}
        \alpha_1 \rho_1 u \\
        \alpha_2 \rho_2 u \\
        \rho u^2 + p \\
        \rho u v \\
        \rho u w \\
        u (E + p) \\
        \alpha_1 u
    \end{bmatrix}, \quad
    \mathbf{G}^c(\mathbf{U}) = \begin{bmatrix}
        \alpha_1 \rho_1 v \\
        \alpha_2 \rho_2 v \\
        \rho v u\\
        \rho v^2 + p \\
        \rho v w \\
        v (E + p) \\
        \alpha_1 v
    \end{bmatrix}, \quad
    \mathbf{H}^c(\mathbf{U}) = \begin{bmatrix}
        \alpha_1 \rho_1 w\\
        \alpha_2 \rho_2 w\\
        \rho w u\\
        \rho w v \\
        \rho w^2 + p \\
        w (E + p) \\
        \alpha_1 w
    \end{bmatrix}.
    \label{eq:convective_fluxes_dim}
\end{align}
Using the convective flux terms in Eq. \eqref{eq:convective_fluxes_dim},
we recast the (non-conservative) volume fraction transport equation,
\begin{align*}
    \frac{\partial \alpha_1}{\partial t} + \mathbf{u} \cdot \nabla \alpha_1 = 0,
\end{align*}
into its equivalent conservative form,
\begin{align}
    \frac{\partial \alpha_1}{\partial t} + \nabla \cdot \left( \alpha_1 \mathbf{u} \right) = \alpha_1 \nabla \cdot \mathbf{u},
    \label{eq:volume_fraction_eq_conservative}
\end{align}
see \cite{Johnsen2006,Coralic2014}.
The resulting source term is accounted for on the right-hand side,
\begin{align}
    \mathbf{S}_{\alpha} = \begin{bmatrix}
        0 \\
        0 \\
        0 \\
        0 \\
        0 \\
        0 \\
        \alpha_1 \nabla \cdot \mathbf{u}
    \end{bmatrix}.
    \label{eq:sourcetermdiffuse}
\end{align}

\subsection{Equation of state}
\label{subsec:equation_of_state}
The system of equations \eqref{eq:DiffConsLaw1} is closed by an equation of state (EOS).
Throughout this work we employ the stiffened gas equation of state \cite{Menikoff1989}.
\begin{align}
    p &= \left( \gamma - 1 \right) \rho e - \gamma p_{\infty} \\
    c &= \sqrt{\frac{\gamma \left( p + p_{\infty} \right)}{\rho}}
    \label{eq:stiffened_gas_eos}
\end{align}
The stiffened gas equation has two parameters: $\gamma$ is the ratio of specific heats,
and $p_{\infty}$ is the background pressure. 
For vanishing background pressures ($p_{\infty} = 0$) the stiffened gas EOS reduces
to the ideal gas law.

For the five-equation model, we consider the mixture of two fluids.
Each fluid is governed by its stiffened gas EOS,
i.e., we have $\left( \gamma_1, p_{1, \infty} \right)$ and $\left( \gamma_2, p_{2, \infty} \right)$.
The isobaric closure \cite{Allaire2002} gives an explicit analytic expression for 
the EOS parameters of the mixture $\left( \gamma, p_{\infty} \right)$.
\begin{align}
    \frac{1}{\gamma - 1} &= \frac{\alpha_1}{\gamma_1 - 1} + \frac{1 - \alpha_1}{\gamma_2 - 1} \\
    \frac{\gamma p_{\infty}}{\gamma - 1} &= \frac{\alpha_1 \gamma_1 p_{1,\infty}}{\gamma_1 - 1}
        + \frac{\left( 1 -\alpha_1 \right) \gamma_2 p_{2,\infty}}{\gamma_2 - 1}
\end{align}

\subsection{Material properties}
\label{subsec:material_properties}
For viscous simulations, we require models for dynamic viscosity and thermal conductivity.
In this work, we use two different models for the dynamic viscosity $\mu$.
The first is a simple power law,
\begin{align}
    \mu = \mu_{ref} \left( \frac{T}{T_{ref}} \right)^{0.7},
    \label{eq:viscosity_power_law}
\end{align}
where $\mu_{ref}$ is the dynamic viscosity at the reference temperature $T_{ref}$.
The second is the well known Sutherland's law \cite{Sutherland1893},
\begin{align}
    \mu = \mu_{ref} \left( \frac{T}{T_{ref}} \right)^{1.5} \frac{T_{ref} + C}{T + C},
    \label{eq:viscosity_sutherland_law}
\end{align}
where $C$ is Sutherland's constant.
In both cases, the thermal conductivity $\lambda$ can be determined
using the assumption of a constant Prandtl number,
\begin{align}
    \lambda = \frac{Pr}{\mu c_p}.
\end{align}
Here, $c_p$ denotes the specific heat capacity at constant pressure.
\section{Numerical models}
\label{sec:numerical_models}
The aforementioned system of PDEs \eqref{eq:DiffConsLaw1} is discretized on a Cartesian grid 
using the finite-volume formulation.
Cell-averaged quantities (denoted by an overbar) are updated using cell face fluxes 
which are calculated from a high-order Godunov-type scheme \cite{Toro2009a}.

%
\subsection{Computational domain, grid, and boundary conditions}
\label{subsec:computational_domain}
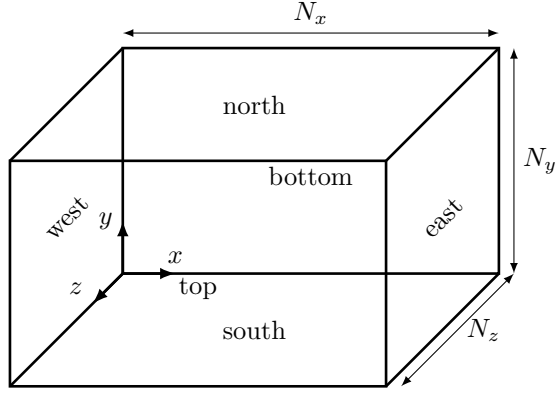
\begin{figure}[!t]
    \centering
    \begin{tikzpicture}
    \node (-0.1, -0.1) {};
    \definecolor{color0}{RGB}{0,101,189}
    \definecolor{color1}{RGB}{227,114,34}
    \definecolor{color2}{RGB}{162,173,0}

    \draw[line width=1pt] (0,0) -- (0,3) -- (5,3) -- (5,0) -- cycle;
    \draw[line width=1pt] (1.5,1.5) -- (1.5,4.5) -- (6.5,4.5) -- (6.5,1.5) -- cycle;
    \draw[line width=1pt] (6.5,1.5) -- (5,0);
    \draw[line width=1pt] (6.5,4.5) -- (5,3);
    \draw[line width=1pt] (0,3) -- (1.5,4.5);
    \draw[line width=1pt] (0,0) -- (1.5,1.5);

    \draw[<->] (1.5,4.7) -- (6.5,4.7) node[midway, above] {$N_x$};
    \draw[<->] (6.7,1.5) -- (6.7,4.5) node[midway, right] {$N_y$};
    \draw[<->] (6.7,1.5) -- (5.2,0) node[midway, right] {$N_z$};


    \draw[line width=1pt, ->] (1.5,1.5) -- (2.2,1.5) node[at end, above] {$x$};
    \draw[line width=1pt, ->] (1.5,1.5) -- (1.5,2.2) node[at end, left] {$y$};
    \draw[line width=1pt, ->] (1.5,1.5) -- (1.1,1.1) node[at end, above left] {$z$};
    
    \draw[draw=none] (0.0,1.5) -- (1.5,3.0) node[midway, sloped] {\textcolor{color1!0!black}{west}};
    \draw[draw=none] (5.0,1.5) -- (6.5,3.0) node[midway, sloped] {\textcolor{color1!0!black}{east}};

    \node[opacity=1.0] at (3.25,0.75) {\textcolor{color2!0!black}{south}};
    \node at (3.25,3.75) {\textcolor{color2!0!black}{north}};

    \node at (2.5,1.3) {\textcolor{color0!0!black}{top}};
    \node[opacity=1.0] at (4,2.8) {\textcolor{color0!0!black}{bottom}};
\end{tikzpicture}
    \caption{Schematic of the computational domain. We denote the axis dircetions .}
    \label{fig:computational_domain}
\end{figure}

%
The computational domain is partitioned in $N_x \times N_y \times N_z$ cuboid finite volumes.
Figure \ref{fig:computational_domain} depicts an
exemplary computational domain including the nomenclature for the boundary locations
at the corresponding axis directions.
The finite-volume cell indexed by $\left( i,j,k \right)$ is associated with the volume 
$x, y, z \in \left[x_{i-\frac{1}{2},j,k}, x_{i+\frac{1}{2},j,k}\right] \times \left[y_{i,j-\frac{1}{2},k}, y_{i,j+\frac{1}{2},k}\right] \times \left[z_{i,j,k-\frac{1}{2}}, z_{i,j,k+\frac{1}{2}}\right]$,
where $i,j,$ and $k$ are the indices in $x$-,$y$-, and $z$-direction, respectively.
The cell faces of cell $\left( i,j,k \right)$ are located at $x_{i \pm \frac{1}{2},j,k}$,
$y_{i,j \pm \frac{1}{2},k}$, and $z_{i,j,k \pm \frac{1}{2}}$,
and the cell center position is given by 
$\left[ \frac{x_{i-\frac{1}{2},j,k} + x_{i+\frac{1}{2},j,k}}{2}, \frac{y_{i,j-\frac{1}{2},k} + y_{i,j+\frac{1}{2},k}}{2}, \frac{z_{i,j,k-\frac{1}{2}} + z_{i,j,k+\frac{1}{2}}}{2}  \right]$.
The cell sizes in the three spatial dimensions are $\Delta x_{i,j,k} = x_{i+\frac{1}{2},j,k} - x_{i-\frac{1}{2},j,k}$,
$\Delta y_{i,j,k} = y_{i,j+\frac{1}{2},k} - y_{i,j-\frac{1}{2},k}$,
and $\Delta z_{i,j,k} = z_{i,j,k+\frac{1}{2}} - z_{i,j,k-\frac{1}{2}}$,
and the cell volume is calculated as $\Delta V_{i,j,k} = \Delta x_{i,j,k} \Delta y_{i,j,k} \Delta z_{i,j,k}$.
JAX-Fluids supports arbitrary one-dimensional mesh-stretching such 
that the cell face positions (and, therefore, the cell sizes) can be functions of the corresponding spatial index.
Exemplary mesh stretchings are given in \ref{sec:appendix_computational_domain}.
If a homogenous grid spacing is used, we indicate cell sizes simply by $\Delta x$, $\Delta y$, and $\Delta z$.
For the isotropic case, we have $\Delta x = \Delta y = \Delta z$.

To evaluate spatial stencils close to and at the outer faces of the computational domain, we make use of halo cells.
In practice, an arbitrary buffer therefore has the shape $[N_x+2N_h,Ny+2N_h,N_z+2N_h]$. The number of halo cells $N_h$
depend on the structure of the utilized spatial stencils, which we will detail in the following subsections.
To enforce symmetry and periodicity at the boundaries, halo cells are inferred from the internal cells
in a straight-forward fashion. To impose Dirichlet boundary conditions, we simply assign the cell center values
of the halo cells to the desired values.

\subsection{High-order Godunov-type finite-volume formulation}
\label{subsec:high_order_godunov}
In this section, we detail the spatial discretization of the convective (inviscid) fluxes
and the diffusive (viscid) fluxes.
The semi-discrete form of the governing equations \eqref{eq:DiffConsLaw1} 
for the cell-averaged state $\bar{\mathbf{U}}_{i,j,k}$ in cell $\left( i,j,k \right)$ is given by
\begin{align}
    \begin{split}
        \frac{\text{d}}{\text{d}t} \bar{\mathbf{U}}_{i,j,k} = 
        & - \frac{1}{\Delta x_{i,j,k}} \left[ \left( \mathbf{F}^c_{i+\frac{1}{2},j,k} - \mathbf{F}^d_{i+\frac{1}{2},j,k} \right) - \left( \mathbf{F}^c_{i-\frac{1}{2},j,k} - \mathbf{F}^d_{i-\frac{1}{2},j,k} \right) \right] \\
        & - \frac{1}{\Delta y_{i,j,k}} \left[ \left( \mathbf{G}^c_{i,j+\frac{1}{2},k} - \mathbf{G}^d_{i,j+\frac{1}{2},k} \right) - \left( \mathbf{G}^c_{i,j-\frac{1}{2},k} - \mathbf{G}^d_{i,j-\frac{1}{2},k} \right) \right] \\
        & - \frac{1}{\Delta z_{i,j,k}} \left[ \left( \mathbf{H}^c_{i,j,k+\frac{1}{2}} - \mathbf{H}^d_{i,j,k+\frac{1}{2}} \right) - \left( \mathbf{H}^c_{i,j,k-\frac{1}{2}} - \mathbf{H}^d_{i,j,k-\frac{1}{2}} \right) \right] \\
        & + \bar{\mathbf{S}}_{i,j,k} = \mathcal{R}_\text{NSE}
       \label{eq:FVD} 
    \end{split}
\end{align}
The convective fluxes are calculated using a high-order shock-capturing scheme 
in combination with the HLLC approximate Riemann solver.
Figure \ref{fig:weno_teno} shows a schematic of reconstruction and flux calculation.
We summarize the procedure step-by-step exemplarily for the flux $\mathbf{F}^c_{i+\frac{1}{2},j,k}$,
i.e., the convective flux in $x$-direction at the cell face $x_{i+\frac{1}{2},j,k}$.
The procedure is applied dimension-by-dimension (method of lines), and the calculation of $\mathbf{G}^c$ and $\mathbf{H}^c$
in $y$- and $z$-directions is analogous.
To improve legibility, we drop the indices $j$ and $k$ for $y$- and $z$-directions, respectively,
for the remainder of this section.
We provide more details in \ref{sec:appendix_spatial_discretization}.
%
\begin{figure}[!t]
    \centering
    \begin{tikzpicture}
    \def\DX{1.25}
    \def\DY{1.0}
    \coordinate (IMM) at (-3,0);

    \coordinate (IM) at ($(IMM) + (\DX,0)$);
    \coordinate (I) at ($(IMM) + (2*\DX,0)$);
    \coordinate (IP) at ($(IMM) + (3*\DX,0)$);
    \coordinate (IPP) at ($(IMM) + (4*\DX,0)$);
    \coordinate (IPPP) at ($(IMM) + (5*\DX,0)$);

    \coordinate (IMM0) at ($(IMM) + (0.0,-\DY)$);
    \coordinate (IM0) at ($(IM) + (0.0,-\DY)$);
    \coordinate (I0) at ($(I) + (0.0,-\DY)$);

    \coordinate (IM1) at ($(IM) + (0.0,-2*\DY)$);
    \coordinate (I1) at ($(I) + (0.0,-2*\DY)$);
    \coordinate (IP1) at ($(IP) + (0.0,-2*\DY)$);

    \coordinate (I2) at ($(I) + (0.0,-3*\DY)$);
    \coordinate (IP2) at ($(IP) + (0.0,-3*\DY)$);
    \coordinate (IPP2) at ($(IPP) + (0.0,-3*\DY)$);

    \coordinate (I3) at ($(I) + (0.0,-4*\DY)$);
    \coordinate (IP3) at ($(IP) + (0.0,-4*\DY)$);
    \coordinate (IPP3) at ($(IPP) + (0.0,-4*\DY)$);
    \coordinate (IPPP3) at ($(IPPP) + (0.0,-4*\DY)$);

    \node at ($(IMM) + (0.0,\DY)$) {(a)};

    \draw (IMM) -- (IPPP);
    \node [circle,fill=black,inner sep=1pt,label=below:\strut$i-2$] at (IMM) {};
    \node [circle,fill=black,inner sep=1pt,label=below:\strut$i-1$] at (IM) {};
    \node [circle,fill=black,inner sep=1pt,label=below:\strut$i$] at (I) {};
    \node [circle,fill=black,inner sep=1pt,label=below:\strut$i+1$] at (IP) {};
    \node [circle,fill=black,inner sep=1pt,label=below:\strut$i+2$] at (IPP) {};
    \node [circle,fill=black,inner sep=1pt,label=below:\strut$i+3$] at (IPPP) {};

    \draw [dashed] ($0.5*(I) + 0.5 *(IP) + (0.0,0.5*\DY)$) node [above] {$i + \frac{1}{2}$} -- ($0.5*(I) + 0.5*(IP) + (0.0,-4.25*\DY)$);
    \node at ($0.5*(I) + 0.5 *(IP) + (-0.4*\DY,0.4*\DY)$) {$-$};
    \node at ($0.5*(I) + 0.5 *(IP) + (0.4*\DY,0.4*\DY)$) {$+$};

    \draw (IMM0) -- (I0);
    \node [circle,fill=black,inner sep=1pt] at (IMM0) {};
    \node [circle,fill=black,inner sep=1pt] at (IM0) {};
    \node [circle,fill=black,inner sep=1pt,label=below right:$S_0$] at (I0) {};
    \draw (IM1) -- (IP1);
    \node [circle,fill=black,inner sep=1pt] at (IM1) {};
    \node [circle,fill=black,inner sep=1pt] at (I1) {};
    \node [circle,fill=black,inner sep=1pt,label=below right:$S_1$] at (IP1) {};
    \draw (I2) -- (IPP2);
    \node [circle,fill=black,inner sep=1pt] at (I2) {};
    \node [circle,fill=black,inner sep=1pt] at (IP2) {};
    \node [circle,fill=black,inner sep=1pt,label=below right:$S_2$] at (IPP2) {};
    \draw (I3) -- (IPPP3);
    \node [circle,fill=black,inner sep=1pt] at (I3) {};
    \node [circle,fill=black,inner sep=1pt] at (IP3) {};
    \node [circle,fill=black,inner sep=1pt] at (IPP3) {};
    \node [circle,fill=black,inner sep=1pt,label=below right:$S_3$] at (IPPP3) {};

    \coordinate (x0) at ($(IPPP) + (1.25*\DX,-4*\DY)$);
    \coordinate (x1) at ($(x0) + (5.5*\DX,0.0)$);
    \coordinate (y0) at ($0.5*(x0)+0.5*(x1)$);
    \coordinate (y1) at ($(y0) + (0.0,4.5*\DY)$);

    \node at ($(x0) + (0.0,5*\DY)$) {(b)};

    \node [circle,fill=black,inner sep=1pt,label=below:\strut$i$] at ($(x0) + (0.75*\DX,0.0)$) {};
    \node [label=below:\strut$i + \frac{1}{2}$] at (y0) {};
    \node [circle,fill=black,inner sep=1pt,label=below:\strut$i+1$] at ($(x1) - (0.75*\DX,0.0)$) {};

    \node at ($(y0) + (-2.5*\DY,1.25*\DY)$) {$\mathbf{U}_{i+\frac{1}{2}}^{-} = \mathbf{U}^L$};
    \node at ($(y0) + (-0.75*\DY,3.0*\DY)$) {$\mathbf{U}^{*L}$};
    \node at ($(y0) + (2.0*\DY,3.0*\DY)$) {$\mathbf{U}^{*R}$};
    \node at ($(y0) + (2.5*\DY,1.25*\DY)$) {$\mathbf{U}_{i+\frac{1}{2}}^{+} = \mathbf{U}^R$};

    \draw[->] (x0) -- (x1) node [below] {$x$};
    \draw[->] (y0) -- (y1) node [left] {$t$};
    \draw[line width=1pt] (y0) -- ($(y0) + (-2.0*\DY,4.0*\DY)$) node [above] {$S^L$};
    \draw[line width=1pt, dashed] (y0) -- ($(y0) + (1.5*\DY,4.0*\DY)$) node [above] {$S^*$};
    \draw[line width=1pt] (y0) -- ($(y0) + (3.5*\DY,4.0*\DY)$) node [above] {$S^R$};

\end{tikzpicture}
    \caption{(a) Schematic of the spatial reconstruction 
    and (b) schematic of the HLLC approximate solution to the resulting Riemann problem.
    (a): The stencils $S_k$ corresponding to the interpolants $p_{k,i+\frac{1}{2}}^{-}$
    for WENO5-Z and TENO6-A reconstruction are shown.
    (b): The corresponding Riemann problem is solved by the HLLC Riemann solver.
    The emanating wave structure ($s^L, s^*, s^R$) and the resulting intermediate start states 
    $\mathbf{U}^{*L}$ and $\mathbf{U}^{*R}$ are visualized.}
    \label{fig:weno_teno}
\end{figure}
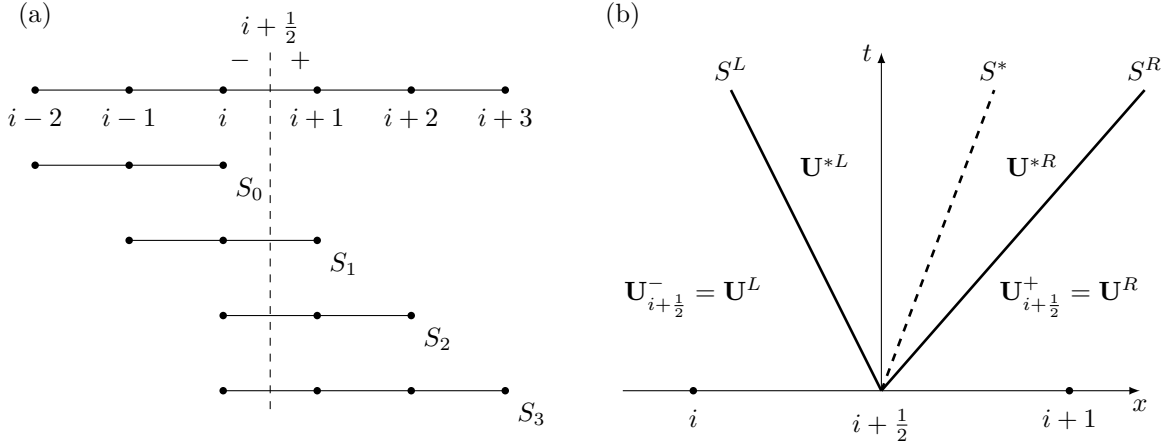
\begin{enumerate}
    \item We approximate the matrix of right eigenvectors $\mathbf{K}$ of the Jacobian
    $\mathbf{A}\left( \mathbf{W} \right) = \left( \partial \mathbf{U} / \partial \mathbf{W} \right)^{-1} \partial \mathbf{F}^c / \partial \mathbf{W}$ 
    at the cell face $x_{i+\frac{1}{2}}$, 
    see \ref{subsec:appendix_char_decomp}.
    The matrix of right eigenvectors is evaluated for the frozen state $\mathbf{W}_{i+\frac{1}{2}}$,
    i.e., $\mathbf{K}_{i+\frac{1}{2}} = \mathbf{K}_{i+\frac{1}{2}} \left( \mathbf{W}_{i+\frac{1}{2}} \right)$.
    Typically, we use an arithmetic average of the neighboring cell center values $\mathbf{W}_{i}$ and $\mathbf{W}_{i+1}$
    to compute $\mathbf{W}_{i+\frac{1}{2}}$.
    Roe averaging is also possible.
    \item The matrix of left eigenvectors $\mathbf{K}_{i+\frac{1}{2}}^{-1}$ 
    computed at the cell face $x_{i+\frac{1}{2}}$ is then used 
    to project the primitive variables in a given reconstruction stencil (see next step) 
    into characteristic space.
    \begin{equation}
        \mathbf{\tilde{W}}_i = \mathbf{K}_{i+\frac{1}{2}}^{-1} \mathbf{W}_i
    \end{equation}
    \item We then apply WENO and TENO shock-capturing schemes for cell face reconstruction in characteristic space.
    In particular, the reconstruction procedure is applied independently for each component of the vector of characteristic variables.
    Given the cell-averaged values $\bar{v}_i$ of a function $v(x)$ within a $K$-point global stencil 
    $S^{K} = \bigcup_{k=0}^{r-1} S_k$,
    WENO/TENO schemes assemble the reconstructed cell face value as a solution-adaptive 
    convex combination of Harten-type interpolation polynomials on candidate sub-stencils $S_k$.
    Figure \ref{fig:weno_teno} shows a schematic of the reconstruction stencils for the cell face
    at $x_{i+\frac{1}{2}}$.
    The interpolated cell face values from the left and from the right $v_{i+\frac{1}{2}}^{\mp}$ are obtained as
    \begin{equation}
        v_{i+\frac{1}{2}}^{\mp} = \sum_{k=0}^{r-1} \omega_{k,i+\frac{1}{2}}^{\mp} p_{k,i+\frac{1}{2}}^{\mp}, \quad
        p_{k,i+\frac{1}{2}}^{\mp} = \sum_{l=0}^{r_k - 1} c_{k,l}^{\mp} \bar{v}_{i-k+l},
    \end{equation}
    where $p_{k,i+\frac{1}{2}}^{\mp}$ are the interpolation polynomials evaluated at $x_{i+\frac{1}{2}}$ and
    $\omega_{k,i+\frac{1}{2}}^{\mp}$ are the corresponding nonlinear solution-adaptive weights.
    In this work, we use WENO5-Z \cite{Borges2008a} and TENO6-A \cite{Fu2019} schemes.
    For WENO5-Z $S^{5} = \left\{ S_0, S_1, S_2 \right\}$ and $r_k = \left\{ 3,3,3 \right\}$,
    and for TENO6-A $S^{6} = \left\{ S_0, S_1, S_2, S_3 \right\}$ and $r_k = \left\{ 3,3,3,4 \right\}$, 
    see Fig. \ref{fig:weno_teno}.
    The coefficients $c_{k,l}^{\mp}$ are grid-dependent, and general expressions for calculating these are given in \cite{Shu1998b}.
    Relations for the interpolation polynomials $p_{k,i+\frac{1}{2}}^{\mp}$ can be found in \cite{Jiang1996,Shu1998b,Borges2008a,Fu2019},
    and definitions for the nonlinear weights $\omega_{k,i+\frac{1}{2}}^{\mp}$ for the WENO5-Z
    and TENO6-A schemes are given in \cite{Borges2008a} and \cite{Fu2019}, respectively.
    \item The cell face reconstructed values are projected back into physical space by 
    $\mathbf{W}_{i+\frac{1}{2}}^{\mp} = \mathbf{K}_{i+\frac{1}{2}} \mathbf{\tilde{W}}_{i+\frac{1}{2}}^{\mp}$,
    and the corresponding conservative states are calculated
    $\mathbf{U}_{i+\frac{1}{2}}^{\mp}=\mathcal{L}_{\mathbf{W}\rightarrow\mathbf{U}}(\mathbf{W}_{i+\frac{1}{2}}^{\mp})$.
    \item When the five-equation diffuse-interface model is used, 
    we combine the aforementioned shock-capturing methods with a THINC-type interface-capturing method \cite{Xiao2005,Xiao2011,Shyue2014}.
    THINC prevents excessive smearing of the interface. 
    THINC reconstruction is used only in interface cells, i.e., cells with $\epsilon_{\alpha} \leq \alpha_{1,i} \leq 1 - \epsilon_{\alpha}$
    and $(\alpha_{1,i+1} - \alpha_{1,i}) (\alpha_{1,i} - \alpha_{1,i-1}) > 0$.
    We use $\epsilon_{\alpha} = 10^{-4}$ in this work.
    In cells away from the interface, we still resort to shock-capturing methods.
    For a given cell $(i)$, THINC assumes the volume fraction field to follow a 
    locally fitted tangent hyperbolic profile
    \begin{align}
        \alpha_{1,i}^{THINC}(x) = \frac{1}{2} \left[ 1 + \tanh \left( \beta_i \left( \frac{x - x_{i-\frac{1}{2}}}{\Delta x_i} + x_{c,i} \right) \right) \right], \quad x \in \left[x_{i-\frac{1}{2}}, x_{i+\frac{1}{2}}\right],
    \end{align}
    where the parameter $\beta_i = \beta \vert n_{x,i} \vert + 0.01$ determines the interface thickness 
    in the direction of reconstruction (here $x$-direction) \cite{Xiao2011}.
    $n_{x,i}$ is the corresponding component of the interface normal $\mathbf{n}$,
    and $x_{c,i} = $ gives the location of the center of the interface.
    We choose $\beta = 1.0$.
    The interpolated cell face values from the left and from the right are
    \begin{align}
        \alpha_{1,i+\frac{1}{2}}^{THINC,-} &= \frac{1}{2} \left[ 1 + \tanh \left( \beta_i \left( 1 + x_{c,i} \right) \right) \right], \\
        \alpha_{1,i+\frac{1}{2}}^{THINC,+} &= \frac{1}{2} \left[ 1 + \tanh \left( \beta_{i+1} x_{c,{i+1}} \right) \right].
    \end{align}
    The reconstructed phase densities $(\alpha_1 \rho_1)_{i+\frac{1}{2}}^{\mp}$ and $(\alpha_2 \rho_2)_{i+\frac{1}{2}}^{\mp}$ are adapted consistently, see \cite{Garrick2017}.
    In the present implementation, velocities and pressure are not affected by the THINC routine 
    and are still reconstructed by shock-capturing schemes.
    \item The local Riemann problem at the cell face $x_{i + \frac{1}{2}}$ is approximately solved by the HLLC Riemann solver \cite{Toro1994,Toro2019},
    $\mathbf{F}^c_{i+\frac{1}{2}} = \mathbf{F}^{\text{HLLC}} \left( \mathbf{U}_{i+\frac{1}{2}}^{-}, \mathbf{U}_{i+\frac{1}{2}}^{+} \right)$, 
    where the two-argument function $\mathbf{F}^{\text{HLLC}} \left( \mathbf{U}^{L}, \mathbf{U}^{R} \right)$ is the numerical flux associated with the HLLC solver.
    States and fluxes left and right of the cell face are denoted by the superscript $K=L,R$.
    The HLLC flux is
    \begin{align}
        \mathbf{F}^{\text{HLLC}} &= 
        \frac{1 + \sign_math \left( s^* \right)}{2} \left( \mathbf{F}^{L} + s^{-} \left( \mathbf{U}^{*L} - \mathbf{U}^{L} \right) \right) \\
        &+ \frac{1 - \sign_math \left( s^* \right)}{2} \left( \mathbf{F}^{R} + s^{+} \left( \mathbf{U}^{*R} - \mathbf{U}^{R} \right) \right), \notag
    \end{align}
    where $\mathbf{F}^{K} = \mathbf{F}^c \left( \mathbf{U}^{K} \right)$, and $\mathbf{U}^{*K}$ is the intermediate star state
    given by 
    \begin{align}
        \mathbf{U}^{*K} = \left( \frac{s^K - u^K}{s^K - s^*} \right) 
        \begin{bmatrix}
            \rho^K \\
            \rho^K s^* \\
            \rho^K v^K \\
            \rho^K w^K \\
            E^K + (s^* - u^K) \left( \rho^K s^* + \frac{p^K}{s^K - u^K} \right)
        \end{bmatrix}
    \end{align}
    for the single-phase Riemann problem and by
    \begin{align}
        \mathbf{U}^{*K} = \left( \frac{s^K - u^K}{s^K - s^*} \right) 
        \begin{bmatrix}
            \left( \alpha_1 \rho_1 \right)^K  \\
            \left( \alpha_2 \rho_2 \right)^K  \\
            \rho^K s^* \\
            \rho^K v^K \\
            \rho^K w^K \\
            E^K + (s^* - u^K) \left( \rho^K s^* + \frac{p^K}{s^K - u^K} \right) \\
            \alpha_1^K
        \end{bmatrix}
    \end{align}
    for the two-phase diffuse-interface Riemann problem, respectively.
    The relations for the wave speeds ($s^{-}, s^{+}, s^L, s^K, s^{*}$)
    are given in \ref{subsec:appendix_hllc}.
    Following \cite{Johnsen2006,Coralic2014}, the normal component of the velocity at the cell face is consistently upwinded according to
    \begin{align}
        u_{i+\frac{1}{2}}^{\text{HLLC}} &= \frac{1 + \sign_math \left( s^* \right)}{2} \left[ u^{L} + s^{-} \left( \frac{s^{L} - u^{L}}{s^{L} - s^{*}} - 1 \right) \right] \\
        &+ \frac{1 - \sign_math \left( s^* \right)}{2} \left[ u^{R} + s^{+} \left( \frac{s^{R} - u^{R}}{s^{R} - s^{*}} - 1 \right) \right], \notag
    \end{align}
\end{enumerate}
The diffusive fluxes $\mathbf{F}^d_{i \pm \frac{1}{2},j,k}$, $\mathbf{G}^d_{i,j \pm \frac{1}{2},k}$,
and $\mathbf{H}^d_{i,j,k \pm \frac{1}{2}}$ are calculated via central finite differences, 
see Sec. 3.4 in \cite{Bezgin2022} for more details regarding the implementation and 
\ref{sec:appendix_overview_numerical_models} for an overview on available finite-difference stencils.
\subsection{Positivity-preserving scheme}
\label{subsec:positivity_preserving}
Negative or zero values for density or internal energy are non-physical.
Additionally, in the DIM, negative phase densities are non-physical,
and the volume fraction has to be bounded between zero and one.
States which fulfill these requirements are said to be physically admissible.
For the stiffened gas EOS, positive internal energy corresponds
to a positive squared speed of sound, see Eq. \eqref{eq:stiffened_gas_eos}.
The squared speed of sound can conveniently be evaluated by the helper variable $\rho c^2 = \gamma (p + p_{\infty})$.
Therefore, we can say that a given flow state is physically admissible if
\begin{align}
    &\text{Single-phase \& SIM: } \left\{ \mathbf{W} \vert \rho > 0, \rho c^2 > 0 \right\}, \\
    &\text{DIM: } \left\{ \mathbf{W} \vert 0 \leq \alpha_1 \leq 1, \alpha_1 \rho_1 > 0, \alpha_2 \rho_2 > 0, \rho c^2 > 0 \right\}.
\end{align}
In general, high-order shock-capturing schemes and 
corresponding high-order HLLC fluxes (see Sec. \ref{subsec:high_order_godunov}) are not positivity-preserving
and may produce physically non-admissible states.
To this end, we use interpolation and flux limiters \cite{Hu2013,Wong2021}
to ensure physically admissible states at all times during the simulation.
The positivity-preserving scheme consists of two successive limiters which we briefly outline below.
We refer to \cite{Hu2013,Wong2021} for more details.
\begin{enumerate}
    \item Interpolation limiter (IL):
    First-order (FO) spatial reconstruction, i.e., $\mathbf{U}_{i+\frac{1}{2}}^{-, \text{FO}} = \mathbf{U}_i$ 
    and $\mathbf{U}_{i+\frac{1}{2}}^{+, \text{FO}} = \mathbf{U}_{i+1}$,
    yields physically admissible reconstructed states given physically admissible cell center values $\mathbf{U}_i$ and $\mathbf{U}_{i+1}$.
    Therefore, we can limit the high-order reconstruction, i.e., WENO/TENO reconstruction (+ THINC for DIM),
    by locally switching to first-order reconstruction.
    For single-phase and sharp-interface simulations, the interpolation limiting process consists of two independent loops.
    Firstly, we check whether the reconstructed density violates positivity.
    If the reconstructed density at a given cell face violates the user-specified tolerance,
    i.e., $\rho_{i+\frac{1}{2}}^{-} < \epsilon_{\rho}$, the interpolation limiter is activated, 
    $\delta_{i+\frac{1}{2}}^{\text{IL},-} = 1$,
    and we locally switch from high-order to first-order reconstruction.
    Otherwise, $\delta_{i+\frac{1}{2}}^{\text{IL},-} = 0$.
    \begin{align}
        \mathbf{U}_{i+\frac{1}{2}}^{-} 
        = \left( 1 - \delta_{i+\frac{1}{2}}^{\text{IL},-} \right) \mathbf{U}_{i+\frac{1}{2}}^{-} 
        + \delta_{i+\frac{1}{2}}^{\text{IL},-} \mathbf{U}_{i+\frac{1}{2}}^{-,\text{FO}}
    \end{align}
    Secondly, this process is repeated for the squared speed of sound with the corresponding tolerance $\epsilon_{\rho c^2}$.
    In the DIM, the limiting process consists of three independent loops over partial densities $\rho_i \alpha_i$ (with $\epsilon_{\rho}$),
    volume fraction $\alpha_1$ (with $\epsilon_{\alpha}$), and squared speed of sound (with $\epsilon_{\rho c^2}$), respectively.
    We obtain physically admissible reconstructed states $\mathbf{U}_{i+\frac{1}{2}}^{-}$ and $\mathbf{W}_{i+\frac{1}{2}}^{-}$, respectively.
    \item Flux limiter (FL): 
    The first-order HLLC flux $\mathbf{F}_{i + \frac{1}{2}}^{\text{HLLC, FO}} = \mathbf{F}^{\text{HLLC}} \left( \mathbf{U}_i, \mathbf{U}_{i+1} \right)$
    is positivity- and boundedness-preserving (e.g., \cite{Wong2021}).
    We use the high-order HLLC flux and perform pseudo-time integration.
    Should the resulting state after integration in pseudo-time violate positivity (or boundedness),
    we can limit the high-order HLLC flux by locally switching to the first-order HLLC flux.
    Similarly to the interpolation limiter, for single-phase and sharp-interface simulations, 
    the flux limiting process consists of independent loops
    over density and squared speed of sound.
    First, we conduct the flux limiting procedure for the density.
    Should the density after pseudo-time integration violate the positivity threshold $\epsilon_{\rho}$,
    we activate the flux limiter, $\delta_{i+\frac{1}{2}}^{\text{FL}} = 1$, and we limit the flux.
    Otherwise, $\delta_{i+\frac{1}{2}}^{\text{FL}} = 0$.
    \begin{align}
        \mathbf{F}_{i+\frac{1}{2}}^{\text{HLLC}} = 
        \left( 1 - \delta_{i+\frac{1}{2}}^{\text{FL}} \right) \mathbf{F}_{i+\frac{1}{2}}^{\text{HLLC}} 
        + \delta_{i+\frac{1}{2}}^{\text{FL}} \mathbf{F}_{i+\frac{1}{2}}^{\text{HLLC,FO}}
    \end{align}
    Second, this process is repeated for the squared speed of sound with $\epsilon_{\rho c^2}$.
    In the DIM, the flux limiting process consists of independent loops 
    over partial densities (with $\epsilon_{\rho}$), volume fraction (with $\epsilon_{\alpha}$), 
    and squared speed of sound (with $\epsilon_{\rho c^2}$).
    Here, the upwinded normal velocities $u^{\text{HLLC}}$ are limited consistently.
    Finally, we obtain a positivity limited flux which we use for time integration. 
\end{enumerate}
Contrary to \cite{Hu2013,Wong2021}, we do not blend high-order and first-order interpolations/fluxes,
but instead we use a binary switch $\delta_{i+\frac{1}{2}}^{\text{IL}/\text{FL}}$.
This simplifies implementation and, in our tests, does not influence solution accuracy.
If multistep Runge-Kutta time integration schemes are used, the limiting procedure is applied at every sub-step.
By default, we activate positivity-preserving limiters for all two-phase calculations.
For single-precision calculations, we set the tolerances as $\epsilon_{\rho} = 10^{-10}, \epsilon_{\rho c^2} = 10^{-8}, \epsilon_{\alpha} = 10^{-10}$,
and we choose $\epsilon_{\rho} = 10^{-12}, \epsilon_{\rho c^2} = 10^{-10}, \epsilon_{\alpha} = 10^{-12}$ for double-precision calculations.
\subsection{Temporal integration}
\label{subsec:temporal_integration}
The semi-discrete system of equations is integrated in time according to
\begin{equation}
    \mathbf{U}^{n+1} = \mathcal{I}\left(\mathbf{U}^n,\mathcal{R}_\text{NSE}(\mathbf{U}^n), \Delta t^n\right),
\end{equation}
where the operator $\mathcal{I}(\cdot)$ denotes an explicit time-integration
scheme. The time step and time step size are denoted by the
superscript $n$ and by $\Delta t$, respectively.
A common choice is the third-order total variation diminishing
Runge-Kutta scheme $\mathcal{I}_\text{RK3}$ \cite{Gottlieb1998a}
\begin{align}
    \mathbf{U}^{(1)}   &= \mathbf{U}^{n} + \Delta t \mathbf{\mathcal{R}_\text{NSE}} \left( \mathbf{U}^{n} \right), \\
    \mathbf{U}^{(2)}   &= \frac{3}{4} \mathbf{U}^{n} + \frac{1}{4} \mathbf{U}^{(1)} + \frac{1}{4} \Delta t \mathbf{\mathcal{R}_\text{NSE}} \left( \mathbf{U}^{(1)} \right), \\
    \mathbf{U}^{n + 1} &= \frac{1}{3} \mathbf{U}^{n} + \frac{2}{3} \mathbf{U}^{(2)} + \frac{2}{3} \Delta t \mathbf{\mathcal{R}_\text{NSE}} \left( \mathbf{U}^{(2)} \right).
\end{align}
We compute the time step size $\Delta t$ according to a
CFL criterion accounting for inviscid and viscous contributions,
as detailed in \cite{Hoppe2022}.
Table \ref{tab:NumericalMethods} gives an overview on
time integration schemes implemented in JAX-Fluids.
\subsection{Single-phase model (SPM)}
\label{subsec:single_phase_model_num}
The solution of the single-phase equations is straightforward.
The steps outlined in the previous Sec. \ref{subsec:high_order_godunov} can be executed
with the appropriate definitions of state and flux vectors given in Sec. \ref{subsec:single_phase_model_phys}.
\subsection{Level-set model (LSM)}
\label{subsec:sharp_interface_model_num}
\begin{figure}[t]
    \centering
    \begin{tikzpicture}

    \definecolor{color0}{RGB}{0,101,189}
    \definecolor{color1}{RGB}{227,114,34}
    \definecolor{color2}{RGB}{162,173,0}

    \coordinate (NULL) at (0,0);
    \coordinate (B) at (5,5);
    \coordinate (C) at ($0.5*(B)$);
    \coordinate (D1) at ($0.125*(B)$);
    \coordinate (D2) at ($0.71*(B)$);

    \coordinate (D11) at ($0.1*(B)$);

    \coordinate (L) at ($1.1*(B|-NULL)+0.96*(B-|NULL)$);

    \filldraw[fill=black!10!white, line width=0.1pt] (NULL) -- (B|-NULL) -- ($(B|-NULL) + 0.16*(B-|NULL)$) -- ($(B|-NULL) + 0.16*(B-|NULL)$) arc (33.5:56.4:15) -- (B-|NULL) -- (NULL);

    \draw[line width=1pt] (0,0) rectangle (B);
    \draw[line width=0.5pt] (C |- NULL) -- (C |- B);
    \draw[line width=0.5pt] (C -| NULL) -- (C -| B);
    \draw[fill=red, draw=red] (C) circle (0.1);

    \draw[dashed] ($(C)+(C-|NULL)$) -- ($(C)+(B-|NULL)$);
    \draw[dashed] ($(C)+(C|-NULL)$) -- ($(C)+(B|-NULL)$);
    \draw[dashed] ($(C)-(C-|NULL)$) -- ($(C)-(B-|NULL)$);
    \draw[dashed] ($(C)-(C|-NULL)$) -- ($(C)-(B|-NULL)$);
    \draw[dashed] ($3*(C|-NULL)-(NULL|-C)$) -- ($-1*(C)$) -- ($3*(C-|NULL)-(NULL-|C)$) -- ($3*(C)$) -- cycle;

    \draw[fill=black, draw=black] (C) circle (0.1) node[below left] {$(i,j,k)$};
    \draw[fill=black, draw=black] ($(C)+(B-|NULL)$) circle (0.1) node[below right] {$(i,j+1,k)$};
    \draw[fill=black, draw=black] ($(C)+(B|-NULL)$) circle (0.1) node[below right] {$(i+1,j,k)$};
    \draw[fill=black, draw=black] ($(C)-(B|-NULL)$) circle (0.1) node[below left] {$(i-1,j,k)$};
    \draw[fill=black, draw=black] ($(C)-(B-|NULL)$) circle (0.1) node[below right] {$(i,j-1,k)$};
    \draw[fill=black, draw=black] ($-1*(C)$) circle (0.1) node[below left] {$(i-1,j-1,k)$};
    \draw[fill=black, draw=black] ($3*(C)$) circle (0.1) node[below right] {$(i+1,j+1,k)$};
    \draw[fill=black, draw=black] ($3*(C-|NULL)-(NULL-|C)$) circle (0.1) node[below left] {$(i-1,j+1,k)$};
    \draw[fill=black, draw=black] ($3*(C|-NULL)-(NULL|-C)$) circle (0.1) node[below right] {$(i+1,j-1,k)$};

    \draw[blue, line width=1pt] ($1.1*(B|-NULL)$) arc (30:60:15); \label{tikz:interface}

    \draw[red, line width=1pt] (B -| D11) -- (D11 -| B) node[at start, below, yshift=-0.5cm, xshift=0.1cm] {$\Delta\Gamma_{i,j,k}$}; \label{tikz:interface_reconstruction}



    
    \coordinate (DIST) at (0.5,0.5);
    \dimline[extension start length=1cm, extension end length=1cm,extension style={black}, label style={above=0.5ex}] {(-0.5,0)}{($(NULL|-B) - (0.5,0)$)}{$A^p_{i-\frac{1}{2},j,k}=1.0$};
    \dimline[extension start length=-1cm, extension end length=-1cm,extension style={black}, label style={below=0.5ex}] {(0.0,-0.5)}{($(NULL-|B) - (0.0,0.5)$)}{$A^p_{i,j-\frac{1}{2},k}=1.0$};
    \dimline[extension start length=0.5cm, extension end length=0.5cm,extension style={black}, label style={above=0.5ex}] {($(NULL|-B) + (DIST -| NULL)$)}{($(NULL|-B) + (DIST -| NULL) + (D11|-NULL)$)}{$A^p_{i,j+\frac{1}{2},k}$};
    \dimline[extension start length=-0.5cm, extension end length=-0.5cm,extension style={black}, label style={below=0.5ex}] {($(NULL-|B) + (DIST |- NULL)$)}{($(NULL-|B) + (DIST|- NULL) + (D11-|NULL)$)}{$A^p_{i+\frac{1}{2},j,k}$};



    \draw[->, line width=1pt] (-1,-1) -- (-1,0) node[left] {$y$};
    \draw[->, line width=1pt] (-1,-1) -- (0,-1) node[below] {$x$};
    \node[inner sep=2, circle, draw=black, line width=1pt] at (-1,-1) {};
    \node[inner sep=1, circle, draw=none, fill=black] at (-1,-1) {};
    \node[below left] at (-1,-1) {$z$};

\end{tikzpicture}
    \caption{Schematic finite-volume discretization for cut cell $(i,j,k)$ on a Cartesian grid.
    The black dots represent the cell centers. 
    The blue line indicates the interface,
    and the red line depicts the linear
    approximation of the interface. 
    The figure illustrates a two-dimensional
    slice in the $(x,y)$-plane.
    (For interpretation of the references to color in this figure legend,
    the reader is referred to the web version of this article.)}
    \label{fig:cut_cell}
\end{figure}
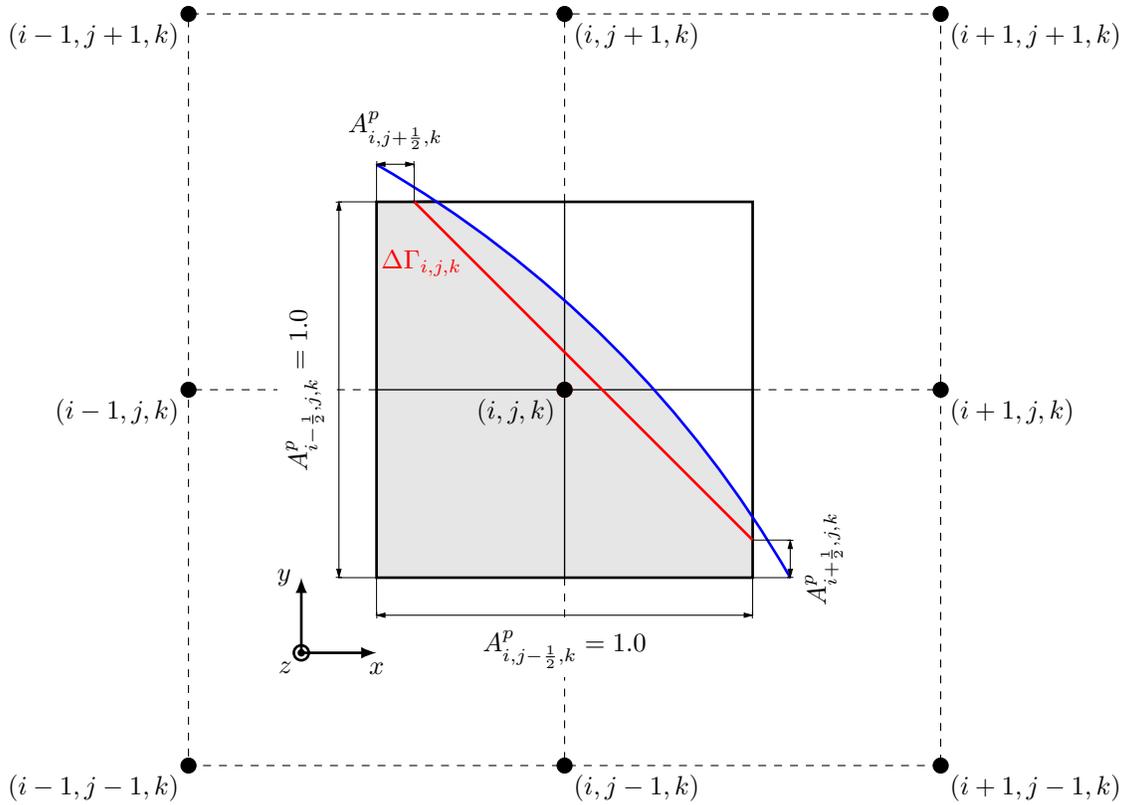

In the level-set method \cite{Osher1988, Sussman1994b}, two distinct phases are separated by a
sharp interface. We track the interface implicitly by a scalar function 
$\phi(\mathbf{x},t)$ that satisfies the signed distance property
$|\nabla \phi| = 1$. The zero
level-set describes the interface location
\begin{equation}
    \Gamma(\mathbf{x},t)=\{\mathbf{x} \ | \ \phi(\mathbf{x},t)=0\}.
\end{equation}
The distinction between the two phases $p\in\{1,2\}$ is given by the sign
of the level-set. Figure \ref{fig:cut_cell} shows a 
finite-volume cell $(i,j,k)$ that contains an interface segment $\Gamma$. 
We refer to such cells as \textit{cut cells}. 
Cells that do not contain an interface segment are referred to as \textit{full cells}.
The gray region in the figure
indicates the volumetric portion $\alpha_{i,j,k}^p$ of the cell that
is occupied by phase $p$ 
(not to be confused with $\alpha_{1,i,j,k}$ which is the volume fraction of fluid/phase 1 in the diffuse-interface method). 
We define apertures
$A_{i\pm\frac{1}{2},j,k}^p$, $A_{i,j\pm\frac{1}{2},k}^p$, and $A_{i,j,k\pm\frac{1}{2}}^p$
as the cell face fractions that are covered by phase $p$.

In the LSM, we solve the Navier-Stokes equations (Eq. \eqref{eq:FVD}) for both phases separately.
For full cells, Eq. \eqref{eq:FVD} remains unchanged.
For cut cells, the following modification to the
equation is made.
\begin{align}
    \begin{split}
        \frac{\text{d}}{\text{d}t} \alpha^p_{i,j,k} \bar{\mathbf{U}}_{i,j,k}^p =
        & - \frac{1}{\Delta x} \left[A^p_{i+\frac{1}{2},j,k} \left(\mathbf{F}^{c,p}_{i+\frac{1}{2},j,k} - \mathbf{F}^{d,p}_{i+\frac{1}{2},j,k}\right) - A^p_{i-\frac{1}{2},j,k} \left(\mathbf{F}^{c,p}_{i-\frac{1}{2},j,k} - \mathbf{F}^{d,p}_{i-\frac{1}{2},j,k} \right) \right] \\
        & - \frac{1}{\Delta y} \left[A^p_{i,j+\frac{1}{2},k} \left(\mathbf{G}^{c,p}_{i,j+\frac{1}{2},k} - \mathbf{G}^{d,p}_{i,j+\frac{1}{2},k}\right) - A^p_{i,j-\frac{1}{2},k} \left(\mathbf{G}^{c,p}_{i,j-\frac{1}{2},k} - \mathbf{G}^{d,p}_{i,j-\frac{1}{2},k} \right) \right] \\
        & - \frac{1}{\Delta z} \left[A^p_{i,j,k+\frac{1}{2}} \left(\mathbf{H}^{c,p}_{i,j,k+\frac{1}{2}} - \mathbf{H}^{d,p}_{i,j,k+\frac{1}{2}}\right) - A^p_{i,j,k-\frac{1}{2}} \left(\mathbf{H}^{c,p}_{i,j,k-\frac{1}{2}} - \mathbf{H}^{d,p}_{i,j,k-\frac{1}{2}} \right) \right] \\
        & + \alpha_{i,j,k} \bar{\mathbf{S}}_{i,j,k}\\
        & - \frac{1}{\Delta x \Delta y \Delta z} \left[\mathbf{X}^{c,p}_{i,j,k} + \mathbf{X}^{d,p}_{i,j,k} \right] = \mathcal{R}_\text{NSE}
        \label{eq:FVD_levelset}
    \end{split}
\end{align}
Here, we weigh the cell-average state $\bar{\mathbf{U}}^p_{i,j,k}$
and cell face fluxes $\mathbf{F}^{c,p}_{\dots},\mathbf{F}^{d,p}_{\dots},
\mathbf{G}^{c,p}_{\dots},\mathbf{G}^{d,p}_{\dots},\mathbf{H}^{c,p}_{\dots},
\mathbf{H}^{d,p}_{\dots}$ with the
volume fraction $\alpha^p_{i,j,k}$ and the corresponding apertures $A^p_{\dots}$, respectively.
The terms $\mathbf{X}^{c,p}$ and $\mathbf{X}^{d,p}$
denote the convective and dissipative interface fluxes.
They read
\begin{equation}
    \mathbf{X}^{c,p} = 
    \begin{bmatrix}
        0 \\
        p_\Gamma \Delta \mathbf{\Gamma}^p \\
        p_\Gamma \Delta \mathbf{\Gamma}^p \cdot \mathbf{u}_\Gamma
    \end{bmatrix}, \qquad
    \mathbf{X}^{d,p} = 
    \begin{bmatrix}
        0 \\
        \mathbf{\tau}_\Gamma^T \Delta \mathbf{\Gamma}^p \\
        (\mathbf{\tau}_\Gamma^T \Delta \mathbf{\Gamma}^p) \cdot \mathbf{u}_\Gamma
    \end{bmatrix}.
    \label{eq:interface_flux}
\end{equation}
Here, $p_\Gamma$ and $\mathbf{u}_\Gamma$ denote the interface pressure
and interface velocity, respectively, and are obtained by solving a two-material
Riemann problem at the interface, as detailed in \ref{app:interface_riemann_problem}.
The projection of the interface segment reads
$\Delta \mathbf{\Gamma}^p = \Delta \Gamma \mathbf{n}_\Gamma^p$,
where the interface normal is given by
$\mathbf{n}_\Gamma = \mathbf{n}_\Gamma^1 = \nabla \phi / |\nabla \phi|$,
and $\Delta \Gamma$
denotes the interface segment.
Interface normals for phases 1 and 2 have
opposite directions $\mathbf{n}_\Gamma^2=-\mathbf{n}_\Gamma^1$.
We compute the interface normal using fourth-order
central finite-differences.
Assuming a linear approximation of the level-set inside a cut cell, we
analytically compute the apertures $A^p$, volume fraction
$\alpha^p$, and interface segment $\Delta \Gamma$,
as described in \ref{app:interface_reconstruction}.

The level-set function is evolved by solving
the level-set advection equation \eqref{eq:levelset_advection}.
We use high-order-upstream-central schemes \cite{Nourgaliev2007} for
the spatial discretization. The temporal integration is performed
with the same scheme as used for the integration of the Navier-Stokes equations (NSE).
In practice, we evolve the level-set function in a narrowband
that extends over a small number of cells around the interface.
It is important to note that, in the current JAX-Fluids version, 
the level-set algorithm requires cut cells and the narrowband to be located
in a homogenous region of the computational grid, i.e., $\Delta x=\Delta y=\Delta z$.

The solution of the level-set advection equation (Eq. \eqref{eq:levelset_advection}) may
violate the signed-distance property of the level-set.
To recover the signed-distance property, we solve the reinitialization equation
\begin{equation}
    \frac{\partial \phi}{\partial \tau} = - \text{sgn}(\phi^0) (|\nabla\phi| - 1) = \mathcal{R}_\text{reinit}
    \label{eq:levelset_reinitialization}
\end{equation}
iteratively in pseudo-time $\tau$ using
a fixed time step size $\Delta \tau=0.25\Delta x$
for a fixed amount of 10 steps. Here,
$\phi^0$ denotes the level-set prior to reinitialization, and
$\text{sgn}(\cdot)$ denotes the sign function.
We discretize the spatial operators by means of third-order
WENO-HJ \cite{Jiang2000} scheme. The temporal evolution is
performed by the second-order Runge-Kutta scheme
(see Sec. \ref{subsec:temporal_integration}). In practice,
we only reinitialize cells where the residual of Eq.
\eqref{eq:levelset_reinitialization} is above the threshold $5\times 10^{-2}$.

The evaluation of stencils close to the interface requires reasonable cell
values of the considered phase on both sides of the interface. We refer
to cells that lie on the opposite side of the interface
as \textit{ghost cells} \cite{Fedkiw1999a}. The ghost cells of one phase 
overlap with the real cells of the other phase.
We perform a constant extrapolation in interface normal
direction to populate the ghost cells with values
from the real cells. 
The extrapolation procedure is performed
on the primitive variables $\mathbf{W}$, and
the extrapolation equation (EE) reads
\begin{equation}
    \frac{\partial \mathbf{W}^p}{\partial \tau} = \mathbf{n}^p_\Gamma \cdot \nabla \mathbf{W} = \mathcal{R}_{\text{ext},\mathbf{W}}.
    \label{eq:extension_prime}
\end{equation}
We solve this equation iteratively in pseudo-time $\tau$
using a fixed time step size $\Delta \tau=0.25\Delta x$ 
and a fixed amount of 10 steps.
The right-hand side is discretized using a first-order upwind
scheme. Temporal integration is performed with a single-step
Euler scheme.

Quantities $\mathbf{Q}_\Gamma$ that are only known at the interface,
i.e., the interface velocity $\mathbf{u}_\Gamma$
and interface pressure $p_\Gamma$, require extrapolation into
the narrowband. We perform a two-way constant extrapolation in
interface normal direction according to
\begin{equation}
    \frac{\partial \mathbf{Q}_\Gamma}{\partial \tau} = -\text{sgn}(\phi)\mathbf{n}_\Gamma \cdot \nabla \mathbf{Q}_\Gamma = \mathcal{R}_{\text{ext},\mathbf{Q}_\Gamma}.
    \label{eq:extension_if}
\end{equation}
We use a fixed time step size $\Delta \tau=0.25\Delta x$ and
a fixed amount of 20 steps
to iteratively solve this equation.
Again, first-order upwind spatial discretization
and single-step Euler integration are used.

The time step size that is computed using the CFL criterion
(see Sec. \ref{subsec:temporal_integration})
based on full cells may be too large for cut cells with 
small volume fractions $\alpha$ and therefore lead
to unstable integration.
Furthermore, as the interface crosses cell faces,
cells are vanishing and new cells are created. This process must be 
accounted for to prevent numerical instability.
We employ a conservative mixing procedure \cite{Hu2006,Lauer2012c}
that treats both of these issues (see \ref{app:mixing}). 

We detail the implementation of the algorithm in Figure \ref{fig:algorithm_haloupdate}.
For further insight on the presented level-set method, the reader is referred
to the implementation of Hoppe et al. \cite{Hoppe2022}.
\subsection{Diffuse-interface model (DIM)}
\label{subsec:diffuse_interface_model_num}
For the five-equation diffuse-interface model, Eq. \eqref{eq:FVD} is solved with the state and flux vectors defined
in Sec. \ref{subsec:diffuse_interface_model_phys}.
Following \cite{Johnsen2006}, the source term in the advection equation (see Eq. \eqref{eq:volume_fraction_eq_conservative}),
\begin{equation*}
    \frac{\partial \alpha_1}{\partial t} + \nabla \cdot \left( \alpha_1 \mathbf{u} \right) = \alpha_1 \nabla \cdot \mathbf{u},
\end{equation*}
is discretized consistently with the convective flux calculation
\begin{align}
    \left( \alpha_1 \nabla \cdot \mathbf{u} \right)_{i,j,k} = \alpha_{1_{i,j,k}} \left[ 
        \frac{u^{\text{HLLC}}_{i+1/2,j,k} - u^{\text{HLLC}}_{i-1/2,j,k}}{\Delta x_{i,j,k}}
        + \frac{v^{\text{HLLC}}_{i,j+1/2,k} - v^{\text{HLLC}}_{i,j-1/2,k}}{\Delta y_{i,j,k}} 
        + \frac{w^{\text{HLLC}}_{i,j,k+1/2} - w^{\text{HLLC}}_{i,j,k-1/2}}{\Delta z_{i,j,k}} 
        \right],
\end{align}
where the velocities used are the ones upwinded by the Riemann solver, see Sec. \ref{subsec:high_order_godunov}.

The THINC reconstruction procedure requires the calculation of the interface normal $\mathbf{n}_{i,j,k}$, see Sec. \ref{subsec:high_order_godunov}.
The interface normal $\mathbf{n}_{i,j,k}$ is computed from a smoothed volume fraction field $\psi$,
\begin{align}
    \psi_{i,j,k} = \frac{ \alpha_{1,i,j,k}^{0.1} }{ \alpha_{1,i,j,k}^{0.1} + \left( 1 - \alpha_{1,i,j,k} \right)^{0.1} },
\end{align}
according to  
\begin{align}
    \mathbf{n}_{i,j,k} = \frac{ \left( \nabla \psi \right)_{i,j,k} }{ \vert \left( \nabla \psi \right)_{i,j,k} \vert },
\end{align}
where we use second-order central differences to evaluate the spatial derivatives.
\begin{table}[t!]
    \footnotesize
    \centering
    \begin{tabular}{r l l}
        \hline
        Notation & Description & Details \\
        \hline
        $\mathcal{I}$ & Explicit time-integration & Section \ref{subsec:temporal_integration} \\
        $\mathcal{L}$ & Transformation between primitive and conservative variables & Section \ref{sec:physical_model} \\
        $\mathcal{R}_\text{NSE}$ & Right-hand side computation of the NSE & Section \ref{subsec:high_order_godunov} \\
        $\mathcal{R}_\text{LSA}$ & Right-hand side computation of the LSA & Section \ref{subsec:sharp_interface_model_num}\\
        $\mathcal{R}_{\text{ext},\mathbf{W}}$ & Right-hand side computation of the EE for primitives & Section \ref{subsec:sharp_interface_model_num} \\
        $\mathcal{R}_{\text{ext},\mathbf{Q}_\Gamma}$ & Right-hand side computation of the EE for interface quantities & Section \ref{subsec:sharp_interface_model_num} \\
        $\mathcal{H}_\text{F},\mathcal{H}_\text{E},\mathcal{H}_\text{V}$ & Face, edge, and vertex halo updates, respectively & Section \ref{subsec:halo_update} \\
        $\mathcal{M}$ & Mixing procedure & \ref{app:mixing}\\
        $\mathcal{Q}_\Gamma$ & Interface quantity computation & \ref{app:interface_riemann_problem}\\
        \hline
    \end{tabular}
    \caption{Notation and description of relevant
    operators/functions that are executed within
    a single Runge-Kutta stage of JAX-Fluids.}
    \label{tab:function_description}
\end{table}

\begin{table}[t!]
    \centering
    \begin{tabular}{r l}
        \hline
        Notation & Description \\
        \hline
        $\mathbf{U}$ & Conservative variables  \\
        $\mathbf{W}$ & Primitive variables  \\
        $\phi$ & Level-set  \\
        $\mathbf{Q}_\Gamma$ & Interface quantities  \\
        \hline
    \end{tabular}
    \caption{Notation and description of relevant buffers.}
    \label{tab:buffer_description}
\end{table}
\section{Parallelization strategy}
\label{sec:parallelization_strategy}
\subsection{JAX-specific aspects}
\label{subsec:jax_specific}
Requirements for the parallelization of the JAX-Fluids CFD code are high-performance and differentiability.
Although there are third-party packages to parallelize JAX code, e.g., \mintinline{python}{mpi4jax} \cite{Hafner2021b},
we use JAX inherent tools to implement our parallelization strategy.
This ensures that differentiability is maintained while
optimal performance is achieved. In particular, we use
\mintinline{python}{jax.pmap}, which expresses single-program multiple-data code,
to transform the compute-intensive functions.
Similarly to \mintinline{python}{jax.jit}, \mintinline{python}{jax.pmap} will compile the
function with the XLA (Accelerated Linear Algebra) (e.g., \cite{Frostig2018}) compiler 
and subsequently execute the function in parallel on the specified devices. 
Furthermore, we use \mintinline{python}{jax.lax.ppermute}
to perform collective permutations of data between devices. 
\mintinline{python}{jax.lax.ppermute}
requires each device to send data to and receive data from exactly one other device.

\subsection{Homogeneous domain decomposition}
\label{subsec:domain_decomposition}
\begin{figure}[!t]
    \centering
    \begin{tikzpicture}

    \coordinate (O) at (0,0);
    \coordinate (H) at (2.5,0.0);
    \coordinate (V) at (0.0,2.5);
    \coordinate (H1) at ($0.5*(H)$);
    \coordinate (V1) at ($0.5*(V)$);

    \coordinate (H2) at (0.75,0.75);
    \draw[line width=1pt,->] ($(-2.5,0)+(H2)$) -- ($(-1.5,0)+(H2)$) node[right] {x};
    \draw[line width=1pt,->] ($(-2.5,0)+(H2)$) -- ($(-2.5,1)+(H2)$) node[right] {y};
    \draw[line width=1pt,->] ($(-2.5,0)+(H2)$) -- ($(-3.0,-0.5)+(H2)$) node[above] {z};

    \node[inner sep=0, anchor=south west] (B111) at (O) {
        \begin{tikzpicture}
            \coordinate (O) at (0,0);
            \coordinate (L) at (1.5,1.5);

            \coordinate (L1) at ($0.5*(L)$);
            \coordinate (L2) at ($3*(L1)$);
        
            \draw[line width=1pt] (O) rectangle (L);
            \draw[line width=1pt] (L1) rectangle (L2);
            \draw[line width=1pt] (L) -- (L2);
            \draw[line width=1pt] (L|-O) -- (L2|-L1);
            \draw[line width=1pt] (O) -- (L1);
            \draw[line width=1pt] (L-|O) -- (L2-|L1);

        \end{tikzpicture}
    };

    \node[inner sep=0, anchor=south west] (B121) at ($(O)+(H)$) {
        \begin{tikzpicture}
            \coordinate (O) at (0,0);
            \coordinate (L) at (1.5,1.5);
        
            \coordinate (L1) at ($0.5*(L)$);
            \coordinate (L2) at ($3*(L1)$);
        
            \draw[line width=1pt] (O) rectangle (L);
            \draw[line width=1pt] (L1) rectangle (L2);
            \draw[line width=1pt] (L) -- (L2);
            \draw[line width=1pt] (L|-O) -- (L2|-L1);
            \draw[line width=1pt] (O) -- (L1);
            \draw[line width=1pt] (L-|O) -- (L2-|L1);

        \end{tikzpicture}
    };

    \node[inner sep=0, anchor=south west] (B211) at ($(O)+(V)$) {
        \begin{tikzpicture}
            \coordinate (O) at (0,0);
            \coordinate (L) at (1.5,1.5);
        
            \coordinate (L1) at ($0.5*(L)$);
            \coordinate (L2) at ($3*(L1)$);
        
            \draw[line width=1pt] (O) rectangle (L);
            \draw[line width=1pt] (L1) rectangle (L2);
            \draw[line width=1pt] (L) -- (L2);
            \draw[line width=1pt] (L|-O) -- (L2|-L1);
            \draw[line width=1pt] (O) -- (L1);
            \draw[line width=1pt] (L-|O) -- (L2-|L1);
        
        \end{tikzpicture}
    };

    \node[inner sep=0, anchor=south west] (B221) at ($(O)+(V)+(H)$) {
        \begin{tikzpicture}
            \coordinate (O) at (0,0);
            \coordinate (L) at (1.5,1.5);
        
            \coordinate (L1) at ($0.5*(L)$);
            \coordinate (L2) at ($3*(L1)$);
        
            \draw[line width=1pt] (O) rectangle (L);
            \draw[line width=1pt] (L1) rectangle (L2);
            \draw[line width=1pt] (L) -- (L2);
            \draw[line width=1pt] (L|-O) -- (L2|-L1);
            \draw[line width=1pt] (O) -- (L1);
            \draw[line width=1pt] (L-|O) -- (L2-|L1);

            \draw[line width=1pt, <->] ([yshift=0.2cm]L1|-L2) -- ([yshift=0.2cm]L2) node[midway, above] {$N_x/S_x$};
            \draw[line width=1pt, <->] ([xshift=0.2cm]L2|-L1) -- ([xshift=0.2cm]L2) node[midway, right] {$N_y/S_y$};
            \draw[line width=1pt, <->] ([xshift=0.2cm]L|-O) -- ([xshift=0.2cm]L2|-L1) node[midway, right, xshift=0.15cm , yshift=0.0cm] {$N_z/S_z$};
        \end{tikzpicture}
    };

\end{tikzpicture}
    \caption{Exemplary domain decomposition with four blocks arranged
    in a $2\times2\times1$ grid.}
    \label{fig:domain_decomposition}
\end{figure}
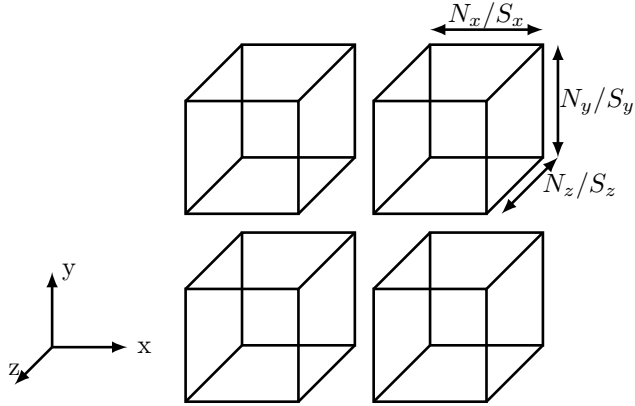

For single device simulations, the computational domain in JAX-Fluids has the shape
$[N_x+2N_h,N_y+2N_h,N_z+2N_h]$, where $N_i$, $i\in {x,y,z}$ represents
the number of cells in the spatial directions.
$N_h$ is the number of halo cells.
We decompose the computational domain into a $S_x\times S_y\times S_z$ grid of equally sized blocks. 
Here, $S_i$, $i\in {x,y,z}$ represents
the number of blocks in the spatial directions.
A single block has the shape $[N_x/S_x+2N_h,N_y/S_y+2N_h,N_z/S_y+2N_h]$.
For multi-device simulations, the shape of the entore computational domain becomes $[S_T,N_x/S_x+2N_h,N_y/S_y+2N_h,N_z/S_y+2N_h]$,
where the leading array axis has length $S_T = S_xS_yS_z$. Figure \ref{fig:domain_decomposition} shows
an exemplary domain decomposition with four blocks arranged in a $2\times2\times1$
grid (halo cells are not shown). The transformation \mintinline{python}{jax.pmap} maps
a function over the leading array axis $S_T$, generating a replication of
the function on each XLA device, and subsequently executing it in parallel.
By \textit{XLA device} one generally refers to any computational device
that can be targeted by the XLA compiler.
In this work, we use the term XLA device synonymously with GPUs and TPUs.

\subsection{Halo update algorithm}
\label{subsec:halo_update}

\begin{figure}[!b]
    \centering
    \input{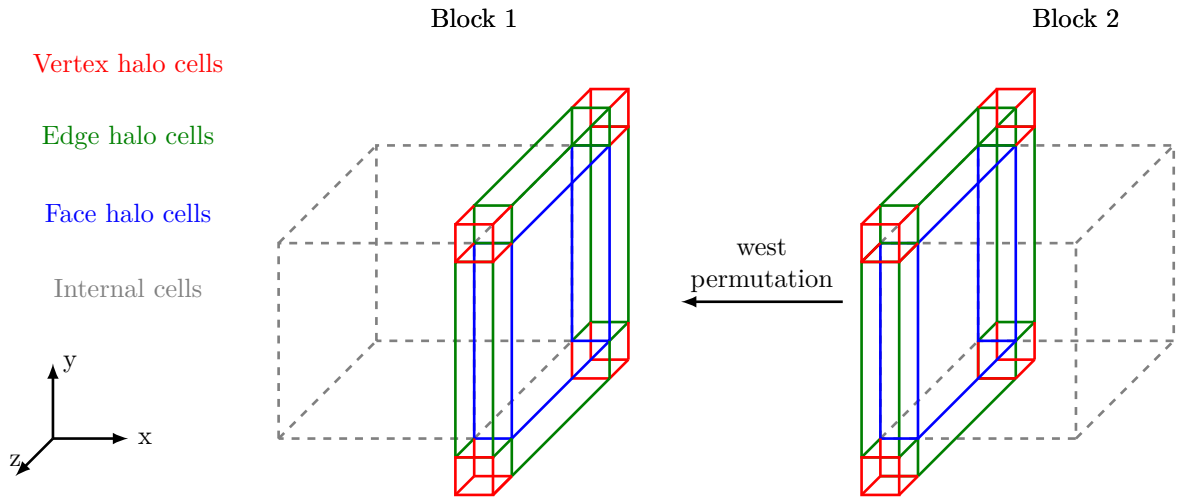}
    \caption{Schematic illustration of two adjacent blocks in $x$-direction within
    the homogeneous domain decomposition during a halo update.
    Depicted are the face (blue), edge (green), and vertex (red) halo cells of block 1
    at the east side of the block. We highlight the corresponding cells in block 2
    that are required for the halo update accordingly. This update entails a west permutation,
    i.e., a permutation of data from block 2 to block 1.}
    \label{fig:halo_update}
\end{figure}

Neighboring blocks in the decomposed domain must communicate data
to update their halo cells. Figure \ref{fig:halo_update} illustrates a 
schematic of two neighboring blocks in $x$-direction, namely block 1 and block 2.
We generally distinguish between internal cells and halo cells.
Furthermore, we define three halo cell regions, i.e., face, edge, and vertex halo cells.
In Fig. \ref{fig:halo_update}, we schematically show the halo update of block 1, i.e., block 1 is receiving
and block 2 is sending data. For the sake of clarity, we only show halo
cells that are located at the east side of block 1. The corresponding cells
in block 2, that are required for the halo update, are highlighted accordingly.
The figure shows that
\begin{itemize}
    \item face halo cells of the receiving block correspond to internal cells of the sending block.
    \item edge halos cells of the receiving block correspond to face halo cells of the sending block.
    \item vertex halo cells of the receiving block correspond to edge halo cells of the sending block.
\end{itemize}
This means that we must first update face, then edge, and lastly vertex halos.
During each of these updates, we iterate over (a subset of) the axis directions
$\in \{east,west,north,south,top,bottom\}$. The axis direction indicates
the location of the halo cells that shall be updated. We perform a collective permutation
with \mintinline{python}{jax.lax.ppermute} in the opposite direction of the
indicated axis direction to receive the required data from neighboring blocks.
In particular, we do the following:
\begin{enumerate}
    \item \textbf{Face halo update:} We iterate over all axis directions
    and update the face halo cells located at the present side of the block.
    \item \textbf{Edge halo update:} We iterate over the axis directions $\in \{east,west,north,south\}$.
    During the $east$ and $west$ iterations, we update all edge halo cells at the present side of the block.
    During the $north$ and $south$ directions, the remaining edge halo cells are updated, preventing duplicate
    updates.
    \item \textbf{Vertex halo update:} We iterate over the axis directions $\in \{east,west\}$
    and update the vertex halo cells located at the present side of the block.
\end{enumerate}
During each of the aforementioned halo updates, we make the following distinction:
\begin{itemize}
    \item \textbf{Inner halo cells:} These are halo cells, that are
    located at the interface of neighboring blocks. They require communication of data.
    \item \textbf{Outer halo cells:} These are halo cells, that are located at the
    outer faces of the computational domain. Here, each device
    updates its halos independently in accordance with the specified boundary
    condition. Periodic boundary conditions pose an exception.
    In this case, if the domain is split into multiple blocks
    in the periodic axis direction,
    nominally outer halo cells are treated like inner halo cells.
\end{itemize}

\SetAlgoLined
\RestyleAlgo{ruled}
\begin{figure}[!t]
\begin{minipage}[t]{0.49\textwidth}

    \footnotesize
    \vspace{0pt}  

    \begin{algorithm}[H]

        \caption{Single-phase and diffuse-interface model.}
        \label{alg:singlephase_diffuse_halo}
        \singlespacing
        \KwData{$\mathbf{U}$, $\mathbf{W}$,  $\Delta t$}
        \texttt{\\}

        $\mathbf{R}_\text{NSE}=\mathcal{R}_\text{NSE}(\mathbf{U}, \mathbf{W})$\;
        $\mathbf{U} = \mathcal{I}(\mathbf{U}, \mathbf{R}_\text{NSE}, \Delta t)$\;
        
        \texttt{\\}
        \color{white}
        $\mathbf{R}_\text{LSA}=\mathcal{R}_\text{LSA}(\phi, \mathbf{Q}_\Gamma$)\;
        $\phi = \mathcal{I}(\phi, \mathbf{R}_\text{LSA}, \Delta t$)\;
        \texttt{\\}

        $\phi = \mathcal{H}_\text{V}(\phi)$\;
        \If{last RK stage}{
            \For{reinitialization steps}{
                $\mathbf{R}_\text{reinit} = \mathcal{R}_\text{reinit}(\phi)$ \;
                $\phi = \mathcal{I}(\phi,\mathbf{R}_\text{reinit},\Delta t)$\;
                $\phi = \mathcal{H}_\text{F}(\phi)$ \;
            }
            $\phi = \mathcal{H}_\text{V}(\phi)$\;
        }

        \texttt{\\}
        $\mathbf{U} = \mathcal{H}_\text{V}(\mathbf{U})$\;
        $\mathbf{U} = \mathcal{M}(\mathbf{U})$\;
        \texttt{\\}

        \color{black}
        $\mathbf{W} = \mathcal{L}_{\mathbf{U}\rightarrow\mathbf{W}}(\mathbf{U})$\;
        \texttt{\\}

        \color{white}
        $\mathbf{W} = \mathcal{H}_\text{F}(\mathbf{W})$\;
        \For{extrapolation steps}{
            $\mathbf{R}_{\text{ext},\mathbf{W}} = \mathcal{R}_{\text{ext},\mathbf{W}}(\mathbf{W})$ \;
            $\mathbf{W} = \mathcal{I}(\mathbf{W},\mathbf{R}_{\text{ext},\mathbf{W}},\Delta t)$\;
            $\mathbf{W} = \mathcal{H}_\text{F}(\mathbf{W})$ \;
        }
        $\mathbf{U} \gets \mathbf{W}$\;
        \texttt{\\}

        \color{black}
        $\mathbf{U}, \mathbf{W} = \mathcal{H}_\text{E}(\mathbf{U}, \mathbf{W})$\;

        \texttt{\\}
        \color{white}
        $\mathbf{Q}_\Gamma = \mathcal{Q}_\Gamma(\mathbf{W}, \phi)$\;
        \For{extrapolation steps}{
            $\mathbf{R}_{\text{ext},\mathbf{Q}_\Gamma} = \mathcal{R}_{\text{ext},\mathbf{Q}_\Gamma}(\mathbf{Q}_\Gamma)$ \;
            $\mathbf{Q}_\Gamma = \mathcal{I}(\mathbf{Q}_\Gamma,\mathbf{R}_{\text{ext},\mathbf{Q}_\Gamma},\Delta t)$\;
            $\mathbf{Q}_\Gamma = \mathcal{H}_\text{F}(\mathbf{Q}_\Gamma)$ \;
        }
    \end{algorithm}
\end{minipage}%
\begin{minipage}[t]{0.49\textwidth}
    \footnotesize
    \vspace{0pt}
    \begin{algorithm}[H]
        \caption{Level-set model. \newline}
        \label{alg:levelset_halo}
        \singlespacing
        \KwData{$\mathbf{U}$, $\mathbf{W}$,
        $\phi$, $\mathbf{Q}_\Gamma$, $\Delta t$}
        \texttt{\\}

        $\mathbf{R}_\text{NSE}=\mathcal{R}_\text{NSE}(\mathbf{U}, \mathbf{W}, \phi, \mathbf{Q}_\Gamma$)\;
        $\mathbf{U} = \mathcal{I}(\mathbf{U}, \mathbf{R}_\text{NSE}, \Delta t)$\;
        
        \texttt{\\}
        \color{blue}
        $\mathbf{R}_\text{LSA}=\mathcal{R}_\text{LSA}(\phi, \mathbf{Q}_\Gamma$)\;
        $\phi = \mathcal{I}(\phi, \mathbf{R}_\text{LSA}, \Delta t$)\;
        \texttt{\\}

        $\phi = \mathcal{H}_\text{V}(\phi)$\;
        \If{last RK stage}{
            \For{reinitialization steps}{
                $\mathbf{R}_\text{reinit} = \mathcal{R}_\text{reinit}(\phi)$ \;
                $\phi = \mathcal{I}(\phi,\mathbf{R}_\text{reinit},\Delta t)$\;
                $\phi = \mathcal{H}_\text{F}(\phi)$ \;
            }
            $\phi = \mathcal{H}_\text{V}(\phi)$\;
        }

        \texttt{\\}
        $\mathbf{U} = \mathcal{H}_\text{V}(\mathbf{U})$\;
        $\mathbf{U} = \mathcal{M}(\mathbf{U})$\;
        \texttt{\\}

        \color{black}
        $\mathbf{W} = \mathcal{L}_{\mathbf{U}\rightarrow\mathbf{W}}(\mathbf{U})$\;
        \texttt{\\}

        \color{blue}
        $\mathbf{W} = \mathcal{H}_\text{F}(\mathbf{W})$\;
        \For{extrapolation steps}{
            $\mathbf{R}_{\text{ext},\mathbf{W}} = \mathcal{R}_{\text{ext},\mathbf{W}}(\mathbf{W})$ \;
            $\mathbf{W} = \mathcal{I}(\mathbf{W},\mathbf{R}_{\text{ext},\mathbf{W}},\Delta t)$\;
            $\mathbf{W} = \mathcal{H}_\text{F}(\mathbf{W})$ \;
        }
        $\mathbf{U} = \mathcal{L}_{\mathbf{W}\rightarrow\mathbf{U}}(\mathbf{W})$\;
        \texttt{\\}

        \color{black}
        $\mathbf{U}, \mathbf{W} = \mathcal{H}_\text{E}(\mathbf{U}, \mathbf{W})$\;

        \texttt{\\}
        \color{blue}
        $\mathbf{Q}_\Gamma = \mathcal{Q}_\Gamma(\mathbf{W},\phi)$\;
        \For{extrapolation steps}{
            $\mathbf{R}_{\text{ext},\mathbf{Q}_\Gamma} = \mathcal{R}_{\text{ext},\mathbf{Q}_\Gamma}(\mathbf{Q}_\Gamma)$ \;
            $\mathbf{Q}_\Gamma = \mathcal{I}(\mathbf{Q}_\Gamma,\mathbf{R}_{\text{ext},\mathbf{Q}_\Gamma},\Delta t)$\;
            $\mathbf{Q}_\Gamma = \mathcal{H}_\text{F}(\mathbf{Q}_\Gamma)$ \;
        }

    \end{algorithm}

\end{minipage}

\caption{Schematic algorithms for a single Runge-Kutta stage illustrating the halo updates.
We compare the single-phase and diffuse-interface model (left) with the level-set model (right).
The color blue highlights parts of the algorithm, which are purely level-set related.
Notation and description of functions and buffers are given
in Table \ref{tab:function_description} and Table \ref{tab:buffer_description},
respectively.}
\label{fig:algorithm_haloupdate}
\end{figure}

The spatial stencil structure of the utilized numerical methods
dictates if halo cell regions are accessed, i.e.,
not all buffers require a halo update for all halo cell regions.
Figure \ref{fig:algorithm_haloupdate} illustrates the 
algorithm of a single Runge-Kutta stage highlighting the positions
and types of halo updates. 
We show the algorithm for the single-phase/diffuse-interface model
\ref{alg:singlephase_diffuse_halo}
and the level-set model \ref{alg:levelset_halo}.
Table \ref{tab:function_description} and Table \ref{tab:buffer_description}
depict the corresponding function and buffer notations.
Algorithm \ref{alg:singlephase_diffuse_halo} starts with
given conservative and primitive variables $\mathbf{U}$ and $\mathbf{W}$.
We evaluate the right-hand side $\mathcal{R}_\text{NSE}$
and subsequently perform a stage integration.
The primitive variables are then
computed from the integrated conservative variables. The stage
ends with an edge halo update of $\mathbf{U}$ and $\mathbf{W}$,
as the spatial stencils that are used for the evaluation of
$\mathcal{R}_\text{NSE}$ do not access the vertex halo cells.
In addition to $\mathbf{U}$ and $\mathbf{W}$,
algorithm \ref{alg:levelset_halo} starts with
given interface quantities $\mathbf{Q}_\Gamma$ and level-set $\phi$.
We evaluate the right-hand sides $\mathcal{R}_\text{NSE}$
and $\mathcal{R}_\text{LSA}$ and perform a stage integration.
In terms of halo updates, the key difference here
are the extrapolation and reinitialization procedures.
These constitute to a large overhead in multi-device simulations,
as they require a face halo 
update in each iteration.

\section{Verification of numerical models}
\label{sec:verification_numerical_models}
In this section, we present verification of the numerical models implemented
in JAX-Fluids. All simulations presented in this section were performed
on GPUs (either NVIDIA A6000 or NVIDIA A100) and TPUs (v3). 
Computations on GPUs were done exclusively with double-precision (float64),
while computations on TPUs were done with single-precision (float32).
\subsection{Convergence test}
\label{subsec:convergence}
\tikzexternaldisable
\definecolor{color0}{RGB}{0,101,189}
\definecolor{color1}{RGB}{227,114,34}
\definecolor{color2}{RGB}{162,173,0}
\begin{figure}[!b]
    \centering
    \begin{tikzpicture}
    

    \begin{groupplot}[
        group style={group size=3 by 1, horizontal sep=1.5cm},
        width=5.2cm, height=5.0cm]
      
    \nextgroupplot[
    tick align=outside,
    tick pos=left,
    legend style={at={(0.95,0.05)},anchor=south east,font=\footnotesize,},
    xtick style={color=black},
    ymin=1e-11, ymax=1,
    xmode=log,
    ymode=log,
    ytick style={color=black},
    xlabel=$\Delta x$,
    ylabel=$\left\Vert  \bar{\rho} - \bar{\rho}_0 \right\Vert _2$,
    title=Single-phase
    ]
    \addplot [line width=1pt, black, forget plot] table[
        x index = {0}, y index = {2}
        ] {figures/convergence/statistics/convergence_fit_singlephase_WENO5-Z.txt}; \label{pgfplots:o5}
    \addplot [line width=1pt, blue, mark=*, mark size=1, mark options={solid}] table[
        x index = {1}, y index = {3}
        ] {figures/convergence/statistics/convergence_singlephase_WENO5-Z.txt}; \label{pgfplots:weno5z}

    \addplot [line width=1pt, black, dashed, forget plot] table[
        x index = {0}, y index = {2}
        ] {figures/convergence/statistics/convergence_fit_singlephase_TENO6-A.txt}; \label{pgfplots:o6}
    \addplot [line width=1pt, red, mark=*, mark size=1, mark options={solid}] table[
        x index = {1}, y index = {3}
        ] {figures/convergence/statistics/convergence_singlephase_TENO6-A.txt}; \label{pgfplots:teno6a}

    \nextgroupplot[
    tick align=outside,
    tick pos=left,
    xtick style={color=black},
    ymin=1e-11, ymax=1,
    xmode=log,
    ymode=log,
    ytick style={color=black},
    xlabel=$\Delta x$,
    title=Level-set
    ]
    \addplot [line width=1pt, black, forget plot] table[
        x index = {0}, y index = {2}
        ] {figures/convergence/statistics/convergence_fit_levelset_WENO5-Z.txt};
    \addplot [line width=1pt, blue, mark=*, mark size=1, mark options={solid}] table[
        x index = {1}, y index = {3}
        ] {figures/convergence/statistics/convergence_levelset_WENO5-Z.txt};

    \addplot [line width=1pt, black, dashed, forget plot] table[
        x index = {0}, y index = {2}
        ] {figures/convergence/statistics/convergence_fit_levelset_TENO6-A.txt};
    \addplot [line width=1pt, red, mark=*, mark size=1, mark options={solid}] table[
        x index = {1}, y index = {3}
        ] {figures/convergence/statistics/convergence_levelset_TENO6-A.txt};

    \nextgroupplot[
    tick align=outside,
    tick pos=left,
    xtick style={color=black},
    ymin=1e-11, ymax=1,
    xmode=log,
    ymode=log,
    ytick style={color=black},
    xlabel=$\Delta x$,
    title=Diffuse-interface
    ]
    \addplot [line width=1pt, black, forget plot] table[
        x index = {0}, y index = {2}
        ] {figures/convergence/statistics/convergence_fit_diffuse_WENO5-Z.txt};
    \addplot [line width=1pt, blue, mark=*, mark size=1, mark options={solid}] table[
        x index = {1}, y index = {3}
        ] {figures/convergence/statistics/convergence_diffuse_WENO5-Z.txt};

    \addplot [line width=1pt, black, dashed, forget plot] table[
        x index = {0}, y index = {2}
        ] {figures/convergence/statistics/convergence_fit_diffuse_TENO6-A.txt};

    \end{groupplot}

\end{tikzpicture}
    \caption{Error convergence for the advection of two Gaussians.
    For single-phase and level-set simulations WENO5-Z (\ref{pgfplots:weno5z}) 
    and TENO6-A (\ref{pgfplots:teno6a}) reconstruction schemes
    were tested. For diffuse-interface simulations, WENO5-Z was tested.
    The solid (\ref{pgfplots:o5}) and dashed (\ref{pgfplots:o6}) black lines 
    denote nominal convergence of $\mathcal{O}(\Delta x^5)$
    and $\mathcal{O}(\Delta x^6)$, respectively.}
    \label{fig:ConvergenceStatistics}
\end{figure}
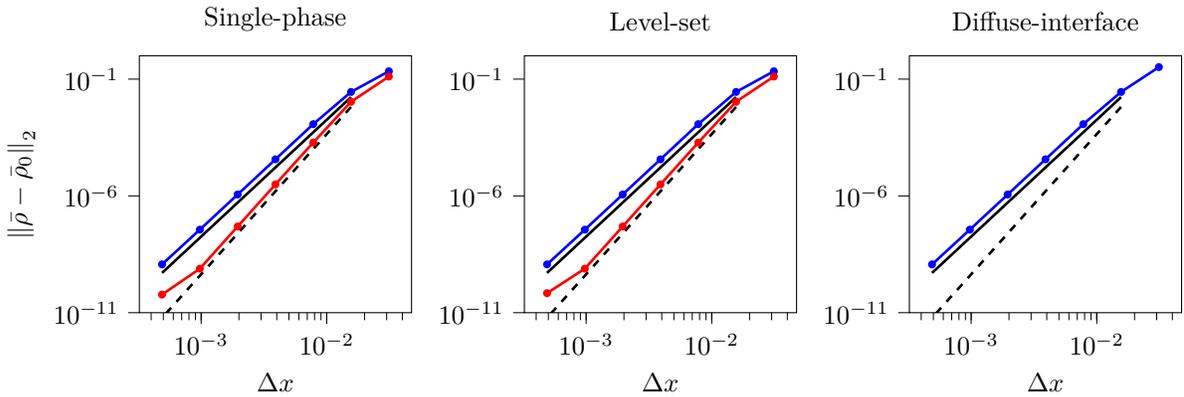
\tikzexternalenable
We perform convergence tests for the inviscid advection of a
one-dimensional density profile with
uniform velocity $u = 1$ and pressure $p = 1$.
The material is an ideal gas with $\gamma=1.4$.
The initial density consists of two Gaussians,
\begin{equation}
    \rho \left( x, t=0 \right) = 1 + 5 \exp \left(-200 (x - 0.5)^2\right) + 5 \exp \left(-200 (x - 1.5)^2\right).
\end{equation}
We initialize cell-averaged values, i.e., for cell $i$ the cell-averaged density reads
\begin{align}
    \bar{\rho}_{0_i} &= \frac{1}{\Delta x_i} \int_{x_{i-1/2}}^{x_{i+1/2}} \rho(x, t=0) \,dx \\
        &= 1 - \frac{5 \sqrt{\pi}}{2 \sqrt{200} \Delta x_i} \left(
            \left[
                erf \left(\sqrt{200} (0.5 - x)\right) \vert_{x_{i-1/2}}^{x_{i+1/2}}
                + erf \left(\sqrt{200} (1.5 - x)\right) \vert_{x_{i-1/2}}^{x_{i+1/2}}
            \right]
        \right)
\end{align}
The computational domain is $x \in \left[0, 2\right]$
with periodic boundary conditions on either side.
The density profile is advected for one flow through time
which corresponds to the final time $t = 2.0$.
For the single-phase model (SPM) and level-set model (LSM), 
we test convergence for WENO5-Z and TENO6-A stencils.
For the diffuse-interface model (DIM), we only test WENO5-Z, 
as the TENO6-A reconstruction occasionally introduces 
over- and undershoots in the volume fraction field which
activate the positivity-preserving algorithm
and, therefore, diminish order of convergence.
We use TVD-RK3 time integration with a fixed time
step size $\Delta t = 10^{-5}$ for all tests. 
The time step size is chosen small enough to ensure
that the error due to numerical time integration 
does not influence the solution.
In Fig. \ref{fig:ConvergenceStatistics},
we visualize the $L_2$-error for successively higher spatial resolutions
(from $N=64$ to $N=4096$ grid points).
The nominal convergence order is achieved for all models.
\subsection{Single-phase simulations}
\label{subsec:single_phase_simulations}

For single-phase simulations, we use a TENO6-A
cell face reconstruction combined with an HLLC Riemann solver.
The TENO6-A reconstruction is complemented by an interpolation limiter.
In this work, we do not use flux limiters for single-phase simulations
as the single-phase cases under investigation do not
feature strong shock discontinuities.
Diffusive fluxes are discretized
using 6th-order central finite-difference approximations.
Temporal evolution is performed with a TVD-RK3 scheme with $CFL = 0.9$.

\subsubsection{Laminar boundary layer}
\label{subsubsec:laminar_bl}
The compressible Blasius boundary layer \cite{White1991-ue} describes the
steady state two-dimensional flow over a flat plate under zero pressure gradient.
We use this test case to verify the implementation of
the  viscous fluxes. We use an ideal gas with $\gamma=1.4$.
The computational domain $(x,y)\in [1.0,1.5]\times[0.0, 0.4]$ is 
discretized using $300\times200$ cells with a uniform
grid in $x$-direction and a stretched mesh in $y$
direction. The stretching parameter is $\beta=2.2$
(see \ref{sec:appendix_computational_domain}).
The boundary conditions are a no-slip adiabatic wall at the south boundary
and zero-gradient extrapolation at north and east boundaries.
At the inlet (west), we impose the self-similar solution,
which is computed by solving the compressible Blasius 
similarity equations. For the given Mach number $Ma_e=2.25$
and Prandtl number $Pr=0.72$, we get normalized velocity $u/u_e$ 
and temperature $T/T_e$ over the self-similar variable
\begin{equation}
    \eta=\frac{u_e}{\sqrt{2\rho_e \mu_e u_e x}}\int_0^y \rho \,dy.
    \label{similarityvariable}
\end{equation}
Here, the index $e$ denotes free-stream conditions.
To impose the Blasius solution at the inlet boundary of the domain, 
we numerically solve Eq. \eqref{similarityvariable} for $y$ and,
subsequently, linearly interpolate the values on the grid at the boundary.
The free stream unit Reynolds number is $Re=u_e\rho_e/\mu_e=10000$.
The temperature dependent dynamic viscosity is
computed using the Sutherland law (see Eq. \eqref{eq:viscosity_sutherland_law}) with
$C=0$, $T_{ref}=T_e$, and $\mu_{ref}=\mu_e$.
\tikzexternaldisable
\begin{figure}[!t]
    \centering
    \begin{tikzpicture}
    
    \begin{groupplot}[
        group style={group size=2 by 1},
        width=5.20cm, height=5cm]
    \nextgroupplot[
    tick align=outside,
    tick pos=left,
    legend style={at={(0.05,0.9)},draw=none,anchor=north west},
    xtick style={color=black},
    ymin=0, ymax=6,
    ytick style={color=black},
    ylabel=$\eta$,
    xlabel=$u/u_e$,
    ]
    \addplot [line width=1pt, black] table[
        x index = {1}, y index = {0}
        ] {figures/laminar_boundarylayer/blasius_present.txt}; \label{pgfplots:blasius_jax}
    \addplot [line width=1pt, black, mark=*, mark size=0.5pt, only marks] table[
        x index = {1}, y index = {0}
    ] {figures/laminar_boundarylayer/blasius_exact.txt}; \label{pgfplots:blasius_ref}
    
    \nextgroupplot[
    tick align=outside,
    tick pos=left,
    xtick style={color=black},
    ymin=0, ymax=6,
    ytick style={color=black},
    xlabel=$T/T_e$,
    ]
    \addplot [line width=1pt, black] table[
        x index = {2}, y index = {0}
        ] {figures/laminar_boundarylayer/blasius_present.txt};
    \addplot [line width=1pt, black, mark=*, mark size=0.5pt, only marks] table[
        x index = {2}, y index = {0}
    ] {figures/laminar_boundarylayer/blasius_exact.txt};
    \end{groupplot}
    
\end{tikzpicture}
    
    \caption{Compressible laminar boundary layer. Normalized velocity $u/u_e$ and
    normalized temperature $T/T_e$ profiles over the self-similar variable
    $\eta$. The solid lines (\ref{pgfplots:blasius_jax}) and markers
    (\ref{pgfplots:blasius_ref}) indicate the JAX-Fluids and Blasius solution,
    respectively.}
    \label{fig:blasius}
\end{figure}
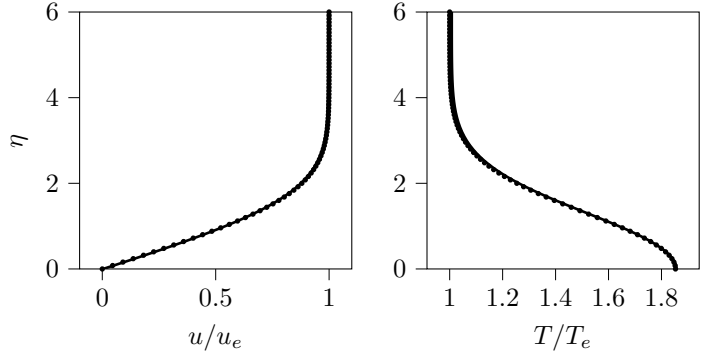
\tikzexternalenable
We evaluate the JAX-Fluids solution at the outlet boundary $x=1.5$ and present
a comparison to the self-similar Blasius solution. Figure \ref{fig:blasius} depicts
normalized velocity $u/u_e$ and temperature $T/T_e$ over the 
self-similar variable $\eta$. We achieve good agreement with the Blasius solution.

\subsubsection{Turbulent channel flow}
\label{subsubsec:turbulent_channel}
\tikzexternaldisable
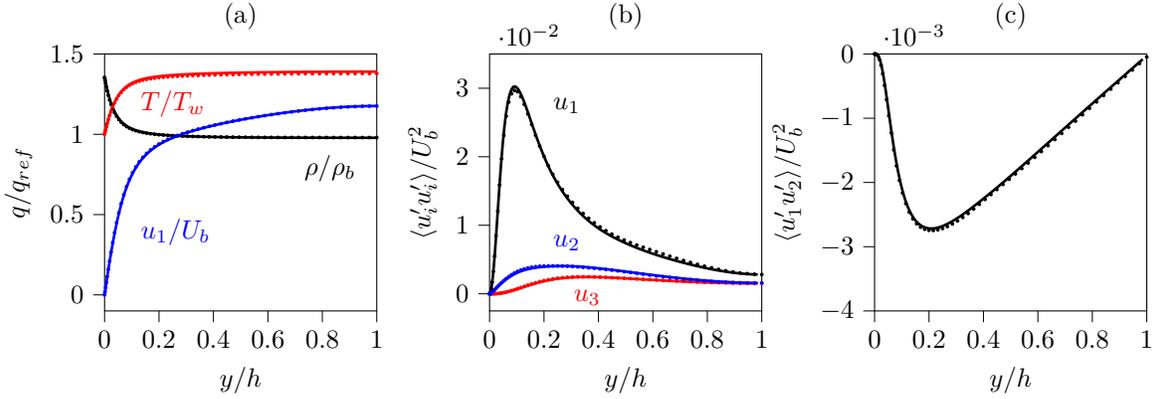
\begin{figure}[!t]
    \centering
    \begin{tikzpicture}
    
    \begin{groupplot}[
        group style={group size=3 by 1, horizontal sep=1.5cm},
        width=5.2cm, height=5cm]
      
    \nextgroupplot[
    tick align=outside,
    tick pos=left,
    xtick style={color=black},
    ymin=-0.1, ymax=1.5,
    xmin=0, xmax=1,
    ytick style={color=black},
    xlabel=$y / h$,
    ylabel=$q/q_{ref}$,
    title=(a)
    ]
    \addplot [line width=1pt, black] table[
        x index = {1}, y index = {2}
        ] {figures/turbulent_channel/mean_present.txt}; \label{pgfplots:ac1r1}
    \addplot [line width=1pt, red] table[
        x index = {1}, y index = {3}
        ] {figures/turbulent_channel/mean_present.txt}; \label{pgfplots:ac1r2}
    \addplot [line width=1pt, blue] table[
        x index = {1}, y index = {4}
        ] {figures/turbulent_channel/mean_present.txt}; \label{pgfplots:ac1r3}
    \addplot [mark=*, only marks, mark size=0.5pt, black] table[
        x index = {0}, y index = {1}
    ] {figures/turbulent_channel/mean_ref.txt}; \label{pgfplots:ac2r1}
    \addplot [mark=*, only marks, mark size=0.5pt, red] table[
        x index = {0}, y index = {2}
    ] {figures/turbulent_channel/mean_ref.txt}; \label{pgfplots:ac2r2}
    \addplot [mark=*, only marks, mark size=0.5pt, blue] table[
        x index = {0}, y index = {3}
    ] {figures/turbulent_channel/mean_ref.txt}; \label{pgfplots:ac2r3}
    \node[anchor=south west, color=black] at (axis cs:0.7,0.65) {$\rho / \rho_b$};
    \node[anchor=south west, color=blue] at (axis cs:0.1,0.25) {$u_1 / U_b$};
    \node[anchor=south west, color=red] at (axis cs:0.1,1.05) {$T / T_w$};

    \nextgroupplot[
    tick align=outside,
    tick pos=left,
    xtick style={color=black},
    ymin=-0.0025, ymax=0.035,
    xmin=0, xmax=1,
    ytick style={color=black},
    xlabel=$y / h$,
    ylabel=$\langle u_i'u_i' \rangle / U_b^2$,
    title=(b)
    ]
    \addplot [line width=1pt, black] table[
        x index = {1}, y index = {2}
        ] {figures/turbulent_channel/reynolds_stresses_present.txt}; \label{pgfplots:bc1r1}
    \addplot [line width=1pt, red] table[
        x index = {1}, y index = {3}
        ] {figures/turbulent_channel/reynolds_stresses_present.txt}; \label{pgfplots:bc1r2}
    \addplot [line width=1pt, blue] table[
        x index = {1}, y index = {4}
        ] {figures/turbulent_channel/reynolds_stresses_present.txt}; \label{pgfplots:bc1r3}
    \addplot [mark=*, only marks, mark size=0.5pt, black] table[
        x index = {0}, y index = {1}
    ] {figures/turbulent_channel/reynolds_stresses_ref.txt}; \label{pgfplots:bc2r1}
    \addplot [mark=*, only marks, mark size=0.5pt, red] table[
        x index = {0}, y index = {2}
    ] {figures/turbulent_channel/reynolds_stresses_ref.txt}; \label{pgfplots:bc2r2}
    \addplot [mark=*, only marks, mark size=0.5pt, blue] table[
        x index = {0}, y index = {3}
    ] {figures/turbulent_channel/reynolds_stresses_ref.txt}; \label{pgfplots:bc2r3}
    \node[anchor=south west, color=black] at (axis cs:0.2,0.025) {$u_1$};
    \node[anchor=south west, color=blue] at (axis cs:0.2,0.005) {$u_2$};
    \node[anchor=south west, color=red] at (axis cs:0.275,-0.003) {$u_3$};

    \nextgroupplot[
    tick align=outside,
    tick pos=left,
    xtick style={color=black},
    ymin=-0.004, ymax=0.0,
    xmin=0, xmax=1,
    ytick style={color=black},
    xlabel=$y / h$,
    ylabel=$\langle u_1'u_2' \rangle / U_b^2$,
    title=(c)
    ]
    \addplot [line width=1pt, black] table[
        x index = {1}, y index = {5}
        ] {figures/turbulent_channel/reynolds_stresses_present.txt}; \label{pgfplots:cc1r1}
    \addplot [mark=*, only marks, mark size=0.5pt, black] table[
        x index = {0}, y index = {4}
    ] {figures/turbulent_channel/reynolds_stresses_ref.txt}; \label{pgfplots:cc2r1}

    \end{groupplot}

\end{tikzpicture}
    \caption{Statistical evaluation of the supersonic turbulent channel flow at $Ma_b = 1.5$ and $Re_b=3000$.
    (a) Normalized mean profiles of density $\rho/\rho_b$, temperature $T/T_w$,
    and streamwise velocity $u_1/U_b$.
    (b) Normalized Reynolds normal stresses $\langle u_i'u_i' \rangle / U_b^2$.
    (c) Normalized Reynolds shear stress $\langle u_1'u_2' \rangle / U_b^2$.
    Reference data is taken from Coleman et al. \cite{Coleman1995}.
    We illustrate the JAX-Fluids solution with solid lines (\ref{pgfplots:cc1r1}). 
    The reference by Coleman et al. \cite{Coleman1995} is depicted with solid markers (\ref{pgfplots:cc2r1}).
    }
    \label{fig:ChannelStatistics}
\end{figure}
\tikzexternalenable
Bi-periodic turbulent channel flows serve as canonical configurations to study wall-bounded turbulence.
Here, we perform direct numerical simulations (DNS) of the turbulent supersonic isothermal-wall channel flow 
of Coleman et al. \cite{Coleman1995,Lechner2001}.
The computational domain is $(x,y,z) \in \left[0, 4 \pi h\right] \times \left[-h, h\right] \times \left[0, 2 \pi h\right]$.
A constant mass flow rate is imposed via a uniform body force in streamwise ($x$) direction.
The bulk density is calculated as $\rho_b = \frac{1}{2h} \int_{-h}^{h} \langle \rho \rangle \,dy$,
and the bulk velocity is calculated as $U_b = \frac{1}{2h} \int_{-h}^{h} \langle \rho u \rangle \,dy$. 
The Mach number based on bulk velocity and speed of sound at the wall $Ma_b = U_b / c_w = 1.5$.
The Reynolds number based on bulk density, bulk velocity, channel half-width, 
and wall viscosity $Re_b = \rho_b U_b h / \mu_w = 3000$.
At the channel walls isothermal, no-slip boundary conditions are imposed 
such that $T = 1$ and $\mathbf{u} = 0$ at $y = \pm h$.
Periodic boundary conditions are applied in streamwise and spanwise direction.
The fluid is assumed to be an ideal gas with constant Prandtl number $Pr = 0.7$ 
and constant ratio of specific heats $\gamma = 1.4$.
The dynamic viscosity follows a temperature dependent power law, see Eq. \eqref{eq:viscosity_power_law}.
The computational domain is discretized by $256 \times 128 \times 128$ cells.
The (DNS) grid is uniform in $x$- and $z$-direction (streamwise and spanwise direction, respectively),
and a hyperbolic-tangent stretching with stretching factor $\beta = 1.8$ is applied in wall normal direction
(see \ref{sec:appendix_computational_domain} for details).
The first grid point is located at $y_1^{+} \approx 0.34$, 
and the tenth point is located at $y_{10}^{+} \approx 8.28$.
The cell sizes in streamwise and spanwise direction are $\Delta x^{+} = \Delta z^{+} = 10.71$,
and the minimum and maximum cell sizes in wall normal direction are $\Delta y_{min}^{+} = 0.69$
and $\Delta y_{max}^{+} = 6.48$, respectively.

We run a precursor simulation of the channel flow on a coarser computational
grid until a statistically steady state is reached.
The simulation is initialized with a streamwise velocity that follows a laminar profile 
superimposed with noise along with constant pressure and density.
Once a statistically steady state is obtained, we interpolate the flow field onto the aforementioned DNS grid.
The DNS is run on two Nvidia A100 GPUs.
Similarly to \cite{Lechner2001}, we wait for 15 characteristic problem times $h / u_{\tau} \approx 18.38$ 
to let initial transients pass. We then collect 300 snapshots 
of the instantaneous flow field over a period of approximately $41 h / u_{\tau}$ to compute flow statistics.

Figure \ref{fig:ChannelStatistics} compares flow statistics to the DNS data of Coleman et al. \cite{Coleman1995}.
We observe excellent quantitative agreement for mean flow profiles and Reynolds stresses
with the cited reference.
Similarly, root-mean-square fluctuations of density and temperature (not shown) are in good agreement with the reference data. 
\subsubsection{Turbulent boundary layer}
\label{subsubsec:turbulent_boundary_layer}
We consider the compressible turbulent boundary layer
case by Pirozzoli et al. \cite{Pirozzoli2004}. Here, a compressible
laminar boundary layer enters the domain, and transition to turbulence
happens by means of blowing and suction within a small region 
close to the inlet boundary.

The computational domain $(x,y,z)\in[4.0,22.0]\times[0.0,0.5]\times[-0.0875,0.0875]$ is
discretized using $2500\times80\times380$ cells. The simulation is run on 2 Nvidia
A100 GPUs. Unless mentioned otherwise,
the spatial coordinates are nondimensional, with $l_{ref}=0.00254 \,\si{m}$ being the
reference length. We impose an adiabatic no-slip wall at the south
boundary and zero gradient at the outlet (east) and spanwise (top, bottom) boundaries.
At the inlet (east) boundary, a self-similar Blasius solution is specified.
In streamwise ($x$) direction, the domain is partitioned into three zones.
The first zone is bound by $x\in[4,7]$ and contains the blowing and suction region
($x_a=4.5$ and $x_b=5.0$). It serves as the transition region, consisting of $700$ cells.
The second zone is that of interest containing fully developed turbulence.
It is bound by $x\in[7,9]$ and consists of $1700$ cells. Within the third zone $x\in[9,22]$
the grid is progressively coarsened using $100$ cells. This serves as a
buffer region minimizing disturbances due to reflections at the outlet boundary.
The grid is uniform in $z$-direction and a tangent stretching 
is applied in $y$-direction with $\beta=3.5$ (see Sec. \ref{subsec:computational_domain}).
Within the well-resolved region, the smallest grid sizes
in streamwise, spanwise, and wall-normal direction are
$\Delta x^+\approx14.2$, $\Delta z^+\approx5.5$, and $\Delta y^+\approx1.0$, respectively.

The laminar-turbulent transition is enforced by blowing and suction within
a small region ($x_a \leq x \leq x_b$) close to the inlet. Here, the wall-normal component of the velocity
at the wall is computed as
\begin{equation}
    v(x,z,t)=Au_ef(x)g(z)h(t),\quad x_a \leq x \leq x_b,
\end{equation}
where $A=0.04$ denotes the amplitude of the disturbance, $u_e$ is the free stream
velocity, and
\begin{align*}
    f(x)&=4\sin\theta(1-\cos\theta)\sqrt{27}, \\
    \theta&=2\pi\frac{x-x_a}{x_b-x_a}, \\
    g(z) &= \sum_{l=1}^{l_{max}} Z_l \sin(2\pi l(z/L_z+\xi_l)), \\
    Z_l &= 1.25Z_{l+1}, \quad \sum_{l=1}^{l_{max}} Z_l = 1, \\
    h(t) &= \sum_{m=1}^{m_{max}} T_m \sin(2\pi m(\beta t + \xi_m)), \\
    T_m &= 1.25T_{m+1}, \quad  \sum_{m=1}^{m_{max}} T_m = 1.
\end{align*}
Here, $x_a$, $x_b$, and $L_z$ denote the start and end locations of the blowing
and suction region and the spanwise domain size, respectively. We use
a frequency of $\beta = 75000 \,\si{Hz}$ for the disturbance, with $m_{max} = 5$
and $l_{max} = 10$ and $\xi_l$ and $\xi_m$ being random numbers between 0 and 1.

The free stream Mach number, temperature, and pressure are $Ma_e=2.25$, $T_e=169.44 \,\si{K}$,
and $p_e=100000 \,\si{Pa}$, respectively. The Prandtl number and free stream unit Reynolds number are 
$Pr=0.72$ and $Re/l_{ref}=\rho_e u_e/\mu_e = 650000$. We use an ideal gas with $\gamma=1.4$.
The temperature dependent dynamic viscosity is computed using the Sutherland law (see Eq. \eqref{eq:viscosity_sutherland_law}) with
$C=0$, $T_{ref}=T_e$, and $\mu_{ref}=\mu_e$.

\tikzexternaldisable
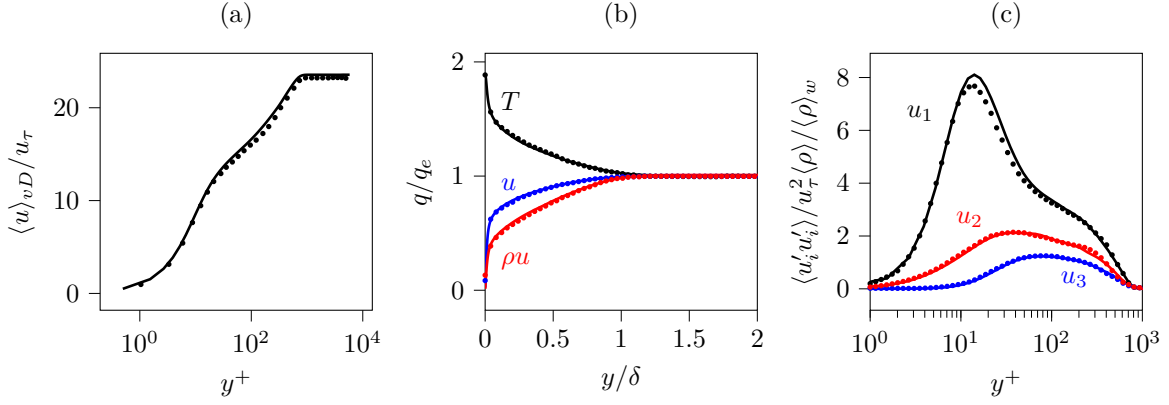
\begin{figure}[!t]
    \centering
    \begin{tikzpicture}
    
    \begin{groupplot}[
        group style={group size=3 by 1, horizontal sep=1.5cm},
        width=5.2cm, height=5cm]
    \nextgroupplot[
    tick align=outside,
    tick pos=left,
    xtick style={color=black},
    ytick style={color=black},
    ylabel=$\langle u \rangle_{vD}/u_\tau$,
    xlabel=$y^+$,
    xmode=log,
    title=(a)
    ]
    \addplot [line width=1pt, black] table[
        x index = {0}, y index = {1}
        ] {figures/turbulent_boundarylayer/u_vd_present.txt}; \label{pgfplots:turbulent_boundarylayer_jaxfluids}
    \addplot [line width=1pt, black, mark=*, mark size=0.5pt, only marks] table[
        x index = {0}, y index = {1}
    ] {figures/turbulent_boundarylayer/u_vd_ref.txt}; \label{pgfplots:turbulent_boundarylayer_ref}
    
    \nextgroupplot[
    tick align=outside,
    tick pos=left,
    xmin=0.0, xmax=2.0,
    xtick style={color=black},
    ytick style={color=black},
    ylabel = $q/q_e$,
    xlabel=$y/\delta$,
    title=(b)
    ]
    \addplot [line width=1pt, black] table[
        x index = {0}, y index = {1}
        ] {figures/turbulent_boundarylayer/state_outer_present.txt};
    \addplot [line width=1pt, blue] table[
        x index = {0}, y index = {2}
        ] {figures/turbulent_boundarylayer/state_outer_present.txt};
    \addplot [line width=1pt, red] table[
        x index = {0}, y index = {3}
        ] {figures/turbulent_boundarylayer/state_outer_present.txt};

    \addplot [line width=1pt, black, mark=*, mark size=0.5pt, only marks] table[
        x index = {0}, y index = {1}
        ] {figures/turbulent_boundarylayer/state_outer_ref.txt};
    \addplot [line width=1pt, blue, mark=*, mark size=0.5pt, only marks] table[
        x index = {0}, y index = {2}
        ] {figures/turbulent_boundarylayer/state_outer_ref.txt};
    \addplot [line width=1pt, red, mark=*, mark size=0.5pt, only marks] table[
        x index = {0}, y index = {3}
        ] {figures/turbulent_boundarylayer/state_outer_ref.txt};

    \node[anchor=south west, color=black] at (axis cs:0.05,1.5) {$T$};
    \node[anchor=south west, color=blue] at (axis cs:0.05,0.8) {$u$};
    \node[anchor=south west, color=red] at (axis cs:0.05,0.1) {$\rho u$};

    \nextgroupplot[
    tick align=outside,
    tick pos=left,
    xmin=1.0, xmax=1000,
    xtick style={color=black},
    ytick style={color=black},
    ylabel=$\langle u'_iu'_i \rangle / u^2_\tau \langle \rho \rangle / \langle \rho \rangle_w $,
    xlabel=$y^+$,
    xmode=log,
    title=(c)
    ]
    \addplot [line width=1pt, black] table[
        x index = {0}, y index = {1}
        ] {figures/turbulent_boundarylayer/reynolds_stresses_present.txt};
    \addplot [line width=1pt, blue] table[
        x index = {0}, y index = {2}
        ] {figures/turbulent_boundarylayer/reynolds_stresses_present.txt};
    \addplot [line width=1pt, red] table[
        x index = {0}, y index = {3}
        ] {figures/turbulent_boundarylayer/reynolds_stresses_present.txt};

    \addplot [line width=1pt, black, mark=*, mark size=0.5pt, only marks] table[
        x index = {0}, y index = {1}
        ] {figures/turbulent_boundarylayer/reynolds_stresses_ref.txt};
    \addplot [line width=1pt, blue, mark=*, mark size=0.5pt, only marks] table[
        x index = {0}, y index = {3}
        ] {figures/turbulent_boundarylayer/reynolds_stresses_ref.txt};
    \addplot [line width=1pt, red, mark=*, mark size=0.5pt, only marks] table[
        x index = {0}, y index = {2}
        ] {figures/turbulent_boundarylayer/reynolds_stresses_ref.txt};

    \node[anchor=south west, color=black] at (axis cs:2,6) {$u_1$};
    \node[anchor=south west, color=red] at (axis cs:7,2) {$u_2$};
    \node[anchor=south west, color=blue] at (axis cs:100,-0.3) {$u_3$};



    
    \end{groupplot}

\end{tikzpicture}
    
    \caption{Statistical evaluation of the turbulent boundary layer flow at $Re_\theta=4000$.
    (a) Normalized van Driest transformed velocity $\langle u \rangle_{vD}/u_\tau$ in inner scaling.
    (b) Normalized mean flow states $\langle T \rangle/T_e$, $\langle u \rangle/u_e$ and $\langle\rho\rangle \langle u\rangle/\rho_e u_e$ in outer scaling.
    (c) Normalized Reynolds normal stresses $\langle u'_iu'_i \rangle / u^2_\tau \langle \rho \rangle / \langle \rho \rangle_w $ in inner scaling.
    We illustrate the JAX-Fluids solution with solid lines (\ref{pgfplots:turbulent_boundarylayer_jaxfluids}).
    The reference by Pirozzoli et al. \cite{Pirozzoli2004} is depicted with markers (\ref{pgfplots:turbulent_boundarylayer_ref}).}
    \label{fig:boundarylayer_statistics}
\end{figure}

\begin{figure}[!b]
    \centering
    \input{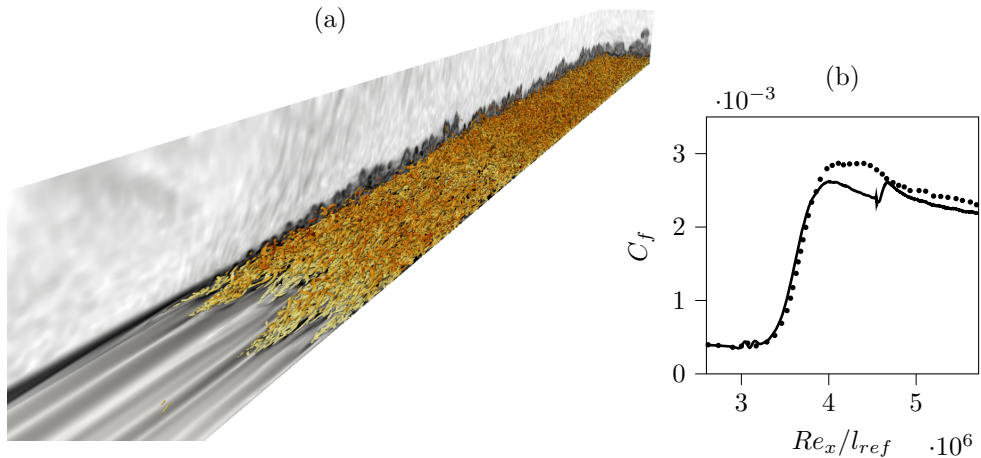}
    \caption{(a) Visualization of the instantaneous flow field showing the transition region.
    The figure shows contours of the numerical schlieren ($\ln \left( \lVert \nabla \rho \rVert \right)$) 
    and isosurfaces of the Q criterion colored by the normalized density $\rho/\rho_e$.
    (b) Skin friction coefficient $C_f$ as a function of the streamwise Reynolds number $Re_x$.
    We illustrate the JAX-Fluids solution with solid lines (\ref{pgfplots:transition_jaxfluids}).
    The reference by Pirozzoli et al. \cite{Pirozzoli2004} is depicted with markers (\ref{pgfplots:transition_ref}).
    }
    \label{fig:boundarylayer_transition}
\end{figure}
\tikzexternalenable

We evaluate the mean flow field at the momentum thickness Reynolds number
$Re_\theta=\rho_e u_e \theta/\mu_e=4000$, where $\theta$ is the momentum thickness. 
To compute averages, we collect around 732 time snapshots over a time
period of approximately 1400$\delta^*$/$u_e$,
where $\delta^*$ denotes the displacement thickness. In the following,
we denote mean quantities by $\langle \cdot \rangle$, and the index $w$ describes 
a quantity at the wall.
Figure \ref{fig:boundarylayer_statistics} (a) shows the wall normal 
profiles of the van-Driest transformed mean velocity
\begin{equation}
    \langle u \rangle_{vD}(y) = \int_0^y \sqrt{ \frac{\langle \rho \rangle (y') }{\langle \rho \rangle_{w}} } \frac{d \langle u \rangle (y')}{dy'}\,dy'
\end{equation}
normalized by the friction velocity
\begin{equation}
    u_\tau = \left. \sqrt{\frac{\langle \mu \rangle}{\langle \rho \rangle} \frac{d\langle u \rangle}{dy} } \right|_{w}.
\end{equation}
Figure \ref{fig:boundarylayer_statistics} (b) illustrates the normalized
temperature $\langle T \rangle/T_e$, streamwise velocity $\langle u \rangle/u_e$, and streamwise momentum
$\langle\rho\rangle \langle u\rangle/\rho_e u_e$.
In Fig. \ref{fig:boundarylayer_statistics} (c) we depict the
normalized Reynolds stresses $\langle u'_iu'_i \rangle / u^2_\tau \langle \rho \rangle / \langle \rho \rangle_w$
in stream ($u_1$), wall normal ($u_2$), and spanwise direction ($u_3$).
We observe good agreement with the reference DNS \cite{Pirozzoli2004}.
Figure \ref{fig:boundarylayer_transition} (a) depicts
a three-dimensional snapshot of the JAX-Fluids DNS solution.
In Fig. \ref{fig:boundarylayer_transition} (b), we display
the skin friction coefficient
\begin{align}
    C_f &=2\tau_w/\rho_e u_e^2, \\
    \tau_w &=\langle \mu\rangle_w \left. \frac{d\langle u\rangle}{dy}\right|_w,
\end{align}
along the streamwise Reynolds number $Re_x=\rho_e u_e x / \mu_e$. We observe
good agreement with the reference \cite{Pirozzoli2004} in terms of transition point. A distinct
jump of $C_f$ at $Re_x/l_{ref}=4.55 \times 10^6$ can be seen. This is the intersection point
between the first and second zone, where the streamwise grid size is
significantly refined. Therefore, this jump is expected. Within the fully developed
region $Re_x/l_{ref}\in[4.55 \times 10^6, 5.85 \times 10^6]$, we again have a good match with the
reference DNS \cite{Pirozzoli2004}.

\subsection{Two-phase simulations}
\label{subsec:two_phase_simulations}
We perform two-phase simulations with the
diffuse-interface model (DIM) and the level-set model (LSM).
The computation of the cell face fluxes is done with a 
WENO5-Z reconstruction combined with an HLLC Riemann solver.
We use interpolation and flux limiters to ensure physically admissible states.
Time integration happens by means of a TVD-RK3 scheme with $CFL = 0.5$.
In all DIM simulations, the volume fraction field $\alpha_1$ is
initialized with $1 - 10^{-8}$ in phase 1 
and with $10^{-8}$ in phase 2.
\subsubsection{Air-helium shock tube problem}
\begin{figure}[!t]
    \centering
    \begin{tikzpicture}

    \begin{groupplot}[
        group style={group size=3 by 1, horizontal sep=1.5cm},
        width=5.2cm, height=5cm]
      
    \nextgroupplot[
    tick align=outside,
    tick pos=left,
    xtick style={color=black},
    ymin=0.0, ymax=1.1,
    xmin=0.0, xmax=1.0,
    ytick style={color=black},
    xlabel=$x$,
    ylabel=$\rho$,
    ytick={0.0,0.5,1.0}
    ]
    \addplot [line width=1pt, black] table[
        x index = {0}, y index = {1}
        ] {figures/shocktube/data/sod_exact.txt}; \label{pgfplots:shocktube_air_helium_exact}
    \addplot [line width=1pt, blue] table[
        x index = {0}, y index = {1}
        ] {figures/shocktube/data/sod_levelset.txt}; \label{pgfplots:shocktube_air_helium_lsm}
    \addplot [line width=1pt, red] table[
        x index = {0}, y index = {1}
        ] {figures/shocktube/data/sod_diffuse.txt}; \label{pgfplots:shocktube_air_helium_dim}

    \nextgroupplot[
    tick align=outside,
    tick pos=left,
    xtick style={color=black},
    ymin=-0.1, ymax=1.1,
    xmin=0.0, xmax=1.0,
    ytick style={color=black},
    ytick={0.0,0.5,1.0},
    xlabel=$x$,
    ylabel=$u$,
    ]
    \addplot [line width=1pt, black] table[
        x index = {0}, y index = {2}
        ] {figures/shocktube/data/sod_exact.txt};
    \addplot [line width=1pt, blue] table[
        x index = {0}, y index = {2}
        ] {figures/shocktube/data/sod_levelset.txt};
    \addplot [line width=1pt, red] table[
        x index = {0}, y index = {2}
        ] {figures/shocktube/data/sod_diffuse.txt};

    \nextgroupplot[
    tick align=outside,
    tick pos=left,
    xtick style={color=black},
    ymin=0.0, ymax=1.1,
    xmin=0.0, xmax=1.0,
    ytick style={color=black},
    ytick = {0.0, 0.5, 1.0},
    xlabel=$x$,
    ylabel=$p$,
    ]
    \addplot [line width=1pt, black] table[
        x index = {0}, y index = {3}
        ] {figures/shocktube/data/sod_exact.txt};
    \addplot [line width=1pt, blue] table[
        x index = {0}, y index = {3}
        ] {figures/shocktube/data/sod_levelset.txt};
    \addplot [line width=1pt, red] table[
        x index = {0}, y index = {3}
        ] {figures/shocktube/data/sod_diffuse.txt};
    \end{groupplot}

\end{tikzpicture}
    \caption{Air-helium shock tube problem visualized at $t = 0.15$. From left to right: density, velocity, and pressure.
    The exact solution (\ref{pgfplots:shocktube_air_helium_exact}) is compared with the
    level-set result (\ref{pgfplots:shocktube_air_helium_lsm}) and the
    diffuse-interface result (\ref{pgfplots:shocktube_air_helium_dim}).}
    \label{fig:AirHeliumShockTube}
\end{figure}
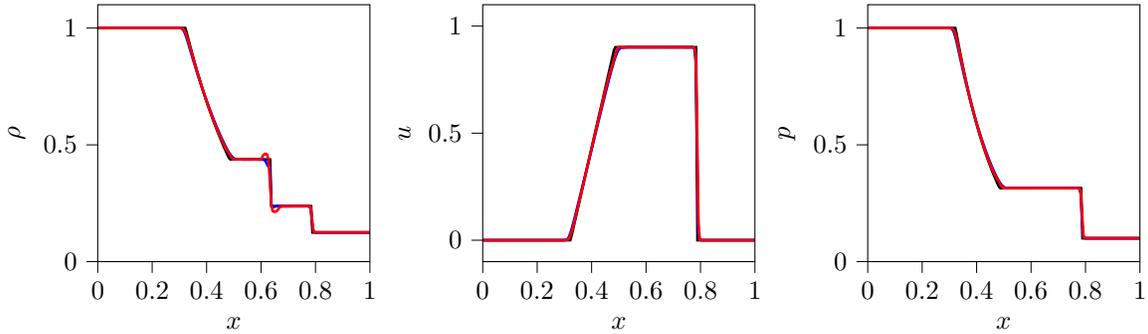

We consider a two-phase gas-gas shock tube problem.
The computational domain $x \in \left[0, 1\right]$ and is discretized by $200$ cells.
We use zero gradient boundary conditions at both boundaries.
An initial discontinuity at $x = 0.5$ separates air on the left from helium on the right.
Initial conditions and material parameters are given in Table \ref{tab:air_helium_shock_tube}.
Figure \ref{fig:AirHeliumShockTube} shows the solutions at time $t = 0.15$.
The reference solution is obtained from an exact Riemann solver \cite{ZEIDAN2007}.
The DIM and LSM results are in very good agreement with the exact solution.
%







\begin{table}[t!]
    \begin{center}
        \footnotesize
        \begin{tabular}{c c c c c c c} 
        \hline
        Material & $x \left[\si{m}\right]$ & $\rho \left[\si{kg/m^3}\right]$ & $u \left[\si{m/s}\right]$ & $p \left[\si{Pa}\right]$ & $\gamma \left[-\right]$ & $p_{\infty} \left[\si{Pa}\right]$ \\
        \hline

        Air     & $x \leq 0.5$    & 1.0     & 0.0 & 1.0 & 1.4 &  0.0 \\
        Helium  & $x > 0.5$       & 0.125   & 0.0 & 0.1 & 1.67 & 0.0 \\
       
        \hline

    \end{tabular}
    \caption{Initial conditions and material parameters for the air-helium shock tube.}
    \label{tab:air_helium_shock_tube}
    \end{center}
\end{table}
\tikzexternaldisable
\tikzexternalenable
\subsubsection{Air-water shock tube problem}
\label{subsubsec:air_water_shock_tube}
We consider a two-phase gas-liquid shock tube problem similarly to \cite{Abgrall2001,Chen2008,Wang2022}.
The high density ratio between the two fluids poses more stringent requirements on the numerical scheme.
The computational domain $x \in \left[0.0, 1.5\right]$ is discretized by $200$ cells.
We use zero gradient boundary conditions at both boundaries.
An initial discontinuity at $x = 0.8$ separates highly compressed water on the left from air on the right.
Initial conditions and material parameters are given in Table \ref{tab:air_water_shock_tube}.
Figure \ref{fig:AirWaterShockTube} shows the solutions at time $t = 0.15 $.
The reference solution is obtained from an exact Riemann solver \cite{ZEIDAN2007}.
The DIM and LSM results are in very good agreement with the exact solution.
\begin{table}[t!]
    \begin{center}
        \footnotesize
        \begin{tabular}{c c c c c c c} 
        \hline
        Material & $x \left[\si{m}\right]$ & $\rho \left[\si{kg/m^3}\right]$ & $u \left[\si{m/s}\right]$ & $p \left[\si{Pa}\right]$ & $\gamma \left[-\right]$ & $p_{\infty} \left[\si{Pa}\right]$ \\
        \hline

        Water   & $x \leq 0.8$    & 1000.0  & 0.0 & $10^{9}$ & 6.12 & $3.43 \times 10^{8}$ \\
        Air     & $x > 0.8$       & 20.0    & 0.0 & $10^{5}$ & 1.4 & 0.0 \\
       
        \hline

    \end{tabular}
    \caption{Initial conditions and material parameters for the air-water shock tube.}
    \label{tab:air_water_shock_tube}
    \end{center}
\end{table}
\tikzexternaldisable
\begin{figure}[!t]
    \centering
    \begin{tikzpicture}

    \definecolor{color0}{RGB}{0,101,189}
    \definecolor{color1}{RGB}{227,114,34}
    \definecolor{color2}{RGB}{162,173,0}
    
    \begin{groupplot}[
        group style={group name=mygroup, group size=3 by 1, horizontal sep=1.5cm},
        width=5.2cm, height=5cm]
      
    \nextgroupplot[
    tick align=outside,
    tick pos=left,
    xtick style={color=black},
    xmin=0.0, xmax=1.5,
    ytick style={color=black},
    xlabel=$x$,
    ylabel=$\rho$,
    ]
    \addplot [line width=1pt, black] table[
        x index = {0}, y index = {1}
        ] {figures/shocktube/data/gasliquid_exact.txt}; \label{pgfplots:shocktube_air_water_exact}
    \addplot [line width=1pt, blue] table[
        x index = {0}, y index = {1}
        ] {figures/shocktube/data/gasliquid_levelset.txt}; \label{pgfplots:shocktube_air_water_lsm}
    \addplot [line width=1pt, red] table[
        x index = {0}, y index = {1}
        ] {figures/shocktube/data/gasliquid_diffuse.txt}; \label{pgfplots:shocktube_air_water_dim}

    \nextgroupplot[
    tick align=outside,
    tick pos=left,
    xtick style={color=black},
    xmin=0.0, xmax=1.5,
    ytick style={color=black},
    xlabel=$x$,
    ylabel=$u$,
    ]
    \addplot [line width=1pt, black] table[
        x index = {0}, y index = {2}
        ] {figures/shocktube/data/gasliquid_exact.txt};
    \addplot [line width=1pt, blue] table[
        x index = {0}, y index = {2}
        ] {figures/shocktube/data/gasliquid_levelset.txt};
    \addplot [line width=1pt, red] table[
        x index = {0}, y index = {2}
        ] {figures/shocktube/data/gasliquid_diffuse.txt};

    \nextgroupplot[
    tick align=outside,
    tick pos=left,
    xtick style={color=black},
    xmin=0.0, xmax=1.5,
    ytick style={color=black},
    xlabel=$x$,
    ylabel=$p$,
    ]
    \addplot [line width=1pt, black] table[
        x index = {0}, y index = {3}
        ] {figures/shocktube/data/gasliquid_exact.txt};
    \addplot [line width=1pt, blue] table[
        x index = {0}, y index = {3}
        ] {figures/shocktube/data/gasliquid_levelset.txt};
    \addplot [line width=1pt, red] table[
        x index = {0}, y index = {3}
        ] {figures/shocktube/data/gasliquid_diffuse.txt};
    \end{groupplot}

    \begin{axis}[
        footnotesize, 
        xmin = 0.4, xmax = 0.55,
        ymin = -2.0e+7, ymax = 4.0e+7,
        at={($(mygroup c3r1.north east)+(-3mm,-3.5mm)$)}, 
        anchor=north east, 
        scale only axis, 
        width=2cm
        ]
    \addplot [line width=1pt, black] table[
        x index = {0}, y index = {3}
        ] {figures/shocktube/data/gasliquid_exact.txt};
    \addplot [line width=1pt, blue] table[
        x index = {0}, y index = {3}
        ] {figures/shocktube/data/gasliquid_levelset.txt};
    \addplot [line width=1pt, red] table[
        x index = {0}, y index = {3}
        ] {figures/shocktube/data/gasliquid_diffuse.txt};
      
    \end{axis}

\end{tikzpicture}
    \caption{Air-water shock tube problem visualized at $t = 0.15$. From left to right: density, velocity, and pressure.
    The exact solution (\ref{pgfplots:shocktube_air_water_exact}) is compared with the
    level-set result (\ref{pgfplots:shocktube_air_water_lsm}) and the
    diffuse-interface result (\ref{pgfplots:shocktube_air_water_dim}).}
    \label{fig:AirWaterShockTube}
\end{figure}
\tikzexternalenable

\subsubsection{Air-helium shock bubble interaction}
\label{subsubsec:air_helium_sbi}
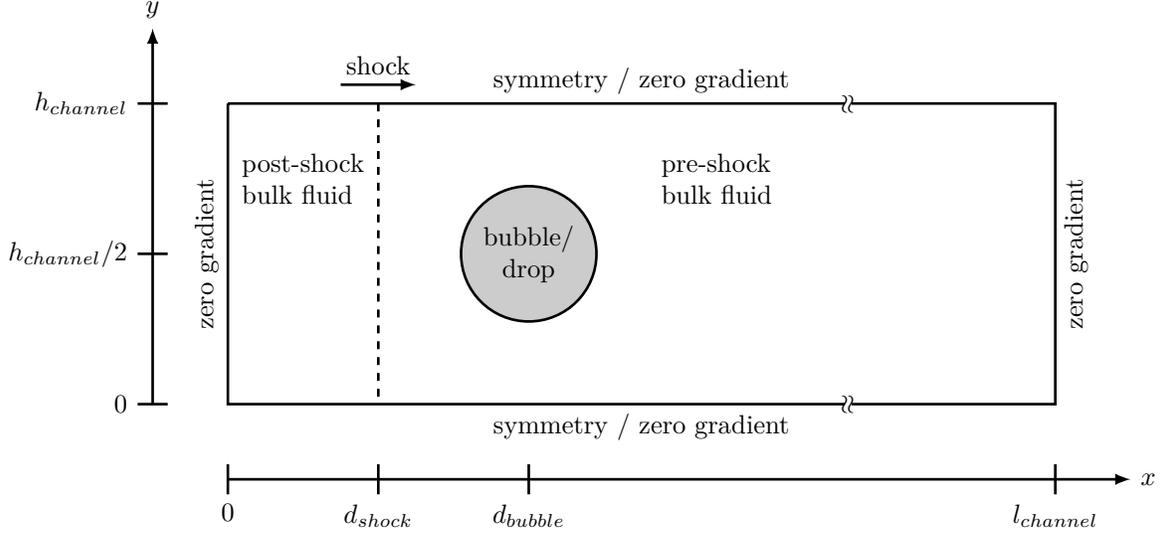
\begin{figure}[!t]
    \centering
    \tikzset{ext/.pic={
\path [fill=white] (-0.2,0)to[bend left](0,0.1)to[bend right](0.2,0.2)to(0.2,0)to[bend left](0,-0.1)to[bend right](-0.2,-0.2)--cycle;
\draw (-0.2,0)to[bend left](0,0.1)to[bend right](0.2,0.2) (0.2,0)to[bend left](0,-0.1)to[bend right](-0.2,-0.2);
}}

\begin{tikzpicture}
    \coordinate (NULL) at (0,0);
    \coordinate (A) at (-4,2);
    \coordinate (B) at (-4,-2);
    \coordinate (C) at (7,-2);
    \coordinate (D) at (7,2);
    \coordinate (center_AB) at ($0.5*(A) + 0.5*(B)$);
    \coordinate (center_BC) at ($0.5*(B) + 0.5*(C)$);
    \coordinate (center_CD) at ($0.5*(C) + 0.5*(D)$);
    \coordinate (center_DA) at ($0.5*(D) + 0.5*(A)$);
    \coordinate (S1) at (-2.0,2);
    \coordinate (S2) at (-2.0,-2);

    \draw[line width=1pt, ->] ($(B) + (-0.0,-1.0)$) -- ($(C) + (1.0,-1.0)$) node[at end, right] {$x$};
    \foreach \x/\y in {-4/0, -2/d_{shock}, 0/d_{bubble}, 7/l_{channel}}
        \draw[line width=1pt] (\x,-2.8) -- (\x,-3.2) node[below]{$\y$};
    \draw[line width=1pt, ->] ($(B) + (-1.0,0.0)$) -- ($(A) + (-1.0,1.0)$) node[at end, above] {$y$};
    \foreach \x/\y in {-2/0, 2/h_{channel}}
        \draw[line width=1pt] (-5.2,\x) node[left]{$\y$} -- (-4.8,\x);
    \draw[line width=1pt] (-5.2,0) node[left]{$h_{channel}/2$} -- (-4.8,0);

    \draw[line width=1pt,fill=black!20!white] (NULL) circle (0.9) node[align=center] {bubble/\\drop};

    \draw[line width=1pt] (A) -- (B) -- (center_BC) -- node[fill=white,rotate=90,inner sep=-1.25pt,outer sep=0,anchor=center]{$\approx$} (C) -- (D) -- node[fill=white,rotate=90,inner sep=-1.25pt,outer sep=0,anchor=center]{$\approx$} (center_DA) -- (A);
    
    \draw[line width=1pt, dashed] (S1) -- (S2);
    \draw[line width=1pt, ->] ($(S1) + (-0.5,0.25)$) -- node[above]{shock} ($(S1) + (0.5,0.25)$);

    \node[align=left] at ($0.5*(A) + 0.5*(S1) + (0.0,-1.0)$) {post-shock\\bulk fluid};
    \node[align=left] at ($0.5*(D) + 0.5*(S1) + (0.0,-1.0)$) {pre-shock\\bulk fluid};

    \node[above,rotate=90] at (center_AB) {zero gradient};
    \node[below] at (center_BC) {symmetry / zero gradient};
    \node[below,rotate=90] at (center_CD) {zero gradient};
    \node[above] at (center_DA) {symmetry / zero gradient};

\end{tikzpicture}
    \caption{Schematic of the computational domain and the initial configuration for 
    the shock bubble interaction and shock drop interaction simulations.}
    \label{fig:shock_bubble_interactions}
\end{figure}

\begin{table}[t!]
    \begin{center}
        \footnotesize
        \begin{tabular}{c c c c c c c c} 
        \hline
        Case & $l_{channel}$ & $h_{channel}$ & $d_{shock}$ & $d_{bubble}$ & $D_{0,bubble}$  & $N_x \times N_y \times N_z$ \\
        \hline

        Helium SBI 2D & 356 & 89 & 60 & 90 & 50 & $4096 \times 1024 \times 1$  \\

        Helium SBI 3D & 356 & 89 & 60 & 90 & 45 & $3072 \times 768 \times 768$ \\

        Water SDI 2D  & 111 & 74 & 20 & 40 & 22 & $3072 \times 2048 \times 1$  \\

        Water SDI 3D  & 111 & 74 & 20 & 40 & 22 & $1536 \times 1024 \times 1024$ \\

        \hline

    \end{tabular}
    \caption{Overview on shock bubble interaction simulations. All length scales are in \si{mm}.}
    \label{tab:OverviewShockBubbleSetups1}
    \end{center}
\end{table}
\begin{table}[t!]
    \begin{center}
        \footnotesize
        \begin{tabular}{c c c c c c c c} 
        \hline
        Case & Incident shock wave & Material & $\rho \left[\si{kg/m^3}\right]$ & $u \left[\si{m/s}\right]$ & $p \left[\si{Pa}\right]$ & $\gamma \left[-\right]$ & $p_{\infty} \left[\si{Pa}\right]$ \\
        \hline

        \multirow{3}{*}{Helium SBI 2D}   & $Ma_S = 1.22$                   & Helium         & 0.1660 & 0.0   & 101325.0 & 1.67                 & 0.0                  \\
                                         & $u_S = 418.746$                & Pre-shock air  & 1.2041 & 0.0   & 101325.0 & \multirow{2}{*}{1.4} & \multirow{2}{*}{0.0} \\
                                         & $t_S = 11.940 \times 10^{-6}$  & Post-shock air & 1.6573 & 114.5 & 159056.0 &                      &                      \\
                                        \\

        \multirow{3}{*}{Helium SBI 3D}   & $Ma_S = 1.25$                   & Helium         & 0.1660 & 0.0   & 101325.0 & 1.67                 & 0.0                  \\
                                         & $u_S = 429.043$                & Pre-shock air  & 1.2041 & 0.0   & 101325.0 & \multirow{2}{*}{1.4} & \multirow{2}{*}{0.0} \\
                                         & $t_S = 17.481 \times 10^{-6}$  & Post-shock air & 1.7201 & 128.7 & 167819.5 &                      &                      \\
                                        \\

        \multirow{3}{*}{Water SDI 2D/3D} & $Ma_S = 2.40$                    & Water          & 1000.0 & 0.0     & 101000.0  & 6.12                 & $3.43 \times 10^{8}$ \\
                                         & $u_S = 834.340$                 & Pre-shock air  & 1.17   & 0.0     & 101000.0  & \multirow{2}{*}{1.4} & \multirow{2}{*}{0.0} \\
                                         & $t_S = 10.787 \times 10^{-6}$   & Post-shock air & 3.7579 & 574.574 & 661886.67 &                      &                      \\

        \hline

    \end{tabular}
    \caption{Initial conditions and material parameters for shock bubble/shock drop interaction simulations.}
    \label{tab:OverviewShockBubbleSetups2}
    \end{center}
\end{table}

The interaction of a shock with a helium bubble
immersed in air is a well-documented 
test case for two-phase flows. 
Experimental data \cite{Haas1987} and
numerical reference simulations \cite{Terashima2009,Hoppe2022}
are available in literature.
We perform two- and three-dimensional simulations of the aforementioned shock bubble interaction (SBI)
and compare results of the level-set (LSM) and the diffuse-interface (DIM) model with experimental data
from Haas et al. \cite{Haas1987} and simulation results from Terashima et al. \cite{Terashima2009}.
The setup of the computational domain
is detailed in Figure \ref{fig:shock_bubble_interactions} and Table \ref{tab:OverviewShockBubbleSetups1}.
The initial conditions and the material properties are listed in Table
\ref{tab:OverviewShockBubbleSetups2}.

\begin{figure}[t!]
    \centering
    \input{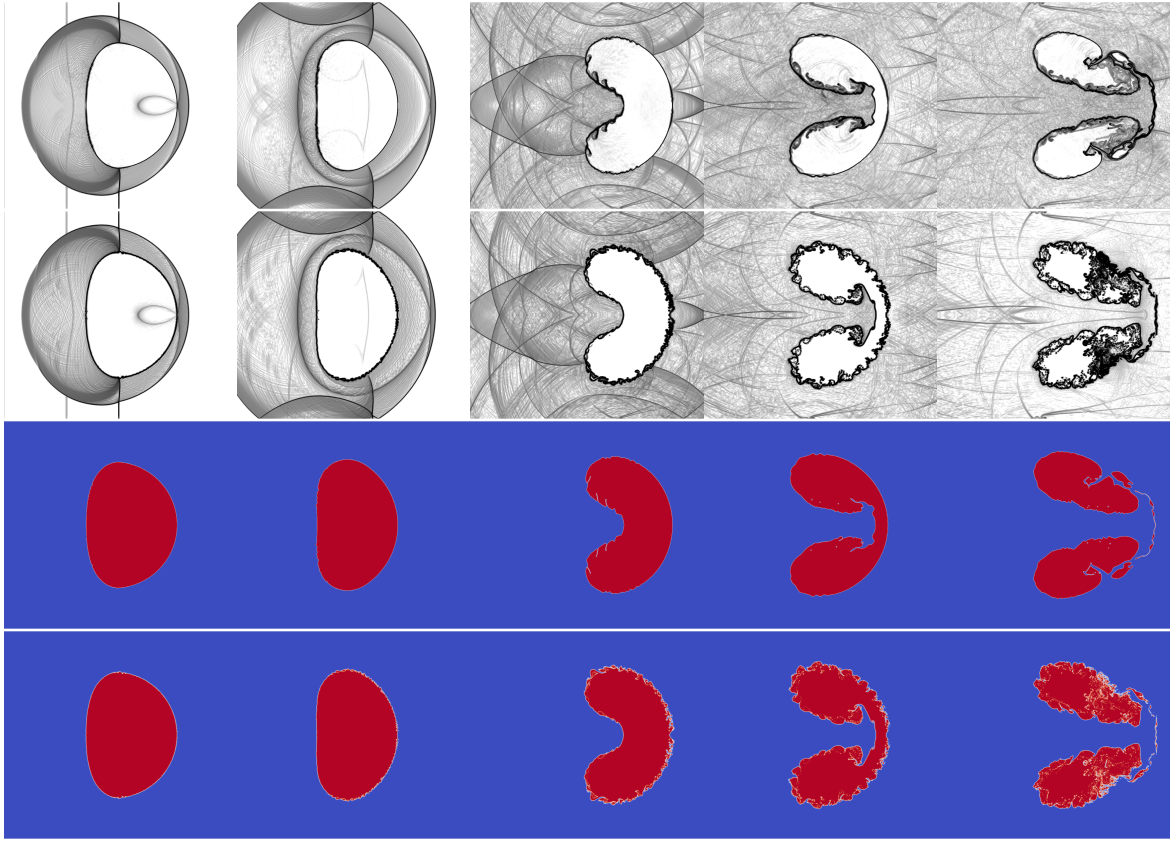}
    \caption{Time series of the two-dimensional air-helium shock bubble interaction.
        From top to bottom:
        Numerical schlieren ($\ln \left( \lVert \nabla \rho \rVert \right)$) from the level-set and diffuse-interface model,
        volume fraction field from level-set and diffuse-interface model.
        In the volume fraction fields, red and blue colors represent
        helium and air, respectively.
        The snapshots are taken at times $t \in \{ 72, 102, 245, 427, 674 \} \,\si{\micro s}$ 
        after the incident shock wave has impacted the helium cylinder.
        (For interpretation of the references to color in this figure legend, 
        the reader is referred to the web version of this article.)
        \textcolor{red}{Note: In the arXiv version, experimental images from \cite{Haas1987} are not shown
        due to limited copyrights.}}
    \label{fig:AirHelium2DFlowfields}
\end{figure}

\begin{figure}[h]
    \centering
    \begin{tikzpicture}

    \begin{groupplot}[
        group style={group size=2 by 1, horizontal sep=1.5cm},
        width=5.2cm, height=5cm]
      
    \nextgroupplot[
    tick align=outside,
    tick pos=left,
    xtick style={color=black},
    ymin=0.0, ymax=1000.0,
    xmin=0.0, xmax=200.0,
    ytick style={color=black},
    xlabel=Position \si{mm},
    ylabel=Time \si{\mu s},
    title=(a)
    ]
    \addplot [line width=1pt, black] table[
        x index = {1}, y index = {0}
        ] {figures/air_helium_shockbubble/2D/statistics/charpoints_SBI_2D_levelset_4096_float64_up.txt}; \label{pgfplots:upstream}
    \addplot [line width=1pt, blue] table[
        x index = {1}, y index = {0}
        ] {figures/air_helium_shockbubble/2D/statistics/charpoints_SBI_2D_levelset_4096_float64_jet.txt}; \label{pgfplots:jet}
    \addplot [line width=1pt, red] table[
        x index = {1}, y index = {0}
        ] {figures/air_helium_shockbubble/2D/statistics/charpoints_SBI_2D_levelset_4096_float64_down.txt}; \label{pgfplots:downstream}
    \addplot [mark=*, only marks, mark size=0.5pt, black] table[
        x index = {0}, y index = {1}
    ] {figures/air_helium_shockbubble/2D/statistics/ref_upstream_2D.txt}; \label{pgfplots:reference_terashima}
    \addplot [mark=*, only marks, mark size=0.5pt, blue] table[
        x index = {0}, y index = {1}
    ] {figures/air_helium_shockbubble/2D/statistics/ref_jet_2D.txt};
    \addplot [mark=*, only marks, mark size=0.5pt, red] table[
        x index = {0}, y index = {1}
    ] {figures/air_helium_shockbubble/2D/statistics/ref_downstream_2D.txt};

    \nextgroupplot[
    tick align=outside,
    tick pos=left,
    xtick style={color=black},
    ymin=0.0, ymax=1000.0,
    xmin=0.0, xmax=200.0,
    ytick style={color=black},
    xlabel=Position \si{mm},
    title=(b)
    ]
    \addplot [line width=1pt, black] table[
        x index = {1}, y index = {0}
        ] {figures/air_helium_shockbubble/2D/statistics/charpoints_SBI_2D_diffuse_4096_float64_up.txt}; 
    \addplot [line width=1pt, blue] table[
        x index = {1}, y index = {0}
        ] {figures/air_helium_shockbubble/2D/statistics/charpoints_SBI_2D_diffuse_4096_float64_jet.txt};
    \addplot [line width=1pt, red] table[
        x index = {1}, y index = {0}
        ] {figures/air_helium_shockbubble/2D/statistics/charpoints_SBI_2D_diffuse_4096_float64_down.txt};
    \addplot [mark=*, only marks, mark size=0.5pt, black] table[
        x index = {0}, y index = {1}
    ] {figures/air_helium_shockbubble/2D/statistics/ref_upstream_2D.txt};
    \addplot [mark=*, only marks, mark size=0.5pt, blue] table[
        x index = {0}, y index = {1}
    ] {figures/air_helium_shockbubble/2D/statistics/ref_jet_2D.txt};
    \addplot [mark=*, only marks, mark size=0.5pt, red] table[
        x index = {0}, y index = {1}
    ] {figures/air_helium_shockbubble/2D/statistics/ref_downstream_2D.txt};

    \end{groupplot}

\end{tikzpicture}
    \caption{Space-time diagram of three characteristic interface points for the level-set method (left)
    and the diffuse-interface method (right). 
    Positions of the upstream point (left-most point of the interface, (\ref{pgfplots:upstream})),
    the jet (left-most point of the interface on the center-line, (\ref{pgfplots:jet})),
    and 
    the downstream point (right-most point of the interface, (\ref{pgfplots:downstream})), 
    are depicted as solid lines.
    Reference values are taken from Terashima et al. \cite{Terashima2009}
    and are depicted as markers (\ref{pgfplots:reference_terashima}).
    (For interpretation of the references to color in this figure legend, 
    the reader is referred to the web version of this article.)
    }
    \label{fig:AirHelium2DStatistics}
\end{figure}
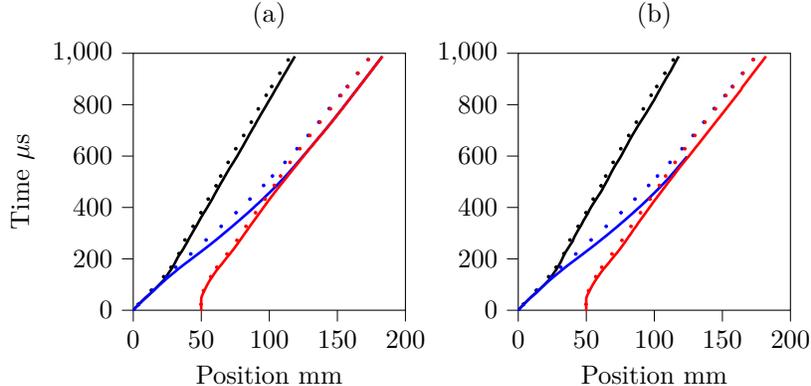

We begin with the two-dimensional test case.
The resolution of the two-dimensional simulations are $4096 \times 1024$,
resulting in approximately $D_0 / \Delta x \approx 575$, where $D_0$ is the initial
bubble diameter.
Simulations were performed in parallel on 4 Nvidia A6000 GPUs.
Figure \ref{fig:AirHelium2DFlowfields} compares snapshots of the flow field between the numerical simulations
and the experiments by Haas et al. \cite{Haas1987}.
The numerical schlieren from the LSM and DIM simulations are in very good agreement with
the experimental schlieren images.
Both schemes accurately predict the propagation of the incident shock wave and the reflected waves.
The macroscopic bubble shape is in good agreement between the LSM and the DIM.
The LSM produces a smoother interface topology while the
DIM generates small scale instabilities at the material interface.
The smoother interface in the level-set case is attributed to the
mixing procedure and first-order reconstruction at the interface.
Inspection of the volume fraction fields reveals that THINC is able to maintain
a constant interface thickness over the duration of the simulation as long as interface features
are resolved on the given grid.
At later times, once finer and finer structures appear, we observe some mixture regions.
The LSM conserves a sharp interface.
Underresolved structures are eliminated by the reinitialization procedure. 
Figure \ref{fig:AirHelium2DStatistics} compares the temporal evolution of characteristic interface points.
Again, both numerical schemes agree very well with the experimental data from \cite{Haas1987}.

\begin{figure}
    \centering
    \input{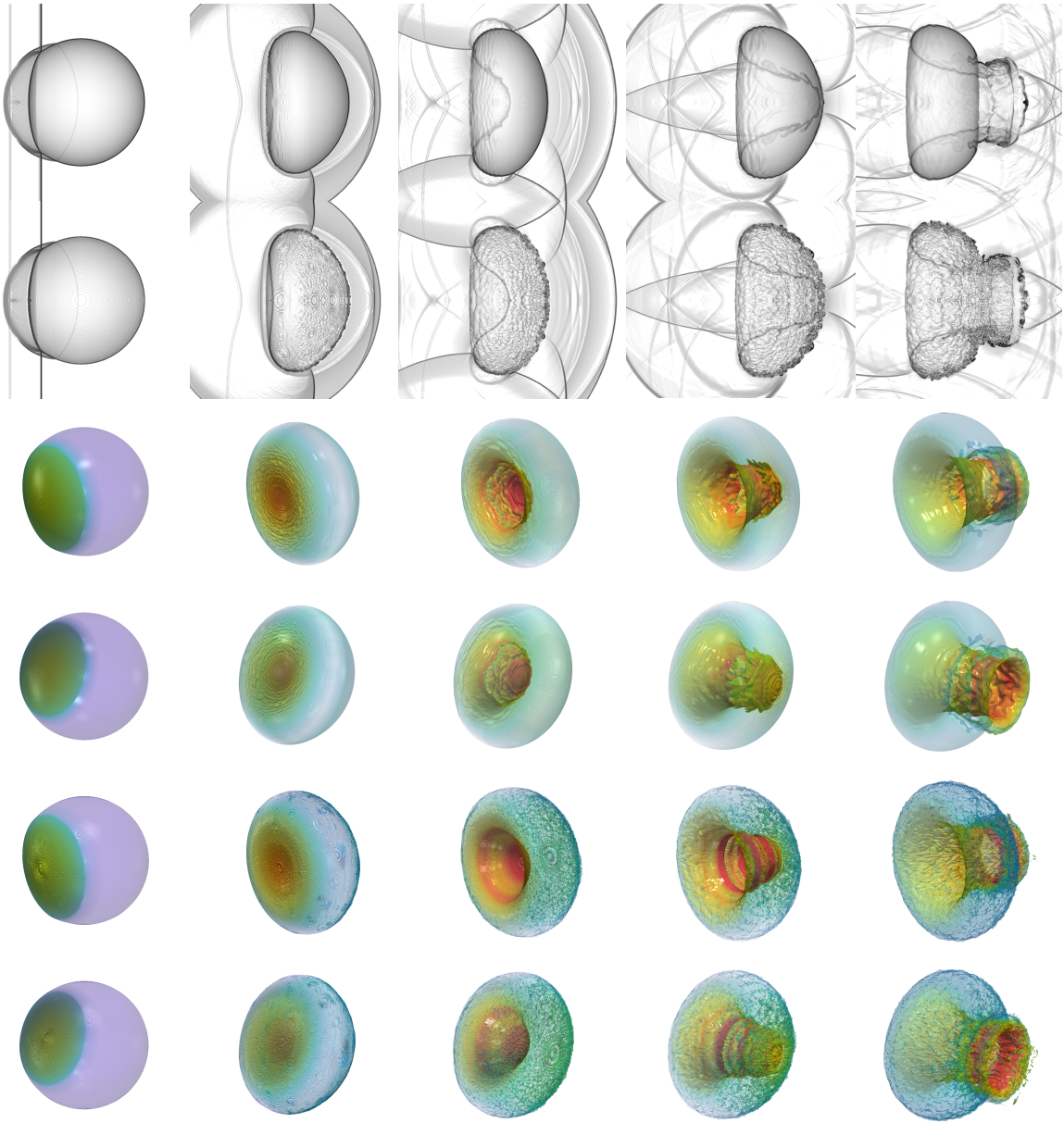}
    \caption{Time series of the three-dimensional air-helium shock bubble interaction.
    The first two rows show numerical schlieren ($\ln \left( \lVert \nabla \rho \rVert \right)$) of the level-set (row 1)
    and diffuse-interface (row 2) model, respectively.
    The next four rows display the interface contour colored by the absolute velocity.
    We illustrate the solution of the level-set (row 3 \& 4) and the diffuse-interface (row 5 \& 6) model, respectively.
    The snapshots are taken at times $t \in \{ 20,82,145,223,350 \} \,\si{\micro s}$ 
    after the incident shock wave has impacted the helium cylinder.
    (For interpretation of the references to color in this figure legend, 
    the reader is referred to the web version of this article.)
    \textcolor{red}{Note: In the arXiv version, experimental images from \cite{Haas1987} are not shown
    due to limited copyrights.}
    }
    \label{fig:AirHelium3DFlowfields}
\end{figure}

We also compare the three-dimensional SBI with the experimental data of Haas et al. \cite{Haas1987}.
The resolutions of the three-dimensional simulations are $3072 \times 768 \times 768$.
The corresponding resolution of the initial bubble diameter is $D_0/\Delta x \approx 388$.
The three-dimensional simulations were done on a TPU v3 256 pod slice. 
Figure \ref{fig:AirHelium3DFlowfields} compares snapshots of the flow field.
The numerical schlieren for both interface models are in good agreement with the experimental data.
Similarly to the 2D results, we observe pronounced interfacial instabilities in the DIM
while the LSM maintains a smoother interface.
\tikzexternaldisable
\definecolor{color0}{RGB}{0,101,189}
\definecolor{color1}{RGB}{227,114,34}
\definecolor{color2}{RGB}{162,173,0}
\tikzexternalenable
%

%

\subsubsection{Air-water shock drop interaction}
\label{subsubsec:air_water_sdi}
As a second, more challenging test case for the two-phase schemes,
we consider the interaction of a Mach 2.4 shock with a water cylinder (in 2D)
and a water drop (in 3D). 
The high density ratio of the fluids involved poses stringent requirements on the 
positivity-preserving routine.
Experimental data are taken from the work of Sembian et al. \cite{Sembian2016a}.
Details on the setup of the computational domain are shown in
Figure \ref{fig:shock_bubble_interactions} and Table \ref{tab:OverviewShockBubbleSetups1}.
The initial conditions and material properties are listed in Table \ref{tab:OverviewShockBubbleSetups2}

\begin{figure}[t!]
    \centering
    \input{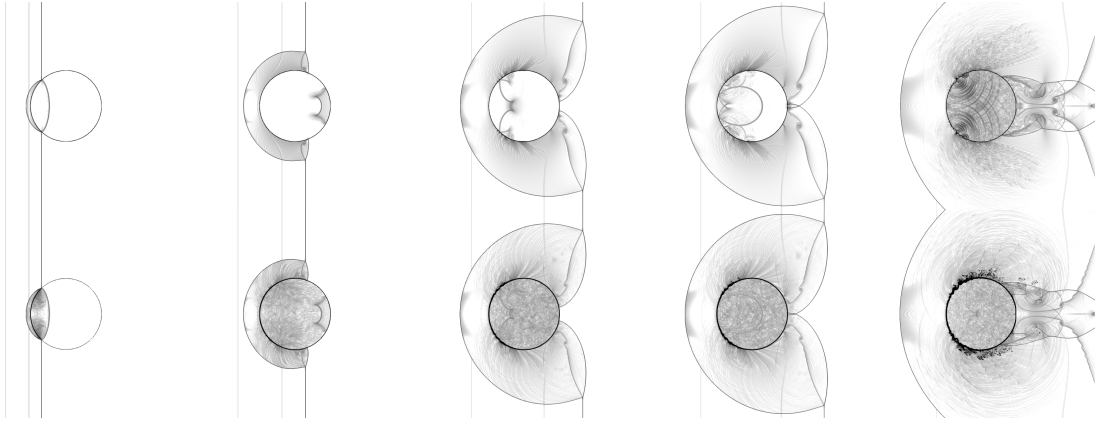}
    \caption{Time series of the two-dimensional air-water shock drop interaction.
    From top to bottom: Numerical schlieren ($\ln \left( \lVert \nabla \rho \rVert \right)$) 
    from level-set and the diffuse-interface model.
    The snapshots are taken at times $t \in \{ 4,17,35,40,67 \} \,\si{\micro s}$ after the incident
    shock wave has impacted the water cylinder.
    \textcolor{red}{Note: In the arXiv version, experimental images from \cite{Sembian2016a} are not shown
    due to limited copyrights.}
    }
    \label{fig:AirWater2DFlowfields}
\end{figure}

\begin{figure}[t!]
    \centering
    \input{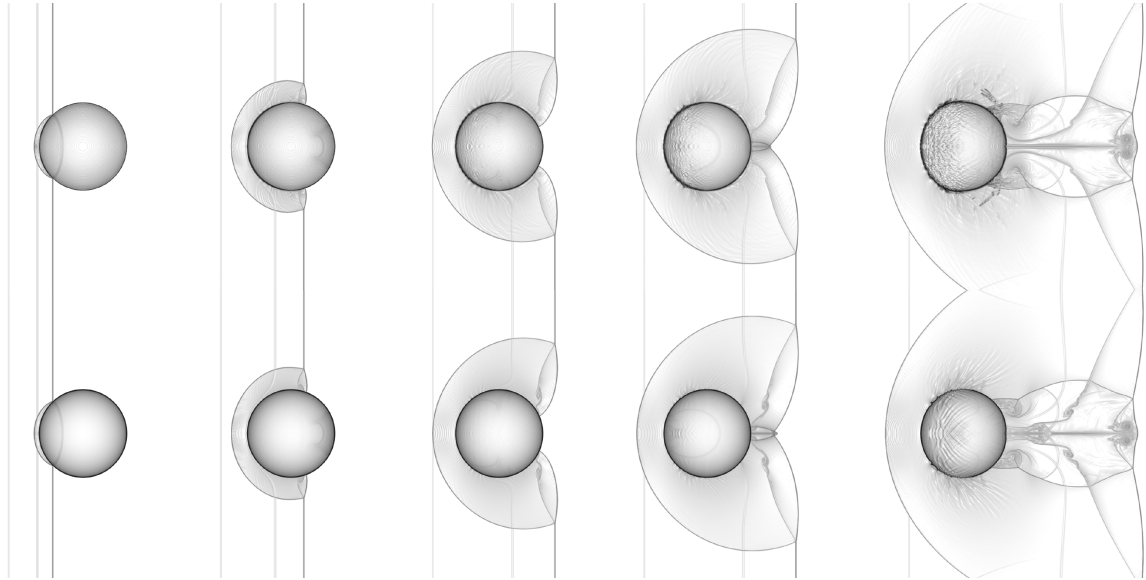}
    \caption{
        Time series of the three-dimensional air-water shock drop interaction.
        From top to bottom: Numerical schlieren ($\ln \left( \lVert \nabla \rho \rVert \right)$)
        from level-set and the diffuse-interface model.
        The snapshots are taken at times $t \in \{ 4,17,35,40,67 \} \,\si{\micro s}$ after the incident
        shock wave has impacted the water cylinder.}
    \label{fig:AirWater3DFlowfields}
\end{figure}

We begin with the two-dimensional test case which was run on 4 Nvidia A6000 GPUs.
The resolution of the initial bubble diameter is $D_0 / \Delta x \approx 608$. 
Figure \ref{fig:AirWater2DFlowfields} shows a comparison between the experimental schlieren images
and our simulations.
The numerical results for the level-set method (LSM) and the diffuse-interface method (DIM) are in excellent 
qualitative agreement with the experimental reference data.
Especially, both interface models capture very well the propagation of incident,
reflected, and transmitted waves.
For later stages, we observe fluid being sheared of the rim of the cylinder in the diffuse-interface simulation.
Figure \ref{fig:AirWater3DFlowfields} shows the numerical schlieren images for the corresponding 3D simulations ($D_0 / \Delta x \approx 304$).
The three-dimensional simulations were done on a TPU v3 256 pod slice.
Again, we observe good agreement between SIM and DIM.

\section{Automatic-differentiation gradients}
\label{sec:verification_gradients}
In this section, we showcase the computation of automatic differentiation gradients in the JAX-Fluids framework.
In particular, we provide verification for settings in which gradients are calculated across
multiple devices.

Figure \ref{fig:run_solver} illustrates two generic 
scenarios for AD-based gradient calculation in JAX-Fluids.
In the first scenario, we are interested in calculating gradients of our quantity of interest with
respect to a parameterized initial condition, see Fig. \ref{fig:run_solver1}.
The function \mintinline{python}{fun()} consists of three steps: calculation of initial conditions 
based on the given parameters, forward pass through the JAX-Fluids solver,
and calculation of the quantity of interest.
If each of these steps is differentiable, we can call the JAX primitive \mintinline{python}{jax.value_and_grad}
to obtain a transformed function which computes the quantity of interest and its derivative 
with respect to the input parameters \mintinline{python}{ic_params}.
Gradients are backpropagated through \mintinline{python}{compute_qoi()}, through the entire \mintinline{python}{feedforward()}
routine which comprises multiple integration steps, and, finally, through the \mintinline{python}{compute_initial_condtion()}
routine.

The scenario depicted in Fig. \ref{fig:run_solver2} is prototypical of training neural network parameters 
within the JAX-Fluids framework.
Here, \mintinline{python}{nn_params} should represent weights and biases 
of a neural network model which is repeatedly being used during forward integration.
Similarly to the aforementioned case, gradients of the quantity of interest (e.g., a user-specified loss function)
with respect to the parameters \mintinline{python}{nn_params} are obtained by using the 
function transformation \mintinline{python}{jax.value_and_grad}.
In \cite{Bezgin2022}, we showcase the training of neural networks within JAX-Fluids.

We use \textit{checkpointing} \cite{Margossian2019} (\mintinline{python}{jax.checkpoint})
to overcome memory bottlenecks related to the length of the trajectory that is unrolled
during the forward pass \mintinline{python}{feedforward()}.
Here, we specify distinct locations (checkpoints) in the computational graph where
the state is stored during the forward pass, instead of storing all intermediate results.
During the backward pass, we recompute the required intermediate results starting from stored checkpoints.
Checkpointing is a trade-off between memory consumption and computation time.
Furthermore, we prevent JAX from unrolling the entire trajectory during
compilation of \mintinline{python}{feedforward()} using \mintinline{python}{jax.lax.scan},
reducing the compilation time
significantly.

\begin{figure}
    \centering
    \begin{subfigure}{0.4\textwidth}
\begin{minted}{python}
import jax
from jaxfluids import SimulationManager

simulation_manager = SimulationManager(
    "case_setup.json", "numerical_setup.json")

def fun(ic_params):
    y0 = compute_initial_condition(ic_params)
    yn = simulation_manager.feedforward(y0, n_steps)
    qoi = compute_qoi(yn)
    return qoi

fun = jax.value_and_grad(fun)
\end{minted}
        \caption{}
        \label{fig:run_solver1}
    \end{subfigure}
    \hfill
    \begin{subfigure}{0.4\textwidth}
\begin{minted}{python}
import jax
from jaxfluids import SimulationManager

simulation_manager = SimulationManager(
    "case_setup.json", "numerical_setup.json")

def fun(nn_params):
    y0 = compute_initial_condition()
    yn = simulation_manager.feedforward(y0, n_steps, nn_params)
    qoi = compute_qoi(yn)
    return qoi

fun = jax.value_and_grad(fun)
\end{minted}
        \caption{}
        \label{fig:run_solver2}
    \end{subfigure}
    \caption{Code snippets illustrating how to obtain automatic differentiation gradients 
    through a simulation with JAX-Fluids. With \mintinline{python}{simulation_manager.feedforward()},
    the forward simulation through the JAX-Fluids solver is called, and the initial condition is
    integrated for \mintinline{python}{n_steps}.}
    \label{fig:run_solver}
    \end{figure}

\subsection{Automatic differentiation gradients for a quasi one-dimensional moving shock}
\label{subsec:gradients_shock}
To verify automatic-differentiation (AD) gradients for parallel
simulations, we investigate the convergence of finite-difference (FD) gradients towards AD gradients.
We consider a two-dimensional domain $(x,y)\in[-0.5, 0.5]\times[-0.5, 0.5]$ discretized with
$512\times 512$ cells and decomposed using two blocks that are oriented in a $1\times2$ grid.
The flow field is initialized with a discontinuity positioned at $x=0.0$ describing
a quasi one-dimensional moving shock with shock Mach number $Ma_S$.
The pre (index l) and post (index r) conditions depend on $Ma_S$ and are related by the
Rankine-Hugoniot conditions \cite{Toro2009a}. Fixing the post shock state
to $p_r = \rho_r = 1.0$ and $u_r = v_r = 0.0$, the initial condition is parameterized solely
by $Ma_S$. For the diffuse-interface (DIM) and level-set (LSM) model, we consider two phases
with the interface placed on the shock discontinuity. The specific heat capacity
ratio for both phases is $\gamma=1.4$.
As the shock propagates into the quiescent domain, the total energy
\begin{align}
    E(t,Ma_S) = \int_{x,y} \left( \rho e + \frac{1}{2} \rho \mathbf{u}\cdot \mathbf{u} \right) \,dx dy
\end{align}
increases. We are interested in the gradient $g$ 
of the total energy increase $\Delta E^n(Ma_S) = E(t^n,Ma_S)-E(t^0,Ma_S)$
with respect to $Ma_S$ after a fixed amount of time steps $n$.
\begin{align}
    g = \frac{\partial \Delta E^n(Ma_S)}{\partial Ma_S}
\end{align}
We compute this derivative using AD, denoted by $g_{AD}$, and second-order central FD
\begin{align}
    g_{FD} = \frac{\Delta E^n (Ma_S + \epsilon_{FD}) - \Delta E^n (Ma_S - \epsilon_{FD})}{2 \epsilon_{FD}},
\end{align}
where $\epsilon_{FD}$ indicates a small number. The computation of $g_{AD}$ requires
the differentiation through the entire (parallel) JAX-Fluids algorithm for multiple time steps.
The calculation of $g_{FD}$ requires two independent forward simulations.

\tikzexternaldisable
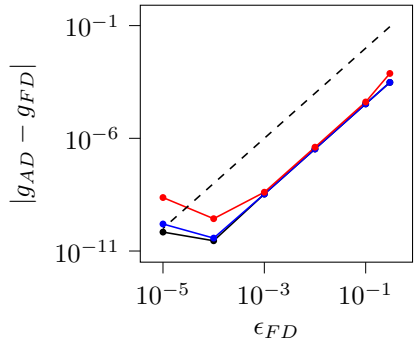
\begin{figure}[t]
    \centering
\begin{tikzpicture}

    \begin{axis}[
    width = 5.201cm,
    height = 5cm,
    log basis x={10},
    log basis y={10},
    tick align=outside,
    tick pos=left,
    x grid style={darkgray176},
    xlabel={$\epsilon_{FD}$},
    xmode=log,
    xtick style={color=black},
    y grid style={darkgray176},
    ylabel={$\vert g_{AD} - g_{FD} \vert$},
    ymode=log,
    ytick style={color=black},
    legend style={at={(0.05,0.95)},draw=none,anchor=north west},
    ]
    \addplot [semithick, mark=*, mark size=1pt, black] table[
        x index = {0}, y index = {1}
    ] {figures/gradients/verification/singlephase_1_2_1.txt}; \label{pgfplotss:ad_singlephase}
    \addplot [semithick, blue, mark=*, mark size=1pt] table[
        x index = {0}, y index = {1}
    ] {figures/gradients/verification/diffuse_1_2_1.txt}; \label{pgfplotss:ad_diffuse}
    \addplot [semithick, red, mark=*, mark size=1pt] table[
        x index = {0}, y index = {1}
    ] {figures/gradients/verification/levelset_1_2_1.txt}; \label{pgfplotss:ad_levelset}
    \addplot [semithick, black, dashed]
    table {%
    0.3 0.09
    0.1 0.01
    0.01 0.0001
    0.001 1e-06
    0.0001 1e-08
    1e-05 1e-10
    }; \label{pgfplotss:ad_nom}
    \end{axis}
    
    \end{tikzpicture}
    
    \caption{Convergence of finite-difference gradients towards
    automatic differentiation gradients. The black (\ref{pgfplotss:ad_singlephase}),
    blue (\ref{pgfplotss:ad_diffuse}), and red (\ref{pgfplotss:ad_levelset})
    lines indicate the single-phase, diffuse-interface and
    level-set model, respectively. The black, dashed line
    (\ref{pgfplotss:ad_nom}) depicts second-order
    convergence $\mathcal{O}\left(\epsilon_{FD}^2\right)$.}
    \label{fig:ad_convergence}
\end{figure}
\tikzexternalenable
We choose $Ma_S=2.0$, $n=40$, and a fixed time step size of $\Delta t=10^{-4}$. The numerical setup for all models
uses a WENO5-Z spatial reconstruction in combination with an HLLC Riemann solver (Sec. \ref{subsec:high_order_godunov}).
The integration is performed using the TVD-RK3 scheme (Sec. \ref{subsec:temporal_integration}).
Figure \ref{fig:ad_convergence} shows the convergence of the absolute 
error $|g_{AD} - g_{FD}|$ with respect to $\epsilon_{FD}$. As expected, we observe second-order convergence for all 
models. Beyond $\epsilon_{FD} \approx 10^{-4}$, the convergence order starts to reduce due to floating point errors.

\subsection{Sensitivities of a helium-air shock bubble interaction}
\label{subsec:gradients_sbi}
In a second test, we showcase the capability of JAX-Fluids to compute automatic differentiation gradients 
through extended simulations run in parallel on multi GPU devices.
In particular, we are interested in sensitivities of initial conditions
on quantities of interest in helium-air shock bubble interaction simulations, 
see Sec. \ref{subsubsec:air_water_sdi}.

Figure \ref{fig:run_solver1} gives a schematic for the computation of sensitivities.
We consider an incident shock wave that interacts with a helium-cylinder which is initially at rest.  
In experimental settings, initial conditions may pose one of the main sources of uncertainty.
For example, in the case at hand, there may be uncertainty about the strength of the incident 
shock or the shape of the initial helium bubble.
We want to understand how variations in these parameters may affect quantities of interest.
We parameterize the strength of the incident shock by the shock Mach number $Ma_S$.
Additionally, we parameterize the shape of the initial helium bubble by its eccentricity $e_b$.
We require that the mass of the helium bubble is invariant with respect to $e_b$,
i.e., that the area of the bubble at $t = 0$ is constant for varying $e_b$.
Therefore, we define the semi-major axis $a_{bubble} = R_{0,bubble} / \left(1 - e_b^2\right)^{0.25}$ and 
the semi-minor axis $b_{bubble} = R_{0,bubble} \left(1 - e_b^2\right)^{0.25}$.
Here, $R_{0,bubble}$ is the radius of the circle ($e_b = 0$) with equivalent surface area.
$a_{bubble}$ is along the $x$-direction for positive $e_b$ and along the $y$-direction for negative $e_b$.
As quantity of interest (QoI), we choose the center of mass drift of the helium bubble 
$\Delta x_{com}^{n}$ in $x$-direction.
Simulations are done with the diffuse-interface model 
(see Secs. \ref{subsec:diffuse_interface_model_phys} and \ref{subsec:diffuse_interface_model_num}).
The center of mass drift of the helium bubble (fluid 1) is calculated as
\begin{align}
    x_{com}(t) &= \frac{ \int_{\Omega} \alpha_1 \rho_1 x\, dx dy }{ \int_{\Omega} \alpha_1 \rho_1 \,dx dy}, \\
    \Delta x_{com}^{n} &= x_{com}(t^{n}) - x_{com}(t=0).
\end{align}
Hence, we are interested in specifying the following sensitivities:
\begin{equation}
    \mathbf{g} = \left[ g_1, g_2 \right]^T = \left[ \frac{\partial \Delta x_{com}^{n}}{\partial Ma_S}, \quad \frac{\partial \Delta x_{com}^{n}}{\partial e_b} \right]^T.
\end{equation}

\tikzexternaldisable
\definecolor{cyan}{RGB}{0,255,255}
\definecolor{magenta}{RGB}{255,0,255}
\definecolor{mediumslateblue128127255}{RGB}{128,127,255}
\begin{figure}[t!]
    \centering
\begin{tikzpicture}


\newcommand{\myLineWidth}{0.75pt}

    \begin{axis}[
        name=plot1, height=5cm, width=5cm,
        tick align=outside,
        tick pos=left,
        xmin=0, xmax=0.5,
        ymin=-0.1, ymax=1.3,
        xlabel={$t / (D_{0,bubble} / u_{ref})$},
        ylabel={$x_{com} / D_{0,bubble}$},
        title=(a)
    ]
    \addplot [draw=black, dotted, line width=0.75pt]
    table[x index={0}, y index={1}]{figures/gradients/verification_2D/2.5e-4/plot_data_4.txt};
    \addplot [draw=black, line width=0.75pt]
    table[x index={0}, y index={2}]{figures/gradients/verification_2D/2.5e-4/plot_data_4.txt};
    \addplot [draw=black, dashed, line width=0.75pt]
    table[x index={0}, y index={3}]{figures/gradients/verification_2D/2.5e-4/plot_data_4.txt};
    
    \addplot [draw=blue, dotted, line width=0.75pt]
    table[x index={0}, y index={4}]{figures/gradients/verification_2D/2.5e-4/plot_data_4.txt};
    \addplot [draw=blue, line width=0.75pt]
    table[x index={0}, y index={5}]{figures/gradients/verification_2D/2.5e-4/plot_data_4.txt};
    \addplot [draw=blue, dashed, line width=0.75pt]
    table[x index={0}, y index={6}]{figures/gradients/verification_2D/2.5e-4/plot_data_4.txt};

    \addplot [draw=red, dotted, line width=0.75pt]
    table[x index={0}, y index={7}]{figures/gradients/verification_2D/2.5e-4/plot_data_4.txt};
    \addplot [draw=red, line width=0.75pt]
    table[x index={0}, y index={8}]{figures/gradients/verification_2D/2.5e-4/plot_data_4.txt};
    \addplot [draw=red, dashed, line width=0.75pt]
    table[x index={0}, y index={9}]{figures/gradients/verification_2D/2.5e-4/plot_data_4.txt};

    \end{axis}

    \begin{axis}[
        name=plot2,
        at={($(plot1.east)+(2.5cm,0)$)},
        anchor=west,
        height=5cm,
        width=6.6cm,
        tick align=outside,
        tick pos=left,
        xmin=-1, xmax=2,
        ymin=-1, ymax=1,
        xlabel={$x / D_{0,bubble}$},
        ylabel={$y / D_{0,bubble}$},
        title=(b)
    ]
        \addplot [draw=gray, dotted, line width=0.75pt]
        table{figures/gradients/verification_2D/2.5e-4/contour-000.dat}; \label{pgfplots:eb0}
        \addplot [draw=gray, line width=0.75pt]
        table{figures/gradients/verification_2D/2.5e-4/contour-001.dat}; \label{pgfplots:eb1}
        \addplot [draw=gray, dashed, line width=0.75pt]
        table{figures/gradients/verification_2D/2.5e-4/contour-002.dat}; \label{pgfplots:eb2}
        
        \addplot [dotted, draw=black, line width=0.75pt]
        table{figures/gradients/verification_2D/2.5e-4/contour-003.dat};
        \addplot [draw=black, line width=0.75pt]
        table{figures/gradients/verification_2D/2.5e-4/contour-004.dat}; \label{pgfplots:Ms0}
        \addplot [dashed, draw=black, line width=0.75pt]
        table{figures/gradients/verification_2D/2.5e-4/contour-005.dat};
        
        \addplot [dotted, draw=blue, line width=0.75pt]
        table{figures/gradients/verification_2D/2.5e-4/contour-006.dat};
        \addplot [draw=blue, line width=0.75pt]
        table{figures/gradients/verification_2D/2.5e-4/contour-007.dat}; \label{pgfplots:Ms1}
        \addplot [dashed, draw=blue, line width=0.75pt]
        table{figures/gradients/verification_2D/2.5e-4/contour-008.dat};
        
        \addplot [dotted, draw=red, line width=0.75pt]
        table{figures/gradients/verification_2D/2.5e-4/contour-009.dat};
        \addplot [draw=red, line width=0.75pt]
        table{figures/gradients/verification_2D/2.5e-4/contour-010.dat}; \label{pgfplots:Ms2}
        \addplot [dashed, draw=red, line width=0.75pt]
        table{figures/gradients/verification_2D/2.5e-4/contour-011.dat};
    \end{axis}

\end{tikzpicture}
    \caption{
        (a) Temporal evolution of the center of mass
        and (b) interface location at final time $t^n = 2.5 \times 10^{-4}$ 
        for various incident shock Mach numbers $Ma_S$ 
        and initial bubble eccentricities $e_b$.
        The shock Mach numbers $Ma_S \in \left\{1.1, 1.3, 1.5\right\}$ are
        depicted as black (\ref{pgfplots:Ms0}), blue (\ref{pgfplots:Ms1}),
        and red (\ref{pgfplots:Ms2}) lines.
        The bubble eccentricities $e_b \in \left\{-0.5, 0.0, 0.5\right\}$ are
        depicted as dotted (\ref{pgfplots:eb0}), solid (\ref{pgfplots:eb1}), 
        and dashed (\ref{pgfplots:eb2}) lines, respectively.
        The gray lines (\ref{pgfplots:eb1}) in (b) visualize the bubble interfaces 
        in the initial configuration ($t = 0$).
        Spatial coordinates are nondimensionalized with the initial bubble diameter $D_{0_bubble} = 5 \times 10^{-2}$.
        Time is nondimensionalized with the timescale $D_{0_bubble} / u_{ref}$ 
        where the reference velocity set to $u_{ref} = 100$.
        (For interpretation of the references to color in this figure legend, 
        the reader is referred to the web version of this article.)
    }
    \label{fig:center_of_mass_drift_interface_shape}
\end{figure}
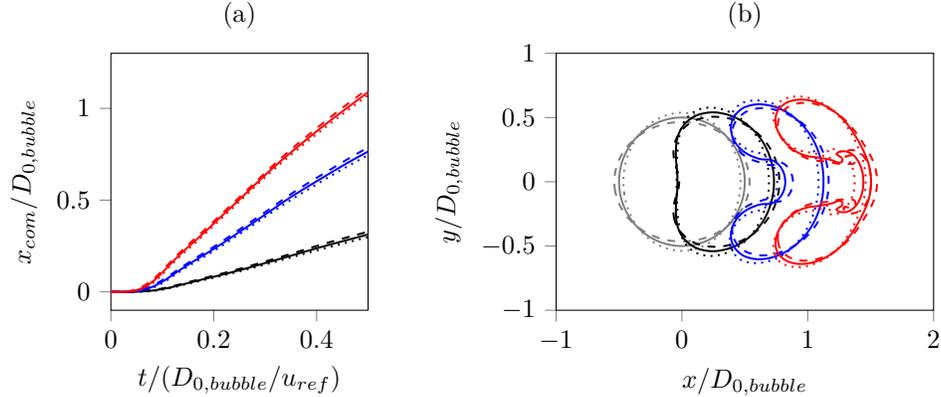
\tikzexternalenable

Except for shock strength and bubble shape, the computational domain and the initial configuration 
are the same as in the air-helium SBI case discussed in Sec. \ref{subsubsec:air_helium_sbi}.
In particular, $R_{0_bubble} = 2.5 \times 10^{-2}$, and the pre-shocked state for air and helium
is given in Table \ref{tab:air_helium_shock_tube}.
The domain is discretized by $512 \times 128$ uniform cells.
We choose a constant time step $\Delta t = 10^{-7}$ and use TVD-RK3 integration.
The center of mass drift is evaluated at $t^n = 10^{-4}$ and $t^n = 2.5 \times 10^{-4}$.
This corresponds to a total of $n=1000$ and $n=2500$ integration steps, respectively,
across which gradients have to be backpropagated.
We use checkpoints after every full integration step such that we do not need to keep
intermediate values of the Runge-Kutta stages in memory.
Each computation is done in parallel on 4 NVIDIA A6000 GPUs.  
The gradients are evaluated with automatic differentiation and finite differences.
Here, we use $\epsilon_{FD} = 10^{-2}$ for the FD gradients.
Gradients are computed for $Ma_S \times e_b \in \left[1.1, 1.5\right] \times \left[-0.5, 0.5\right]$
which is discretized uniformly with $21 \times 21$ points.
We highlight two important aspects with respect to the computation of the gradients:
(i) In order to compute gradients with central finite differences,
we need to run two simulations
per point of the $Ma_S-e_b$ grid per component of the gradient.
For automatic differentiation gradients, however, a single forward run with subsequent backward pass
is sufficient for each $(Ma_S, e_b)$.
(ii) The computational effort associated with the AD gradients only increases slightly
for each additional sensitivity that we want to compute.
This is because the computational graphs for the forward simulation (\mintinline{python}{feedforward()}) 
and the computation of the quantity of interest (\mintinline{python}{compute_qoi()}) remain unchanged,
and only the backpropagation through \mintinline{python}{compute_initial_condition()}
has to account for additional sensitivities.

\begin{figure}[t!]
    \centering
    \begin{tikzpicture}
\pgfmathdeclarefunction{lg10}{1}{%
    \pgfmathparse{ln(#1)/ln(10)}%
}

\pgfplotsset{
    colormap={spectralr}{rgb255=(94,79,162) rgb255=(68,112,177) rgb255=(59,146,184) rgb255=(89,180,170)
        rgb255=(126,203,164) rgb255=(166,219,164) rgb255=(202,233,157) rgb255=(232,246,156) 
        rgb255=(247,252,179) rgb255=(254,245,175) rgb255=(254,227,145) rgb255=(253,200,119) 
        rgb255=(252,170,95) rgb255=(247,131,77) rgb255=(236,97,69) rgb255=(218,70,76)
        rgb255=(190,36,73) rgb255=(158,1,66)}}

\begin{groupplot}[
    group style={
        group size=2 by 2,
        horizontal sep=3cm,
        vertical sep=2cm
    }, 
    width=5.5cm, 
    height=5.5cm
    ]
    \nextgroupplot[
        tick align=outside,
        tick pos=left,
        title={(a)},
        xtick style={color=black},
        ylabel={$e_b$},
        ytick style={color=black},
        enlargelimits=false,
        colormap name=spectralr,
        colorbar,
        colorbar style={
            ytick style={
                color=black,
            }
        }
    ]
    \addplot [
        matrix plot*,
        mesh/cols=21,
        mesh/rows=21,
        point meta=explicit,
        mesh/ordering=y varies,
    ] table[meta=value] {figures/gradients/verification_2D/1e-4/plot_data_0.txt};

    \nextgroupplot[
        tick align=outside,
        tick pos=left,
        title={(b)},
        xtick style={color=black},
        ytick style={color=black},
        enlargelimits=false,
        colormap name=spectralr,
        colorbar,
        colorbar style={
            ytick style={
                color=black,
            }
        }
    ]
    \addplot [
        matrix plot*,
        mesh/cols=21,
        mesh/rows=21,
        point meta=explicit,
        mesh/ordering=y varies,
    ] table[meta=value] {figures/gradients/verification_2D/1e-4/plot_data_2.txt};

    \nextgroupplot[
        tick align=outside,
        tick pos=left,
        title={(c)},
        xlabel={$Ma_S$},
        xtick style={color=black},
        ylabel={$e_b$},
        ytick style={color=black},
        enlargelimits=false,
        colormap name=spectralr,
        colorbar,
        colorbar style={
            ytick={-7,-5,-3},
            yticklabel={$10^{\pgfmathprintnumber{\tick}}$},
            ytick style={
                color=black,
            }
        },
        point meta min=-7.096910013008056,
        point meta max=-2.9208187539523752,
    ]
    \addplot [
        matrix plot*,
        mesh/cols=21,
        mesh/rows=21,
        point meta=explicit,
        mesh/ordering=y varies,
    ] table[point meta={lg10(\thisrow{value})}] {figures/gradients/verification_2D/1e-4/plot_data_1.txt};

    \nextgroupplot[
        tick align=outside,
        tick pos=left,
        title={(d)},
        xlabel={$Ma_S$},
        xtick style={color=black},
        ytick style={color=black},
        enlargelimits=false,
        colormap name=spectralr,
        colorbar,
        colorbar style={
            ytick={-7,-5,-3},
            yticklabel={$10^{\pgfmathprintnumber{\tick}}$},
            ytick style={
                color=black,
            }
        },
        point meta min=-7.096910013008056,
        point meta max=-2.9208187539523752,
    ]
    \addplot [
        matrix plot*,
        mesh/cols=21,
        mesh/rows=21,
        mesh/ordering=y varies,
    ] table[point meta={lg10(\thisrow{value})}] {figures/gradients/verification_2D/1e-4/plot_data_3.txt};
\end{groupplot}
\end{tikzpicture}
    \caption{
        Automatic differentiation gradients of the center of mass drift $\Delta x_{com}^{n}$ at $t^n = 10^{-4}$
        with respect to incident shock Mach number $Ma_S$ and initial bubble eccentricity $e_b$.
        (a): $ g_{1,AD} = \left( \frac{\partial \Delta x_{com}^{n}}{\partial Ma_S} \right)_{AD} $,
        (b): $ g_{2,AD} = \left( \frac{\partial \Delta x_{com}^{n}}{\partial e_b} \right)_{AD} $,
        (c): $ \vert g_{1,AD} - g_{1,FD} \vert / \vert g_{1,FD} \vert_{\infty} $,
        (d): $ \vert g_{2,AD} - g_{2,FD} \vert / \vert g_{2,FD} \vert_{\infty} $.
    }
    \label{fig:gradient_shock_bubble_1e-4}
\end{figure}
\begin{figure}[t!]
    \centering
    \begin{tikzpicture}
\pgfmathdeclarefunction{lg10}{1}{%
    \pgfmathparse{ln(#1)/ln(10)}%
}

\pgfplotsset{
    colormap={spectralr}{rgb255=(94,79,162) rgb255=(68,112,177) rgb255=(59,146,184) rgb255=(89,180,170)
        rgb255=(126,203,164) rgb255=(166,219,164) rgb255=(202,233,157) rgb255=(232,246,156) 
        rgb255=(247,252,179) rgb255=(254,245,175) rgb255=(254,227,145) rgb255=(253,200,119) 
        rgb255=(252,170,95) rgb255=(247,131,77) rgb255=(236,97,69) rgb255=(218,70,76)
        rgb255=(190,36,73) rgb255=(158,1,66)}}

\begin{groupplot}[
    group style={
        group size=2 by 2,
        horizontal sep=3cm,
        vertical sep=2cm
    }, 
    width=5.5cm, 
    height=5.5cm
    ]
    \nextgroupplot[
        tick align=outside,
        tick pos=left,
        title={(a)},
        xtick style={color=black},
        ylabel={$e_b$},
        ytick style={color=black},
        enlargelimits=false,
        colormap name=spectralr,
        colorbar,
        colorbar style={
            ytick style={
                color=black,
            }
        }
    ]
    \addplot [
        matrix plot*,
        mesh/cols=21,
        mesh/rows=21,
        point meta=explicit,
        mesh/ordering=y varies,
    ] table[meta=value] {figures/gradients/verification_2D/2.5e-4/plot_data_0.txt};

    \nextgroupplot[
        tick align=outside,
        tick pos=left,
        title={(b)},
        xtick style={color=black},
        ytick style={color=black},
        enlargelimits=false,
        colormap name=spectralr,
        colorbar,
        colorbar style={
            ytick style={
                color=black,
            }
        }
    ]
    \addplot [
        matrix plot*,
        mesh/cols=21,
        mesh/rows=21,
        point meta=explicit,
        mesh/ordering=y varies,
    ] table[meta=value] {figures/gradients/verification_2D/2.5e-4/plot_data_2.txt};

    \nextgroupplot[
        tick align=outside,
        tick pos=left,
        title={(c)},
        xlabel={$Ma_S$},
        xtick style={color=black},
        ylabel={$e_b$},
        ytick style={color=black},
        enlargelimits=false,
        colormap name=spectralr,
        colorbar,
        colorbar style={
            ytick={-5,-4,-3,-2},
            yticklabel={$10^{\pgfmathprintnumber{\tick}}$},
            ytick style={
                color=black,
            }
        },
        point meta min=-5.096910013008056,
        point meta max=-1.9208187539523752,
    ]
    \addplot [
        matrix plot*,
        mesh/cols=21,
        mesh/rows=21,
        point meta=explicit,
        mesh/ordering=y varies,
    ] table[point meta={lg10(\thisrow{value})}] {figures/gradients/verification_2D/2.5e-4/plot_data_1.txt};

    \nextgroupplot[
        tick align=outside,
        tick pos=left,
        title={(d)},
        xlabel={$Ma_S$},
        xtick style={color=black},
        ytick style={color=black},
        enlargelimits=false,
        colormap name=spectralr,
        colorbar,
        colorbar style={
            ytick={-7,-5,-3},
            yticklabel={$10^{\pgfmathprintnumber{\tick}}$},
            ytick style={
                color=black,
            }
        },
        point meta min=-7.096910013008056,
        point meta max=-2.9208187539523752,
    ]
    \addplot [
        matrix plot*,
        mesh/cols=21,
        mesh/rows=21,
        mesh/ordering=y varies,
    ] table[point meta={lg10(\thisrow{value})}] {figures/gradients/verification_2D/2.5e-4/plot_data_3.txt};
\end{groupplot}
\end{tikzpicture}
    \caption{
        Automatic differentiation gradients of the center of mass drift $\Delta x_{com}^{n}$ at $t^n = 2.5 \times 10^{-4}$
        with respect to incident shock Mach number $Ma_S$ and initial bubble eccentricity $e_b$.
        (a): $ g_{1,AD} = \left( \frac{\partial \Delta x_{com}^{n}}{\partial Ma_S} \right)_{AD} $,
        (b): $ g_{2,AD} = \left( \frac{\partial \Delta x_{com}^{n}}{\partial e_b} \right)_{AD} $,
        (c): $ \vert g_{1,AD} - g_{1,FD} \vert / \vert g_{1,FD} \vert_{\infty} $,
        (d): $ \vert g_{2,AD} - g_{2,FD} \vert / \vert g_{2,FD} \vert_{\infty} $.
    }
    \label{fig:gradient_shock_bubble_2.5e-4}
\end{figure}

Figure \ref{fig:center_of_mass_drift_interface_shape} visualizes the nondimensional center of mass drift (a)
and the shape of the interface ($\alpha_1 = 0.5$ contour) at $t^n = 2.5 \times 10^{-4}$ (b)
for different bubble eccentricities and different incident shock Mach numbers.
The strength of the incident shock strongly influences the drift of the helium bubble.
The variation in the initial bubble geometry is propagated through the simulation.  
Figures \ref{fig:gradient_shock_bubble_1e-4} and \ref{fig:gradient_shock_bubble_2.5e-4} show 
the gradients $g_1 = \frac{\partial \Delta x_{com}^{n}}{\partial Ma_S}$
and $g_2 = \frac{\partial \Delta x_{com}^{n}}{\partial e_b}$ obtained by AD and relative errors with respect to
the corresponding FD gradients for $t^n = 10^{-4}$ and $t^n = 2.5 \times 10^{-4}$, respectively.
At both time instances, AD gradients are in very good agreement with their FD counterparts.
We conclude that JAX-Fluids allows the parallel evaluation of AD gradients through extended integration trajectories.

\section{Parallel performance on GPU and TPU systems}
\label{sec:parallel_performance}
We evaluate the parallel efficiency of JAX-Fluids by performing
weak scaling tests of the single-phase (SPM),
diffuse-interface (DIM), and level-set model (LSM). The numerical setup consists of 
the WENO5-Z spatial reconstruction scheme combined with an 
HLLC approximate Riemann solver (Sec. \ref{subsec:high_order_godunov}).
The time integration is performed
with the TVD-RK3 scheme (Sec. \ref{subsec:temporal_integration}).

Within the homogeneous domain decomposition (Sec. \ref{subsec:domain_decomposition}), 
each XLA device allocates and processes a single block.
For the scaling runs, we define a block as
a cube of size $1.0\times 1.0\times 1.0$ discretized with $N\times N\times N$
cells. We initialize a homogeneous flow field in each block, where
the pressure and density are unity and the velocity is zero.
For the DIM and the LSM, we consider two phases that are separated
by a spherical interface with radius 0.25. The sphere is placed
at the center of the block.
Periodic boundary conditions are imposed.
We perform $M=200$ integration steps and measure the performance
using the mean wall clock time
\begin{equation}
    \Delta \overline{T} = \frac{1}{M}\sum_{n=1}^M \Delta T^n.
\end{equation}
Here, $\Delta T^n$ denotes the wall clock time of time step $n$.

\begin{table}[!b]
    \centering
\begin{tabular}{c c c c c }
\hline
& \multicolumn{2}{c}{TPU v3} & \multicolumn{2}{c}{NVIDIA A100 GPU}\\
Nodes & XLA devices & Decomposition & XLA devices & Decomposition\\
\hline
1 & 8 & $2\times2\times2$ & - & - \\
2 & - & - & 8 & $2\times2\times2$ \\
4 & 32 & $4\times4\times2$ & 16 & $4\times2\times2$ \\
8 & 64 & $4\times4\times4$ & 32 & $4\times4\times2$ \\
16 & 128 & $8\times4\times4$ & 64 & $4\times4\times4$ \\
32 & 256 & $8\times8\times4$ & 128 & $8\times4\times4$ \\
64 & 512 & $8\times8\times8$ & 256 & $8\times8\times4$ \\
128 & 1024 & $16\times8\times8$ & 512 & $8\times8\times8$ \\
\hline
\end{tabular}
\caption{Domain decomposition for weak scaling runs.
(The Google TPU Cloud did not support a TPU v3 pod slice at the time of the scaling runs.)}
\label{tab:scaling_decomposition}
\end{table}

The weak scaling tests are performed on a Google TPU v3 pod and on the
JUWELS Booster module at Juelich Supercomputing Centre (JSC). A single node of a 
TPU v3 consists of 8 TPU v3 cores, each having access to 16GB of RAM.
A JUWELS Booster node entails 4 NVIDIA A100 graphics cards
with 40GB of RAM each. This means
that a single node of the TPU and GPU system has 8 and 4 XLA devices, respectively.
It is important to note that the computations on the GPU are performed
using double precision, whereas on the TPU, we use single precision.
This is because the TPU v3 hardware does not have native support for double precision operations. 
TPU v3 systems emulate double precision arithmetic in a very
costly fashion, reducing the overall performance significantly.

\begin{table}[!t]
    \centering
\begin{tabular}{c c c c}
\hline
XLA Device & SPM &  DIM & LSM \\
\hline
NVIDIA A100 GPU & $320^3$ & $288^3$ & $232^3$ \\
TPU v3 & $240^3$ & $184^3$ & $152^3$ \\
\hline
\end{tabular}
\caption{Number of cells $N^3$ per XLA device.}
\label{tab:weak_scaling_resolution}
\end{table}

During the weak scaling runs, we successively increase the number
of XLA devices (blocks) from 8 up to 1024 on the TPU cluster 
and from 8 up to 512 on the GPU cluster.
We start with 8 XLA devices to ensure that communication occurs
in each spatial direction from the get-go, i.e., we start with a domain decomposition
of $2\times2\times2$. Therefore, we begin the scaling
with two nodes for the GPU system and with a single node for the TPU system.
Table \ref{tab:scaling_decomposition}
illustrates an overview of the domain decomposition
in accordance with the number of nodes. As the table shows,
we did not perform a run on a TPU v3 pod slice with 16 cores,
as this specific number of cores is not provided by the Google TPU Cloud.
Table \ref{tab:weak_scaling_resolution}
depicts the resolution of a single block depending on
the model and the system. Here, we aim to allocate as
much memory as possible on each device, as this ensures
that the devices work at their peak efficiency.

\begin{table}[!b]
    \centering
\begin{tabular}{c c c c}
\hline
XLA Device & SPM &  DIM & LSM \\
\hline
NVIDIA A100 GPU & 58.01 &  124.59 & 332.45 \\
TPU v3 & 152.73 &  304.56 & 1777.99 \\
\hline
\end{tabular}
\caption{Mean wall clock time per cell
$\Delta \overline{T}_8/N^3$ for 8 XLA devices in $\si{ns}$.}
\label{tab:reference_performance}
\end{table}

\begin{figure}[!t]
    \centering
\begin{tikzpicture}

\definecolor{darkgray176}{RGB}{176,176,176}
\definecolor{green01270}{RGB}{0,127,0}

\begin{groupplot}[group style={group size=2 by 1}, width=5.2cm, height=5cm]
\nextgroupplot[
log basis x={2},
tick align=outside,
tick pos=left,
title={TPU v3},
xlabel={nodes},
ylabel={$\omega$},
xmin=0.784584097896751, xmax=163.143760296866,
xmode=log,
xtick style={color=black},
xtick={1,2,4,8,16,32,64,128},
xticklabels={1,2,4,8,16,32,64,128},
ytick={0.6,0.7,0.8,0.9,1.0},
yticklabels={0.6,0.7,0.8,0.9,1.0},
legend style={at={(0.05,0.05)},draw=none,anchor=south west},
ymin=0.75, ymax=1.05,
]

\addplot [semithick, mark=*, mark size=1pt, black] table[
    x index = {0}, y index = {3}
] {figures/scaling/singlephase_tpu.txt}; \label{pgfplotss:scaling_singlephase}
\addplot [semithick, blue, mark=*, mark size=1pt] table[
    x index = {0}, y index = {3}
] {figures/scaling/diffuse_tpu.txt}; \label{pgfplotss:scaling_diffuse}
\addplot [semithick, red, mark=*, mark size=1pt] table[
    x index = {0}, y index = {3}
] {figures/scaling/levelset_tpu.txt}; \label{pgfplotss:scaling_levelset}

\nextgroupplot[
log basis x={2},
tick align=outside,
tick pos=left,
title={NVIDIA A100 GPU},
xlabel={nodes},
xmin=0.784584097896751, xmax=163.143760296866,
xmode=log,
ytick={0.6,0.7,0.8,0.9,1.0},
yticklabels={0.6,0.7,0.8,0.9,1.0},
xtick style={color=black},
xtick={1,2,4,8,16,32,64,128},
xticklabels={1,2,4,8,16,32,64,128},
ymin=0.75, ymax=1.05,
]



\addplot [semithick, mark=*, mark size=1pt, black] table[
    x index = {0}, y index = {3}
] {figures/scaling/singlephase_gpu1.txt};
\addplot [semithick, blue, mark=*, mark size=1pt] table[
    x index = {0}, y index = {3}
] {figures/scaling/diffuse_gpu1.txt};
\addplot [semithick, red, mark=*, mark size=1pt] table[
    x index = {0}, y index = {3}
] {figures/scaling/levelset_gpu1.txt};

\end{groupplot}
\end{tikzpicture}
    \caption{Weak scaling. The black (\ref{pgfplotss:scaling_singlephase}),
    blue (\ref{pgfplotss:scaling_diffuse}), and red (\ref{pgfplotss:scaling_levelset})
    lines indicate the single-phase, diffuse-interface, and
    level-set model, respectively.}
    \label{fig:weak_scaling}
\end{figure}
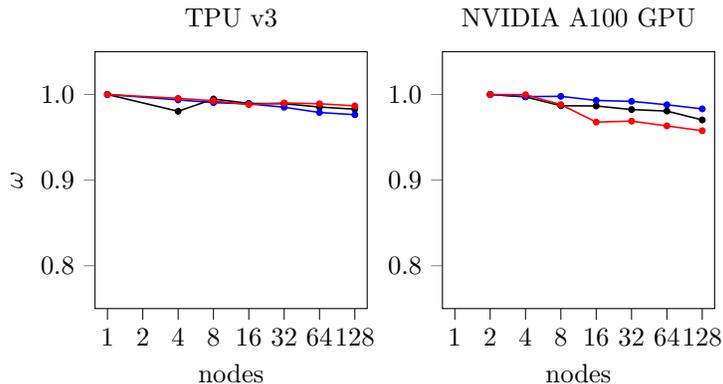

We measure the mean wall clock time per time step
$\Delta \overline{T}_{S_{XLA}}$ for each number of XLA devices $S_{XLA}$.
The weak scaling represents the normalized mean wall clock time
$\omega=\Delta \overline{T}_{S_{XLA}}/\Delta \overline{T}_8$
over the amount of XLA devices (nodes). Figure \ref{fig:weak_scaling} depicts
the weak scaling for both the TPU and GPU system. We observe a very good
scaling of $\omega > 0.95$ up to 128 nodes for all models on either system.
The largest simulation (SPM, 128 nodes, GPU system) has 16B cells.
In terms of scaling efficiency, the TPU v3 pod is slightly
better than the NVIDIA A100 GPU system. We depict
the reference performance $\Delta \overline{T}_{8}$
in Table \ref{tab:reference_performance}. The NVIDIA A100 system
clearly outperforms the TPU v3. This is expected, as JAX-Fluids
kernels generally operate in the memory bound regime \cite{Bezgin2022}, and
the NVIDIA A100 has roughly twice as much memory bandwidth compared to the
TPU v3.

\tikzexternaldisable
\tikzexternalenable

\section{Conclusion}
\label{sec:conclusion}
In this work, we have presented the second version of
the open-source JAX-Fluids computational fluid dynamics (CFD) solver.
Designed as a differentiable solver for compressible single- and two-phase flows,
JAX-Fluids facilitates research at the intersection of conventional CFD
and machine learning.
Potential research directions for utilizing JAX-Fluids may include 
data-driven surrogate modeling, data assimilation, uncertainty quantification, and flow control.

With the release of JAX-Fluids 2.0, we propel differentiable CFD towards high performance computing (HPC).
The main achievements of our work are summarized as follows.
\begin{enumerate}
    \item \textbf{JAX Primitives-based Parallelization:}
    We decompose the computational domain into multiple homogenous blocks
    and subsequently distribute them across available XLA devices using \mintinline{python}{jax.pmap}.
    Data is communicated by collective permutations \mintinline{python}{jax.lax.ppermute}.
    The parallelization strategy is validated on GPU and TPU clusters. 
    In particular, we have performed weak scaling runs on up to 512 NVIDIA A100 GPUs 
    and up to 1024 TPU v3 cores.
    We achieve an excellent weak scaling of $>0.95$ for all numerical models.
    \item \textbf{Automatic Differentiation (AD) through Parallel Simulations:}
    The above-mentioned parallelization strategy allows for seamless gradient calculation
    in distributed computations. Gradients from automatic differentiation are validated against finite-difference
    gradients, and we showcase stable AD through trajectories composed of several thousand
    integration steps.
    \item \textbf{Diffuse-Interface Two-phase Model:}
    A five-equation diffuse-interface model for two-phase flows was added.
    The diffuse-interface model complements the existing level-set based sharp-interface
    model and allows users to choose between two popular modeling approaches for 
    two-phase flows.
\end{enumerate}
Furthermore, the updated JAX-Fluids package incorporates positivity-preserving limiters,
support for stretched Cartesian meshes,
refactored I/O routines for increased flexibility and ease of use, 
performance optimizations, and an updated list of numerical discretization schemes.
Newly introduced features to JAX-Fluids have been thoroughly verified by canonical single- and two-phase
test cases.
In particular, we have shown results for turbulent boundary layer and channel flows, 
air-helium shock bubble interactions, and air-water shock drop interactions.
Good agreement with reference data from literature was achieved.

JAX-Fluids is subject of continuous development, and
we highlight two of many areas which we actively pursue.
\begin{enumerate}
    \item \textbf{Multi-Physics Framework:}
    JAX-Fluids is evolving towards a comprehensive multi-physics framework.
    This includes, amongst others, the extension to reactive multi-component flows, particle laden
    flows, phase change phenomena, and fluid-structure interaction.
    \item \textbf{Adaptive Multiresolution:}
    The current version of JAX-Fluids employs uniform and stretched Cartesian meshes. 
    Adaptive multiresolution allocates computational resources
    in regions of interest thereby increasing computational efficiency \cite{Harten1994,Harten1995}. 
    The implementation of adaptive multiresolution
    in differentiable and just-in-time compiled domain-specific languages like JAX
    is subject of ongoing research.
\end{enumerate}

\section*{Acknowledgements}
\label{sec:acknowledgements}
We thank Steffen J. Schmidt, Spencer H. Bryngelson, Qing Wang, and Yi-fan Chen 
for fruitful discussions throughout this work.
TPU computing resources granted by Google's TRC program are gratefully acknowledged.
We also acknowledge GPU computing resources at Juelich Supercomputing Centre (JSC).  
\appendix


\section{Overview Numerical Models}
\label{sec:appendix_overview_numerical_models}
Table \ref{tab:NumericalMethods} gives an overview on numerical methods implemented in the JAX-Fluids library.
\begin{table}[t!]
    \begin{center}
        \footnotesize
        \begin{tabular}{c c c} 
        \hline
        Time Integration    & Euler & \\
                            & TVD-RK2 \cite{Gottlieb1998a} & \\
                            & TVD-RK3 \cite{Gottlieb1998a} & \\
        \hline

        Flux Function/Riemann Solver    & Lax-Friedrichs (LxF) & According to \cite{Toro1994} \\
                                        & Local Lax-Friedrichs (LLxF, Rusanov) & According to \cite{Toro1994}\\
                                        & HLL/HLLC/HLLC-LM \cite{Harten1983a,Toro1994,Toro2009a,Toro2019,Fleischmann2020} & Signal speed estimates see below \\
                                        & AUSM+ \cite{Liou1996} & \\     
                                        & Componentwise LLxF & Flux-splitting formulation \\
                                        & Roe \cite{Roe1981} & Flux-splitting formulation \\
        \hline
        
        Signal Speed Estimates  & Arithmetic    & \\
                                & Davis \cite{Davis1988}         & \\
                                & Einfeldt \cite{Einfeldt1988a}     & \\
                                & Toro \cite{Toro1994}  & \\

        \hline

        Spatial Reconstruction  & WENO1 \cite{Jiang1996}                                                & \\
                                & WENO3-JS/Z/N/F3+/NN \cite{Jiang1996,Acker2016a,Gande2020,Bezgin2021b} & \\
                                & WENO5-JS/Z \cite{Jiang1996,Borges2008a}                               & \\
                                & WENO6-CU/CUM  \cite{Hu2010,Hu2011}                                    & \\
                                & WENO7-JS \cite{Balsara2000}                                           & \\
                                & WENO9-JS \cite{Balsara2000}                                           & \\
                                & TENO5/TENO5-A \cite{Fu2016,Fu2019}                                    & \\
                                & TENO6/TENO6-A \cite{Fu2016,Fu2019}                                    & \\
                                & MUSCL \cite{vanLeer1979}                                              & Slope limiters as summarized by \cite{Toro2009a} \\
                                & THINC \cite{Xiao2005,Xiao2011,Shyue2014}                         & For five-equation model only \\
                                & 2nd-/4th-/6th-order central                                      & For dissipative terms only \\

        \hline

        Spatial Derivatives & 2nd-/4th-/6th-order central & \\
                            & HOUC-3/5/7 \cite{Nourgaliev2007}& \\
        \hline
        Level-set reinitialization  & First-order \cite{Russo2000} & \\
                                   & HJ-WENO \cite{Jiang2000} & \\

        \hline
        Ghost fluid extension & First-order upwind \cite{Hu2006} & \\

        \hline

        LES Modules & ALDM \cite{Hickel2014b} & \\

        \hline 

        Equation of State   & Ideal gas & \\
                            & Stiffened gas \cite{Menikoff1989} & \\
                            & Tait \cite{Fedkiw1999a}& \\

        \hline Boundary Conditions  & Periodic & \\
                                    & Zero gradient & E.g., used for outflow boundaries \\
                                    & Dirichlet & \\
                                    & No-slip wall \\

        \hline
    \end{tabular}
    \caption{Overview on numerical methods available in JAX-Fluids.}
    \label{tab:NumericalMethods}
    \end{center}
\end{table}

\section{Mesh stretching}
\label{sec:appendix_computational_domain}
JAX-Fluids employs a Cartesian mesh which supports arbitrary
one-dimensional mesh stretching. Cell face positions and resulting cell sizes
can be strictly monotonous functions of the corresponding spatial index.
In this work, we use a hyperbolic mesh stretching for the
turbulent channel simulation in Sec. \ref{subsubsec:turbulent_channel}.
Cell face positions in $y$-direction are given as
\begin{equation}
    y_{i,j-\frac{1}{2},k} = \frac{y_{N_y} - y_{0}}{2} \cdot \tanh \left( \beta \left(2 j / N_j - 1 \right)  \right) / \tanh \left( \beta \right), j \in \left[0, N_y\right],
\end{equation}

For the boundary layer simulations in Secs. \ref{subsubsec:laminar_bl} and \ref{subsubsec:turbulent_boundary_layer},
we use a hyperbolic stretching given by
\begin{equation}
    y_{i,j-\frac{1}{2},k} = \left( y_{N_y} - y_{0} \right) \cdot \tanh \left( \beta \left(j / N_j - 1 \right)  \right) / \tanh \left( \beta \right) + y_{N_y}, j \in \left[0, N_y\right],
\end{equation}

In both cases, $\beta$ is the stretching parameter.
\section{Spatial discretization}
\label{sec:appendix_spatial_discretization}

\subsection{Characteristic decomposition}
\label{subsec:appendix_char_decomp}
Neglecting viscous terms and source terms, the quasi-linear form of 
the conservation equation \eqref{eq:DiffConsLaw1} can be reexpressed as
\begin{equation}
    \frac{\partial \mathbf{W}}{\partial t} 
    + \mathbf{A} \left( \mathbf{W} \right) \frac{\partial \mathbf{W}}{\partial x} 
    + \mathbf{B} \left( \mathbf{W} \right) \frac{\partial \mathbf{W}}{\partial y} 
    + \mathbf{C} \left( \mathbf{W} \right) \frac{\partial \mathbf{W}}{\partial z} 
    = 0,
    \label{eq:primitive_form}
\end{equation}
where
$\mathbf{A} \left( \mathbf{W} \right) = \left( \partial \mathbf{U} / \partial \mathbf{W} \right)^{-1} \partial \mathbf{F}^{c} / \partial \mathbf{W}$,
$\mathbf{B} \left( \mathbf{W} \right) = \left( \partial \mathbf{U} / \partial \mathbf{W} \right)^{-1} \partial \mathbf{G}^{c} / \partial \mathbf{W}$, 
and
$\mathbf{C} \left( \mathbf{W} \right) = \left( \partial \mathbf{U} / \partial \mathbf{W} \right)^{-1} \partial \mathbf{H}^{c} / \partial \mathbf{W}$ 
are Jacobian matrices associated with
the three spatial dimensions.
We can find an eigendecomposition for each of these Jacobian matrices. 
E.g., we decompose $\mathbf{A} \left( \mathbf{W} \right)$ as
\begin{equation}
    \mathbf{A} \left( \mathbf{W} \right) = \mathbf{K} \mathbf{\Lambda} \mathbf{K}^{-1},
\end{equation}
where $\mathbf{K}$ is a matrix whose columns are the right eigenvectors of $\mathbf{A}$, and 
$\mathbf{\Lambda}$ is a matrix whose diagonal elements are the corresponding eigenvalues.
For the single-phase Navier-Stokes equations, the Jacobian matrix and the matrix of right
eigenvectors are given as
\begin{align}
    \mathbf{A} = \begin{bmatrix}
        u & \rho     & 0 & 0 & 0                \\
        0 & u        & 0 & 0 & \frac{1}{\rho}   \\
        0 & 0        & u & 0 & 0                \\
        0 & 0        & 0 & u & 0                \\
        0 & \rho c^2 & 0 & 0 & u                
    \end{bmatrix}, \quad 
    \mathbf{K} = \begin{bmatrix}
        \rho     & 1 & 0 & 0 & \rho \\
        -c       & 0 & 0 & 0 & c    \\
        0        & 0 & 1 & 0 & 0    \\
        0        & 0 & 0 & 1 & 0    \\
        \rho c^2 & 0 & 0 & 0 & \rho c^2                
    \end{bmatrix}.
\end{align}
For the five-equation diffuse-interface equations, we find analogous expressions.
\begin{equation}
    \mathbf{A} = \begin{bmatrix}
        u & 0 & \alpha_1 \rho_1   & 0 & 0 & 0               & 0 \\
        0 & u & \alpha_2 \rho_2   & 0 & 0 & 0               & 0 \\
        0 & 0 & u                 & 0 & 0 & \frac{1}{\rho}  & 0 \\
        0 & 0 & 0                 & u & 0 & 0               & 0 \\
        0 & 0 & 0                 & 0 & u & 0               & 0 \\
        0 & 0 & \rho c^2          & 0 & 0 & u               & 0 \\
        0 & 0 & 0                 & 0 & 0 & 0               & u
    \end{bmatrix}, \quad 
    \mathbf{K} = \begin{bmatrix}
        \alpha_1 \rho_1  & 1 & 0 & 0 & 0 & 0 & \alpha_1 \rho_1 \\
        \alpha_2 \rho_2  & 0 & 1 & 0 & 0 & 0 & \alpha_2 \rho_2 \\
        -c               & 0 & 0 & 0 & 0 & 0 & c               \\
        0                & 0 & 0 & 1 & 0 & 0 & 0               \\
        0                & 0 & 0 & 0 & 1 & 0 & 0               \\
        \rho c^2         & 0 & 0 & 0 & 0 & 0 & \rho c^2        \\
        0                & 0 & 0 & 0 & 0 & 1 & 0             
    \end{bmatrix}
\end{equation}

\subsection{HLLC approximate Riemann solver}
\label{subsec:appendix_hllc}
We specify the wave speeds used in the HLLC Riemann solver, see Sec. \ref{subsec:high_order_godunov}.
Following Einfeldt et al. \cite{Einfeldt1988a}, the wave speeds of left and right wave are
\begin{align}
    s^- = \min (0, s^L), \\
    s^+ = \max (0, s^R),
\end{align}
and
\begin{align}
    s^L = \min (\bar{u} - \bar{c}, u^L - c^L), \\
    s^R = \max (\bar{u} + \bar{c}, u^R + c^R).
\end{align}
The wave speed for the intermediate wave is chosen following Batten et al. \cite{Batten1997}
\begin{align}
    s^* = \frac{p^R - p^L + \rho^L u^L (s^L - u^L) - \rho^R u^R (s^R - u^R)}{\rho^L (s^L - u^L) - \rho^R (s^R - u^R)}.
\end{align}
\section{Level-set model}

\subsection{Interface Riemann Problem}
\label{app:interface_riemann_problem}
To compute the interface pressure $p_\Gamma$ and interface velocity $\mathbf{u}_\Gamma$,
we solve a two-material Riemann problem at the interface. The linearized solution
reads \cite{Hu2004}
\begin{align}
    \mathbf{u}_\Gamma &= \frac{\rho_1c_1 \mathbf{u}_1\cdot \mathbf{n}_\Gamma + \rho_2c_2 \mathbf{u}_2\cdot \mathbf{n}_\Gamma + p_2 - p_1 - \sigma \kappa}{\rho_1c_1 + \rho_2c_2} \mathbf{n}_\Gamma, \notag \\
    p_\Gamma &= \frac{\rho_1c_1(p_2) + \rho_2c_2p_1 + \rho_1c_1\rho_2c_2(\mathbf{u}_2\cdot \mathbf{n}_\Gamma - \mathbf{u}_1\cdot \mathbf{n}_\Gamma)}{\rho_1c_1 + \rho_2c_2}, \notag
\end{align}
where $\mathbf{n}_\Gamma$, $\rho$, and $c$ denote the interface normal, the density,
and the speed of sound, respectively. The subscripts 1 and 2
indicate the two distinct phases.

\subsection{Computation of apertures and volume fraction}
\label{app:interface_reconstruction}

\begin{figure}[!h]
    \centering
    \begin{tikzpicture}
        \draw[thick, black] (0,0) -- (0,2) node[at end, circle, draw=blue] {} -- (2,2) node[at end, circle, draw=blue] {} -- (2,0) node[at end, circle, draw=blue] {} -- (0,0) node[at end, circle, draw=blue, fill=blue] {};
        \draw[thick, dashed] (1.0,0.0) -- (0.0,1.0);
        \draw[thick, black] (3,0) -- (3,2) node[at end, circle, draw=blue, fill=blue] {} -- (5,2) node[at end, circle, draw=blue] {} -- (5,0) node[at end, circle, draw=blue] {} -- (3,0) node[at end, circle, draw=blue, fill=blue] {};
        \draw[thick, dashed] (3.8,0.0) -- (4.2,2.0);
        \draw[thick, black] (6,0) -- (6,2) node[at end, circle, draw=blue, fill=blue] {} -- (8,2) node[at end, circle, draw=blue] {} -- (8,0) node[at end, circle, draw=blue, fill=blue] {} -- (6,0) node[at end, circle, draw=blue] {};
        \draw[thick, dashed] (6.0,0.5) -- (7.0,2.0);
        \draw[thick, dashed] (7.2,0.0) -- (8.0,0.8);
    \end{tikzpicture}
    \caption{Three typical aperture shapes. Solid and hollow blue circles indicate
    positive and negative level-set values, respectively.}
    \label{fig:cut_cell_face}
\end{figure}
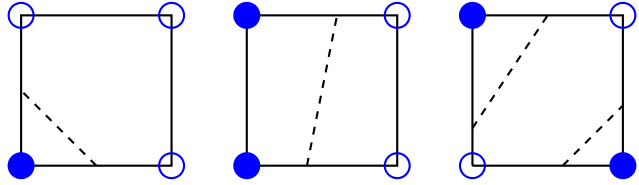

We assume a linear approximation of the level-set within each cell.
Figure \ref{fig:cut_cell} shows a cut cell $(i,j,k)$ with the linear interface
segment $\Delta \Gamma_{i,j,k}$ depicted in red. To compute
the apertures, we first interpolate the level-set at the vertex points of the cells.
The sign of the level-set values at the four vertex points of a cell face
determines the general shape of the aperture. A sign change 
along the edge of a cell face implies an intersection
of the interface with the edge.
Figure \ref{fig:cut_cell_face} shows three typical aperture shapes.
In total, there are $2^4=16$ different sign combinations of the level-set
values along the vertices. Hence, there are 16 different
aperture shapes. The area of each of these shapes is computed
analytically (trapezoid, triangle, etc.).
The interface segment $\Delta \Gamma$ is computed from the apertures as follows.
\begin{align}
    \Delta \Gamma_{i,j,k} &=  \left[ \left((A_{i+\frac{1}{2},j,k} - A_{i-\frac{1}{2},j,k})\Delta y\Delta z\right)^2 + \left((A_{i,j+\frac{1}{2},k} - A_{i,j-\frac{1}{2},k})\Delta x\Delta z\right)^2 \right. \notag \\
    &+ \left. \left((A_{i,j,k+\frac{1}{2}} - A_{i,j,k-\frac{1}{2}})\Delta x\Delta y\right)^2 \right]^\frac{1}{2} \notag
\end{align}
Geometrical reconstruction with seven pyramids yields the volume fraction $\alpha$.
\begin{align}
    \alpha_{i,j,k} &= \frac{1}{3} \frac{1}{\Delta x \Delta y \Delta z} \left[ A_{i+\frac{1}{2},j,k} \Delta y \Delta z \frac{1}{2} \Delta x +  A_{i-\frac{1}{2},j,k} \Delta y \Delta z \frac{1}{2} \Delta x  + A_{i,j+\frac{1}{2},k} \Delta x \Delta z \frac{1}{2} \Delta y \right. \notag \\
    &+ \left. A_{i,j-\frac{1}{2},k} \Delta x \Delta z \frac{1}{2} \Delta y + A_{i,j,k-\frac{1}{2}} \Delta x \Delta y \frac{1}{2} \Delta z + A_{i,j,k+\frac{1}{2}} \Delta x \Delta y \frac{1}{2} \Delta z + \Delta \Gamma_{i,j,k} \phi_{i,j,k}  \vphantom{\frac12}\right] \notag
\end{align}
The described approach yields the volume fraction and apertures with respect to the positive phase.

\subsection{Conservative mixing procedure}
\label{app:mixing}
The presented level-set method (see Sec. \ref{subsec:sharp_interface_model_num})
is not consistent when the interface crosses a cell face within a single time step,
i.e., when new cells are created or cells have vanished. 
Furthermore, small cut cells may lead to an unstable
integration using the time step restriction that is based on a full cell.
We apply a mixing procedure \cite{Hu2006,Lauer2012c} that
deals with both of these problems.
There are three types of cells that require mixing
\begin{enumerate}
    \item Cells for which $\alpha = 0$ after integration but $\alpha \neq 0$ before (vanished cells).
    \item Cells for which $\alpha > 0$ after integration but $\alpha = 0$ before (newly created cells).
    \item Cells for which $\alpha < \alpha_{\text{mix}}$ after integration (small cells).
\end{enumerate}
Here, $\alpha_\text{mix}$ is a specified threshold.
We identify neighboring, target (subscript trg) cells
from the interface normal.
One in each spatial direction $x$, $y$, and $z$, one in
each plane $xy$, $xz$, and $yz$, and one in $xyz$.
Seven mixing weights are computed as
\begin{align}
    \beta_x &= |\mathbf{n}_\Gamma\cdot \mathbf{i}|^2 \alpha_\text{trg}, \notag \\
    \beta_y &= |\mathbf{n}_\Gamma\cdot \mathbf{j}|^2 \alpha_\text{trg}, \notag \\
    \beta_z &= |\mathbf{n}_\Gamma\cdot \mathbf{k}|^2 \alpha_\text{trg}, \notag \\
    \beta_{xy} &= |\left(\mathbf{n}_\Gamma\cdot \mathbf{i}\right) \left(\mathbf{n}_\Gamma\cdot \mathbf{j}\right) | \alpha_\text{trg}, \\
    \beta_{xz} &= |\left(\mathbf{n}_\Gamma\cdot \mathbf{i}\right) \left(\mathbf{n}_\Gamma\cdot \mathbf{k}\right) | \alpha_\text{trg}, \notag \\
    \beta_{yz} &= |\left(\mathbf{n}_\Gamma\cdot \mathbf{j}\right) \left(\mathbf{n}_\Gamma\cdot \mathbf{k}\right) | \alpha_\text{trg}, \notag \\
    \beta_{xyz} &= |\left(\mathbf{n}_\Gamma\cdot \mathbf{i}\right) \left(\mathbf{n}_\Gamma\cdot \mathbf{j}\right) \left(\mathbf{n}_\Gamma\cdot \mathbf{k}\right) |^{2/3} \alpha_\text{trg}, \notag
\end{align}
where $\mathbf{i}$, $\mathbf{j}$, and $\mathbf{k}$ are the unit vectors in $x$, $y$, and $z$ direction.
We normalize the mixing weights so that $\sum_\text{trg} \beta_\text{trg} =1$, where $\text{trg}\in\{x,y,z,xy,xz,yz,xyz\}$.
Subsequently, the mixing flux $\mathbf{M}_\text{trg}$ is computed as
\begin{equation}
    \mathbf{M}_\text{trg} = \frac{\beta_\text{trg}}{\alpha \beta_\text{trg} + \alpha_\text{trg}} \left[(\alpha_\text{trg}\bar{\mathbf{U}}_\text{trg})\alpha - (\alpha \bar{\mathbf{U}})\alpha_\text{trg} \right].
\end{equation}
The conservative variables are then updated according to
\begin{align}
    \alpha \bar{\mathbf{U}} &= \left( \alpha \bar{\mathbf{U}} \right)^* + \sum_\text{trg}\mathbf{M}_\text{trg}, \\
    \alpha_\text{trg} \bar{\mathbf{U}}_\text{trg} &= \left( \alpha_\text{trg} \bar{\mathbf{U}}_\text{trg} \right)^* - \sum_\text{trg}\mathbf{M}_\text{trg}.
\end{align}
Here, $\alpha \bar{\mathbf{U}}$ and $\alpha_\text{trg} \bar{\mathbf{U}}_\text{trg}$
denote the conservative variables of the cells that require mixing 
and the conservative variables of the corresponding target cells, respectively.
Star-quantities denote conservative variables before mixing. The mixing
procedure is performed on the integrated conservative variables.
%

\section*{Data Availability Statement}
JAX-Fluids is available under the GNU
GPLv3 license at \url{https://github.com/tumaer/JAXFLUIDS}.

\newpage
\bibliography{BIB}

\begin{thebibliography}{10}
\expandafter\ifx\csname url\endcsname\relax
  \def\url#1{\texttt{#1}}\fi
\expandafter\ifx\csname urlprefix\endcsname\relax\def\urlprefix{URL }\fi
\expandafter\ifx\csname href\endcsname\relax
  \def\href#1#2{#2} \def\path#1{#1}\fi

\bibitem{Brunton2020a}
S.~L. Brunton, B.~R. Noack, P.~Koumoutsakos, Machine learning for fluid
  mechanics, Annual Review of Fluid Mechanics 52 (2020) 477--508.
\newblock \href {https://doi.org/10.1146/annurev-fluid-010719-060214}
  {\path{doi:10.1146/annurev-fluid-010719-060214}}.

\bibitem{Vinuesa2022}
R.~Vinuesa, S.~L. Brunton,
  \href{https://www.nature.com/articles/s43588-022-00264-7}{Enhancing
  computational fluid dynamics with machine learning}, Nature Computational
  Science 2 (2022) 358--366.
\newblock \href {https://doi.org/10.1038/s43588-022-00264-7}
  {\path{doi:10.1038/s43588-022-00264-7}}.
\newline\urlprefix\url{https://www.nature.com/articles/s43588-022-00264-7}

\bibitem{Baydin2017}
A.~G. Baydin, B.~A. Pearlmutter, J.~M. Siskind, Automatic differentiation in
  machine learning: a survey, Journal of Machine Learning Research 18 (2017)
  5595--5637.

\bibitem{Abadi}
M.~Abadi, P.~Barham, J.~Chen, Z.~Chen, A.~Davis, J.~Dean, M.~Devin,
  S.~Ghemawat, G.~Irving, M.~Isard, M.~Kudlur, J.~Levenberg, R.~Monga,
  S.~Moore, D.~G. Murray, B.~Steiner, P.~Tucker, V.~Vasudevan, P.~Warden,
  M.~Wicke, Y.~Yu, X.~Zheng, G.~Brain,
  \href{https://tensorflow.org.}{Tensorflow: A system for large-scale machine
  learning tensorflow: A system for large-scale machine learning}, Proceedings
  of the 12th USENIX Symposium on Operating Systems Design and Implementation
  (2016).
\newline\urlprefix\url{https://tensorflow.org.}

\bibitem{Paszke2019}
A.~Paszke, S.~Gross, F.~Massa, A.~Lerer, J.~Bradbury, G.~Chanan, T.~Killeen,
  Z.~Lin, N.~Gimelshein, L.~Antiga, A.~Desmaison, A.~Köpf, E.~Yang, Z.~DeVito,
  M.~Raison, A.~Tejani, S.~Chilamkurthy, B.~Steiner, L.~Fang, J.~Bai,
  S.~Chintala, Pytorch: An imperative style, high-performance deep learning
  library, Vol.~32, 2019.

\bibitem{jax2018github}
J.~Bradbury, R.~Frostig, P.~Hawkins, M.~J. Johnson, C.~Leary, D.~Maclaurin,
  G.~Necula, A.~Paszke, J.~Vander\{P\}las, S.~Wanderman-\{M\}ilne, Q.~Zhang,
  \{JAX\}: composable transformations of \{P\}ython+\{N\}um\{P\}y programs
  (2018).

\bibitem{Schoenholz2020}
S.~S. Schoenholz, E.~D. Cubuk, Jax, m.d. a framework for differentiable
  physics, Vol. 2020-December, Neural information processing systems
  foundation, 2020.

\bibitem{Kochkov2021}
D.~Kochkov, J.~A. Smith, A.~Alieva, Q.~Wang, M.~P. Brenner, S.~Hoyer,
  \href{http://arxiv.org/abs/2102.01010}{{Machine learning accelerated
  computational fluid dynamics}} (2021) 1--13\href
  {http://arxiv.org/abs/2102.01010} {\path{arXiv:2102.01010}}.
\newline\urlprefix\url{http://arxiv.org/abs/2102.01010}

\bibitem{Wang2022}
Q.~Wang, M.~Ihme, Y.~F. Chen, J.~Anderson,
  \href{http://arxiv.org/abs/2108.11076
  http://dx.doi.org/10.1016/j.cpc.2022.108292}{{A TensorFlow simulation
  framework for scientific computing of fluid flows on tensor processing
  units}}, Computer Physics Communications 274 (aug 2022).
\newblock \href {http://arxiv.org/abs/2108.11076} {\path{arXiv:2108.11076}},
  \href {https://doi.org/10.1016/j.cpc.2022.108292}
  {\path{doi:10.1016/j.cpc.2022.108292}}.
\newline\urlprefix\url{http://arxiv.org/abs/2108.11076
  http://dx.doi.org/10.1016/j.cpc.2022.108292}

\bibitem{Ataei2023}
M.~Ataei, H.~Salehipour, \href{http://arxiv.org/abs/2311.16080}{Xlb: A
  differentiable massively parallel lattice boltzmann library in python} (11
  2023).
\newline\urlprefix\url{http://arxiv.org/abs/2311.16080}

\bibitem{Xue2022}
T.~Xue, S.~Liao, Z.~Gan, C.~Park, X.~Xie, W.~K. Liu, J.~Cao,
  \href{http://arxiv.org/abs/2212.00964
  http://dx.doi.org/10.1016/j.cpc.2023.108802}{Jax-fem: A differentiable
  gpu-accelerated 3d finite element solver for automatic inverse design and
  mechanistic data science} (12 2022).
\newblock \href {https://doi.org/10.1016/j.cpc.2023.108802}
  {\path{doi:10.1016/j.cpc.2023.108802}}.
\newline\urlprefix\url{http://arxiv.org/abs/2212.00964
  http://dx.doi.org/10.1016/j.cpc.2023.108802}

\bibitem{Bezgin2022}
D.~A. Bezgin, A.~B. Buhendwa, N.~A. Adams,
  \href{http://arxiv.org/abs/2203.13760}{{JAX-FLUIDS: A fully-differentiable
  high-order computational fluid dynamics solver for compressible two-phase
  flows}}, Computer Physics Communications 282 (2022) 108527.
\newblock \href {http://arxiv.org/abs/2203.13760} {\path{arXiv:2203.13760}},
  \href {https://doi.org/10.1016/j.cpc.2022.108527}
  {\path{doi:10.1016/j.cpc.2022.108527}}.
\newline\urlprefix\url{http://arxiv.org/abs/2203.13760}

\bibitem{Frostig2018}
R.~Frostig, M.~J. Johnson, C.~Leary, Compiling machine learning programs via
  high-level tracing, 2018.

\bibitem{Harris2020}
C.~R. Harris, K.~J. Millman, S.~J. van~der Walt, R.~Gommers, P.~Virtanen,
  D.~Cournapeau, E.~Wieser, J.~Taylor, S.~Berg, N.~J. Smith, R.~Kern, M.~Picus,
  S.~Hoyer, M.~H. van Kerkwijk, M.~Brett, A.~Haldane, J.~F. del Río, M.~Wiebe,
  P.~Peterson, P.~Gérard-Marchant, K.~Sheppard, T.~Reddy, W.~Weckesser,
  H.~Abbasi, C.~Gohlke, T.~E. Oliphant,
  \href{https://doi.org/10.1038/s41586-020-2649-2}{Array programming with
  numpy} (2020).
\newblock \href {https://doi.org/10.1038/s41586-020-2649-2}
  {\path{doi:10.1038/s41586-020-2649-2}}.
\newline\urlprefix\url{https://doi.org/10.1038/s41586-020-2649-2}

\bibitem{Osher1988}
S.~Osher, J.~A. Sethian, Fronts propagating with curvature-dependent speed:
  Algorithms based on hamilton-jacobi formulations, Journal of Computational
  Physics 79 (1988) 12--49.
\newblock \href {https://doi.org/10.1016/0021-9991(88)90002-2}
  {\path{doi:10.1016/0021-9991(88)90002-2}}.

\bibitem{Sussman1994b}
M.~Sussman, {A level set approach for computing solutions to incompressible
  two-phase flow}, Journal of Computational Physics 114~(1) (1994) 146--159.
\newblock \href {https://doi.org/10.1006/jcph.1994.1155}
  {\path{doi:10.1006/jcph.1994.1155}}.

\bibitem{Hoppe2022}
N.~Hoppe, J.~M. Winter, S.~Adami, N.~A. Adams,
  \href{https://linkinghub.elsevier.com/retrieve/pii/S0010465521003581}{{ALPACA
  - a level-set based sharp-interface multiresolution solver for conservation
  laws}}, Computer Physics Communications 272 (2022) 108246.
\newblock \href {https://doi.org/10.1016/j.cpc.2021.108246}
  {\path{doi:10.1016/j.cpc.2021.108246}}.
\newline\urlprefix\url{https://linkinghub.elsevier.com/retrieve/pii/S0010465521003581}

\bibitem{Allaire2002}
G.~Allaire, S.~Clerc, S.~Kokh,
  \href{https://linkinghub.elsevier.com/retrieve/pii/S0021999102971433}{A
  five-equation model for the simulation of interfaces between compressible
  fluids}, Journal of Computational Physics 181 (2002) 577--616.
\newblock \href {https://doi.org/10.1006/jcph.2002.7143}
  {\path{doi:10.1006/jcph.2002.7143}}.
\newline\urlprefix\url{https://linkinghub.elsevier.com/retrieve/pii/S0021999102971433}

\bibitem{Perigaud2005}
G.~Perigaud, R.~Saurel,
  \href{https://linkinghub.elsevier.com/retrieve/pii/S0021999105001853}{A
  compressible flow model with capillary effects}, Journal of Computational
  Physics 209 (2005) 139--178.
\newblock \href {https://doi.org/10.1016/j.jcp.2005.03.018}
  {\path{doi:10.1016/j.jcp.2005.03.018}}.
\newline\urlprefix\url{https://linkinghub.elsevier.com/retrieve/pii/S0021999105001853}

\bibitem{Coralic2014}
V.~Coralic, T.~Colonius,
  \href{https://linkinghub.elsevier.com/retrieve/pii/S0021999114004082}{{Finite-volume
  WENO scheme for viscous compressible multicomponent flows}}, Journal of
  Computational Physics 274 (2014) 95--121.
\newblock \href {https://doi.org/10.1016/j.jcp.2014.06.003}
  {\path{doi:10.1016/j.jcp.2014.06.003}}.
\newline\urlprefix\url{https://linkinghub.elsevier.com/retrieve/pii/S0021999114004082}

\bibitem{Wong2021}
M.~L. Wong, J.~B. Angel, M.~F. Barad, C.~C. Kiris,
  \href{https://linkinghub.elsevier.com/retrieve/pii/S0021999121004642}{A
  positivity-preserving high-order weighted compact nonlinear scheme for
  compressible gas-liquid flows}, Journal of Computational Physics 444 (2021)
  110569.
\newblock \href {https://doi.org/10.1016/j.jcp.2021.110569}
  {\path{doi:10.1016/j.jcp.2021.110569}}.
\newline\urlprefix\url{https://linkinghub.elsevier.com/retrieve/pii/S0021999121004642}

\bibitem{Beck2019}
A.~Beck, D.~Flad, C.~D. Munz, {Deep neural networks for data-driven LES closure
  models}, Journal of Computational Physics 398 (2019).
\newblock \href {https://doi.org/10.1016/j.jcp.2019.108910}
  {\path{doi:10.1016/j.jcp.2019.108910}}.

\bibitem{Sirignano2020}
J.~Sirignano, J.~F. MacArt, J.~B. Freund, Dpm: A deep learning pde augmentation
  method with application to large-eddy simulation, Journal of Computational
  Physics 423 (12 2020).
\newblock \href {https://doi.org/10.1016/j.jcp.2020.109811}
  {\path{doi:10.1016/j.jcp.2020.109811}}.

\bibitem{Bezgin2021a}
D.~A. Bezgin, S.~J. Schmidt, N.~A. Adams,
  \href{https://linkinghub.elsevier.com/retrieve/pii/S0021999121002199}{A
  data-driven physics-informed finite-volume scheme for nonclassical
  undercompressive shocks}, Journal of Computational Physics 437 (2021) 110324.
\newblock \href {https://doi.org/10.1016/j.jcp.2021.110324}
  {\path{doi:10.1016/j.jcp.2021.110324}}.
\newline\urlprefix\url{https://linkinghub.elsevier.com/retrieve/pii/S0021999121002199}

\bibitem{Raissi2019}
M.~Raissi, P.~Perdikaris, G.~E. Karniadakis, Physics-informed neural networks:
  A deep learning framework for solving forward and inverse problems involving
  nonlinear partial differential equations, Journal of Computational Physics
  378 (2019) 686--707.
\newblock \href {https://doi.org/10.1016/j.jcp.2018.10.045}
  {\path{doi:10.1016/j.jcp.2018.10.045}}.

\bibitem{Buhendwa2021c}
A.~B. Buhendwa, S.~Adami, N.~A. Adams,
  \href{https://doi.org/10.1016/j.mlwa.2021.100029}{Inferring incompressible
  two-phase flow fields from the interface motion using physics-informed neural
  networks}, Machine Learning with Applications 4 (2021) 100029.
\newblock \href {https://doi.org/10.1016/j.mlwa.2021.100029}
  {\path{doi:10.1016/j.mlwa.2021.100029}}.
\newline\urlprefix\url{https://doi.org/10.1016/j.mlwa.2021.100029}

\bibitem{Jagtap2022}
A.~D. Jagtap, Z.~Mao, N.~Adams, G.~E. Karniadakis,
  \href{https://linkinghub.elsevier.com/retrieve/pii/S0021999122004648}{Physics-informed
  neural networks for inverse problems in supersonic flows}, Journal of
  Computational Physics 466 (2022) 111402.
\newblock \href {https://doi.org/10.1016/j.jcp.2022.111402}
  {\path{doi:10.1016/j.jcp.2022.111402}}.
\newline\urlprefix\url{https://linkinghub.elsevier.com/retrieve/pii/S0021999122004648}

\bibitem{Karnakov2022}
P.~Karnakov, S.~Litvinov, P.~Koumoutsakos,
  \href{http://arxiv.org/abs/2205.04611}{{Optimizing a DIscrete Loss (ODIL) to
  solve forward and inverse problems for partial differential equations using
  machine learning tools}} (2022) 1--7\href {http://arxiv.org/abs/2205.04611}
  {\path{arXiv:2205.04611}}.
\newline\urlprefix\url{http://arxiv.org/abs/2205.04611}

\bibitem{Johnsen2006}
E.~Johnsen, T.~Colonius,
  \href{https://linkinghub.elsevier.com/retrieve/pii/S0021999106002014}{Implementation
  of weno schemes in compressible multicomponent flow problems}, Journal of
  Computational Physics 219 (2006) 715--732.
\newblock \href {https://doi.org/10.1016/j.jcp.2006.04.018}
  {\path{doi:10.1016/j.jcp.2006.04.018}}.
\newline\urlprefix\url{https://linkinghub.elsevier.com/retrieve/pii/S0021999106002014}

\bibitem{Menikoff1989}
R.~Menikoff, B.~J. Plohr, The riemann problem for fluid flow of real materials,
  Reviews of Modern Physics 61 (1989) 75--130.
\newblock \href {https://doi.org/10.1103/RevModPhys.61.75}
  {\path{doi:10.1103/RevModPhys.61.75}}.

\bibitem{Sutherland1893}
W.~Sutherland, Lii. the viscosity of gases and molecular force, The London,
  Edinburgh, and Dublin Philosophical Magazine and Journal of Science 36 (1893)
  507--531.
\newblock \href {https://doi.org/10.1080/14786449308620508}
  {\path{doi:10.1080/14786449308620508}}.

\bibitem{Toro2009a}
E.~F. Toro, Riemann solvers and numerical methods for fluid dynamics a
  practical introduction, 3rd Edition, Springer Verlag, 2009.

\bibitem{Borges2008a}
R.~Borges, M.~Carmona, B.~Costa, W.~S. Don, An improved weighted essentially
  non-oscillatory scheme for hyperbolic conservation laws, Journal of
  Computational Physics 227 (2008) 3191--3211.
\newblock \href {https://doi.org/10.1016/j.jcp.2007.11.038}
  {\path{doi:10.1016/j.jcp.2007.11.038}}.

\bibitem{Fu2019}
L.~Fu, X.~Y. Hu, N.~A. Adams,
  \href{https://arc.aiaa.org/doi/10.2514/1.J057370}{Improved five- and
  six-point targeted essentially nonoscillatory schemes with adaptive
  dissipation}, AIAA Journal 57 (2019) 1143--1158.
\newblock \href {https://doi.org/10.2514/1.J057370}
  {\path{doi:10.2514/1.J057370}}.
\newline\urlprefix\url{https://arc.aiaa.org/doi/10.2514/1.J057370}

\bibitem{Shu1998b}
C.-W. Shu, \href{http://link.springer.com/10.1007/BFb0096355}{Essentially
  non-oscillatory and weighted essentially non-oscillatory schemes for
  hyperbolic conservation laws}, 1998, pp. 325--432.
\newblock \href {https://doi.org/10.1007/BFb0096355}
  {\path{doi:10.1007/BFb0096355}}.
\newline\urlprefix\url{http://link.springer.com/10.1007/BFb0096355}

\bibitem{Jiang1996}
G.~S. Jiang, C.~W. Shu, Efficient implementation of weighted eno schemes,
  Journal of Computational Physics 126 (1996) 202--228.
\newblock \href {https://doi.org/10.1006/jcph.1996.0130}
  {\path{doi:10.1006/jcph.1996.0130}}.

\bibitem{Xiao2005}
F.~Xiao, Y.~Honma, T.~Kono,
  \href{https://onlinelibrary.wiley.com/doi/10.1002/fld.975}{A simple algebraic
  interface capturing scheme using hyperbolic tangent function}, International
  Journal for Numerical Methods in Fluids 48 (2005) 1023--1040.
\newblock \href {https://doi.org/10.1002/fld.975} {\path{doi:10.1002/fld.975}}.
\newline\urlprefix\url{https://onlinelibrary.wiley.com/doi/10.1002/fld.975}

\bibitem{Xiao2011}
F.~Xiao, S.~Ii, C.~Chen,
  \href{https://linkinghub.elsevier.com/retrieve/pii/S0021999111003615}{Revisit
  to the thinc scheme: A simple algebraic vof algorithm}, Journal of
  Computational Physics 230 (2011) 7086--7092.
\newblock \href {https://doi.org/10.1016/j.jcp.2011.06.012}
  {\path{doi:10.1016/j.jcp.2011.06.012}}.
\newline\urlprefix\url{https://linkinghub.elsevier.com/retrieve/pii/S0021999111003615}

\bibitem{Shyue2014}
K.-M. Shyue, F.~Xiao,
  \href{https://linkinghub.elsevier.com/retrieve/pii/S0021999114001831}{An
  eulerian interface sharpening algorithm for compressible two-phase flow: The
  algebraic thinc approach}, Journal of Computational Physics 268 (2014)
  326--354.
\newblock \href {https://doi.org/10.1016/j.jcp.2014.03.010}
  {\path{doi:10.1016/j.jcp.2014.03.010}}.
\newline\urlprefix\url{https://linkinghub.elsevier.com/retrieve/pii/S0021999114001831}

\bibitem{Garrick2017}
D.~P. Garrick, W.~A. Hagen, J.~D. Regele,
  \href{https://linkinghub.elsevier.com/retrieve/pii/S0021999117303674}{An
  interface capturing scheme for modeling atomization in compressible flows},
  Journal of Computational Physics 344 (2017) 260--280.
\newblock \href {https://doi.org/10.1016/j.jcp.2017.04.079}
  {\path{doi:10.1016/j.jcp.2017.04.079}}.
\newline\urlprefix\url{https://linkinghub.elsevier.com/retrieve/pii/S0021999117303674}

\bibitem{Toro1994}
E.~F. Toro, M.~Spruce, W.~Speares,
  \href{http://link.springer.com/article/10.1007/BF01414629
  https://link.springer.com/content/pdf/10.1007%2FBF01414629.pdf}{Restoration
  of the contact surface in the hll-riemann solver}, Shock waves 4 (1994)
  25–34.
\newline\urlprefix\url{http://link.springer.com/article/10.1007/BF01414629
  https://link.springer.com/content/pdf/10.1007%2FBF01414629.pdf}

\bibitem{Toro2019}
E.~F. Toro, \href{https://doi.org/10.1007/s00193-019-00912-4}{The hllc riemann
  solver}, Shock Waves 29 (2019) 1065--1082.
\newblock \href {https://doi.org/10.1007/s00193-019-00912-4}
  {\path{doi:10.1007/s00193-019-00912-4}}.
\newline\urlprefix\url{https://doi.org/10.1007/s00193-019-00912-4}

\bibitem{Hu2013}
X.~Y. Hu, N.~A. Adams, C.-W. Shu,
  \href{https://linkinghub.elsevier.com/retrieve/pii/S0021999113000557}{Positivity-preserving
  method for high-order conservative schemes solving compressible euler
  equations}, Journal of Computational Physics 242 (2013) 169--180.
\newblock \href {https://doi.org/10.1016/j.jcp.2013.01.024}
  {\path{doi:10.1016/j.jcp.2013.01.024}}.
\newline\urlprefix\url{https://linkinghub.elsevier.com/retrieve/pii/S0021999113000557}

\bibitem{Gottlieb1998a}
S.~Gottlieb, C.-W. Shu,
  \href{http://www.ams.org/mcom/1998-67-221/S0025-5718-98-00913-2/}{Total
  variation diminishing runge-kutta schemes}, Mathematics of computation of the
  American Mathematical Society 67 (1998) 73--85.
\newline\urlprefix\url{http://www.ams.org/mcom/1998-67-221/S0025-5718-98-00913-2/}

\bibitem{Nourgaliev2007}
R.~R. Nourgaliev, T.~G. Theofanous,
  \href{http://linkinghub.elsevier.com/retrieve/pii/S0021999106005511}{High-fidelity
  interface tracking in compressible flows: Unlimited anchored adaptive level
  set}, Journal of Computational Physics 224 (2007) 836--866.
\newblock \href {https://doi.org/10.1016/j.jcp.2006.10.031}
  {\path{doi:10.1016/j.jcp.2006.10.031}}.
\newline\urlprefix\url{http://linkinghub.elsevier.com/retrieve/pii/S0021999106005511}

\bibitem{Jiang2000}
G.~S. Jiang, D.~Peng, \href{http://www.siam.org/journals/ojsa.php}{Weighted eno
  schemes for hamilton-jacobi equations}, SIAM Journal on Scientific Computing
  21 (2000) 2126--2143.
\newblock \href {https://doi.org/10.1137/S106482759732455X}
  {\path{doi:10.1137/S106482759732455X}}.
\newline\urlprefix\url{http://www.siam.org/journals/ojsa.php}

\bibitem{Fedkiw1999a}
R.~P. Fedkiw, T.~Aslam, B.~Merriman, S.~Osher, A non-oscillatory eulerian
  approach to interfaces in multimaterial flows (the ghost fluid method),
  Journal of Computational Physics 152 (1999) 457--492.
\newblock \href {https://doi.org/10.1006/jcph.1999.6236}
  {\path{doi:10.1006/jcph.1999.6236}}.

\bibitem{Hu2006}
X.~Y. Hu, B.~C. Khoo, N.~A. Adams, F.~L. Huang,
  \href{http://linkinghub.elsevier.com/retrieve/pii/S0021999106001926}{A
  conservative interface method for compressible flows}, J. Comput. Phys. 219
  (2006) 553--578.
\newblock \href {https://doi.org/10.1016/j.jcp.2006.04.001}
  {\path{doi:10.1016/j.jcp.2006.04.001}}.
\newline\urlprefix\url{http://linkinghub.elsevier.com/retrieve/pii/S0021999106001926}

\bibitem{Lauer2012c}
E.~Lauer, X.~Y. Hu, S.~Hickel, N.~A. Adams,
  \href{http://linkinghub.elsevier.com/retrieve/pii/S0045793012002885}{Numerical
  modelling and investigation of symmetric and asymmetric cavitation bubble
  dynamics}, Computers \& Fluids 69 (2012) 1--19.
\newblock \href {https://doi.org/10.1016/j.compfluid.2012.07.020}
  {\path{doi:10.1016/j.compfluid.2012.07.020}}.
\newline\urlprefix\url{http://linkinghub.elsevier.com/retrieve/pii/S0045793012002885}

\bibitem{Hafner2021b}
D.~Häfner, F.~Vicentini,
  \href{https://joss.theoj.org/papers/10.21105/joss.03419}{mpi4jax: Zero-copy
  mpi communication of jax arrays}, Journal of Open Source Software 6 (2021)
  3419.
\newblock \href {https://doi.org/10.21105/joss.03419}
  {\path{doi:10.21105/joss.03419}}.
\newline\urlprefix\url{https://joss.theoj.org/papers/10.21105/joss.03419}

\bibitem{White1991-ue}
F.~M. White, {Viscous Fluid Flow}, 2nd Edition, McGraw Hill Higher Education,
  Maidenhead, England, 1991.

\bibitem{Coleman1995}
G.~N. Coleman, J.~Kim, R.~D. Moser,
  \href{https://www.cambridge.org/core/product/identifier/S0022112095004587/type/journal_article}{A
  numerical study of turbulent supersonic isothermal-wall channel flow},
  Journal of Fluid Mechanics 305 (1995) 159--183.
\newblock \href {https://doi.org/10.1017/S0022112095004587}
  {\path{doi:10.1017/S0022112095004587}}.
\newline\urlprefix\url{https://www.cambridge.org/core/product/identifier/S0022112095004587/type/journal_article}

\bibitem{Lechner2001}
R.~Lechner, J.~Sesterhenn, R.~Friedrich,
  \href{http://www.tandfonline.com/doi/abs/10.1088/1468-5248/2/1/001}{Turbulent
  supersonic channel flow}, Journal of Turbulence 2 (2001) N1.
\newblock \href {https://doi.org/10.1088/1468-5248/2/1/001}
  {\path{doi:10.1088/1468-5248/2/1/001}}.
\newline\urlprefix\url{http://www.tandfonline.com/doi/abs/10.1088/1468-5248/2/1/001}

\bibitem{Pirozzoli2004}
S.~Pirozzoli, F.~Grasso, T.~B. Gatski, {Direct numerical simulation and
  analysis of a spatially evolving supersonic turbulent boundary layer at M =
  2.25}, Physics of Fluids 16~(3) (2004) 530--545.
\newblock \href {https://doi.org/10.1063/1.1637604}
  {\path{doi:10.1063/1.1637604}}.

\bibitem{ZEIDAN2007}
D.~ZEIDAN, A.~SLAOUTI, E.~ROMENSKI, E.~F. TORO,
  \href{https://www.worldscientific.com/doi/abs/10.1142/S0219876207000984}{Numerical
  solution for hyperbolic conservative two-phase flow equations}, International
  Journal of Computational Methods 04 (2007) 299--333.
\newblock \href {https://doi.org/10.1142/S0219876207000984}
  {\path{doi:10.1142/S0219876207000984}}.
\newline\urlprefix\url{https://www.worldscientific.com/doi/abs/10.1142/S0219876207000984}

\bibitem{Abgrall2001}
R.~Abgrall, S.~Karni,
  \href{https://linkinghub.elsevier.com/retrieve/pii/S0021999100966853}{Computations
  of compressible multifluids}, Journal of Computational Physics 169 (2001)
  594--623.
\newblock \href {https://doi.org/10.1006/jcph.2000.6685}
  {\path{doi:10.1006/jcph.2000.6685}}.
\newline\urlprefix\url{https://linkinghub.elsevier.com/retrieve/pii/S0021999100966853}

\bibitem{Chen2008}
H.~Chen, S.~M. Liang, Flow visualization of shock/water column interactions,
  Shock Waves 17 (2008) 309--321.
\newblock \href {https://doi.org/10.1007/s00193-007-0115-9}
  {\path{doi:10.1007/s00193-007-0115-9}}.

\bibitem{Haas1987}
J.-F. Haas, B.~Sturtevant,
  \href{http://www.journals.cambridge.org/abstract_S0022112087002003}{Interaction
  of weak shock waves with cylindrical and spherical gas inhomogeneities},
  Journal of Fluid Mechanics 181 (1987) 41.
\newblock \href {https://doi.org/10.1017/S0022112087002003}
  {\path{doi:10.1017/S0022112087002003}}.
\newline\urlprefix\url{http://www.journals.cambridge.org/abstract_S0022112087002003}

\bibitem{Terashima2009}
H.~Terashima, G.~Tryggvason,
  \href{https://linkinghub.elsevier.com/retrieve/pii/S0021999109000898}{A
  front-tracking/ghost-fluid method for fluid interfaces in compressible
  flows}, Journal of Computational Physics 228 (2009) 4012--4037.
\newblock \href {https://doi.org/10.1016/j.jcp.2009.02.023}
  {\path{doi:10.1016/j.jcp.2009.02.023}}.
\newline\urlprefix\url{https://linkinghub.elsevier.com/retrieve/pii/S0021999109000898}

\bibitem{Sembian2016a}
S.~Sembian, M.~Liverts, N.~Tillmark, N.~Apazidis,
  \href{http://dx.doi.org/10.1063/1.4948274}{Plane shock wave interaction with
  a cylindrical water column}, Physics of Fluids 28 (2016).
\newblock \href {https://doi.org/10.1063/1.4948274}
  {\path{doi:10.1063/1.4948274}}.
\newline\urlprefix\url{http://dx.doi.org/10.1063/1.4948274}

\bibitem{Margossian2019}
C.~C. Margossian,
  \href{https://wires.onlinelibrary.wiley.com/doi/10.1002/widm.1305}{A review
  of automatic differentiation and its efficient implementation}, WIREs Data
  Mining and Knowledge Discovery 9 (7 2019).
\newblock \href {https://doi.org/10.1002/widm.1305}
  {\path{doi:10.1002/widm.1305}}.
\newline\urlprefix\url{https://wires.onlinelibrary.wiley.com/doi/10.1002/widm.1305}

\bibitem{Harten1994}
A.~Harten,
  \href{https://linkinghub.elsevier.com/retrieve/pii/S0021999184711995}{Adaptive
  multiresolution schemes for shock computations}, Journal of Computational
  Physics 115 (1994) 319--338.
\newblock \href {https://doi.org/10.1006/jcph.1994.1199}
  {\path{doi:10.1006/jcph.1994.1199}}.
\newline\urlprefix\url{https://linkinghub.elsevier.com/retrieve/pii/S0021999184711995}

\bibitem{Harten1995}
A.~Harten, \href{http://doi.wiley.com/10.1002/cpa.3160481201
  http://onlinelibrary.wiley.com/doi/10.1002/cpa.3160481201/abstract}{Multiresolution
  algorithms for the numerical solution of hyperbolic conservation laws},
  Communications on Pure and Applied Mathematics 48 (1995) 1305--1342.
\newblock \href {https://doi.org/10.1002/cpa.3160481201}
  {\path{doi:10.1002/cpa.3160481201}}.
\newline\urlprefix\url{http://doi.wiley.com/10.1002/cpa.3160481201
  http://onlinelibrary.wiley.com/doi/10.1002/cpa.3160481201/abstract}

\bibitem{Harten1983a}
A.~Harten, P.~D. Lax, B.~van Leer,
  \href{http://epubs.siam.org/doi/10.1137/1025002}{On upstream differencing and
  godunov-type schemes for hyperbolic conservation laws}, SIAM Review 25 (1983)
  35--61.
\newblock \href {https://doi.org/10.1137/1025002} {\path{doi:10.1137/1025002}}.
\newline\urlprefix\url{http://epubs.siam.org/doi/10.1137/1025002}

\bibitem{Fleischmann2020}
N.~Fleischmann, S.~Adami, N.~A. Adams,
  \href{https://doi.org/10.1016/j.jcp.2020.109762}{A shock-stable modification
  of the hllc riemann solver with reduced numerical dissipation}, Journal of
  Computational Physics 423 (2020) 109762.
\newblock \href {https://doi.org/10.1016/j.jcp.2020.109762}
  {\path{doi:10.1016/j.jcp.2020.109762}}.
\newline\urlprefix\url{https://doi.org/10.1016/j.jcp.2020.109762}

\bibitem{Liou1996}
M.~S. Liou, A sequel to ausm: Ausm+, Journal of Computational Physics 129
  (1996) 364--382.
\newblock \href {https://doi.org/10.1006/jcph.1996.0256}
  {\path{doi:10.1006/jcph.1996.0256}}.

\bibitem{Roe1981}
P.~L. Roe, Approximate riemann solvers, parameter vectors, and difference
  schemes, Journal of Computational Physics 43 (1981) 357--372.
\newblock \href {https://doi.org/10.1016/0021-9991(81)90128-5}
  {\path{doi:10.1016/0021-9991(81)90128-5}}.

\bibitem{Davis1988}
S.~F. Davis, \href{http://epubs.siam.org/doi/10.1137/0909030}{Simplified
  second-order godunov-type methods}, SIAM Journal on Scientific and
  Statistical Computing 9 (1988) 445--473.
\newblock \href {https://doi.org/10.1137/0909030} {\path{doi:10.1137/0909030}}.
\newline\urlprefix\url{http://epubs.siam.org/doi/10.1137/0909030}

\bibitem{Einfeldt1988a}
B.~Einfeldt, On godunov-type methods for gas dynamics, SIAM Journal on
  Numerical Analysis 25 (1988) 294--318.
\newblock \href {https://doi.org/10.1137/0725021} {\path{doi:10.1137/0725021}}.

\bibitem{Acker2016a}
F.~Acker, R.~B. Borges, B.~Costa,
  \href{http://dx.doi.org/10.1016/j.jcp.2016.01.038}{An improved weno-z
  scheme}, Journal of Computational Physics 313 (2016) 726--753.
\newblock \href {https://doi.org/10.1016/j.jcp.2016.01.038}
  {\path{doi:10.1016/j.jcp.2016.01.038}}.
\newline\urlprefix\url{http://dx.doi.org/10.1016/j.jcp.2016.01.038}

\bibitem{Gande2020}
N.~R. Gande, A.~A. Bhise, Modified third and fifth order weno schemes for
  inviscid compressible flows, Numerical Algorithms (2020).
\newblock \href {https://doi.org/10.1007/s11075-020-01039-9}
  {\path{doi:10.1007/s11075-020-01039-9}}.

\bibitem{Bezgin2021b}
D.~A. Bezgin, S.~J. Schmidt, N.~A. Adams, {WENO3-NN: A maximum-order
  three-point data-driven weighted essentially non-oscillatory scheme}, Journal
  of Computational Physics 452 (2022) 110920.
\newblock \href {https://doi.org/10.1016/j.jcp.2021.110920}
  {\path{doi:10.1016/j.jcp.2021.110920}}.

\bibitem{Hu2010}
X.~Y. Hu, Q.~Wang, N.~A. Adams,
  \href{http://linkinghub.elsevier.com/retrieve/pii/S0021999110004560}{An
  adaptive central-upwind weighted essentially non-oscillatory scheme}, Journal
  of Computational Physics 229 (2010) 8952--8965.
\newblock \href {https://doi.org/10.1016/j.jcp.2010.08.019}
  {\path{doi:10.1016/j.jcp.2010.08.019}}.
\newline\urlprefix\url{http://linkinghub.elsevier.com/retrieve/pii/S0021999110004560}

\bibitem{Hu2011}
X.~Hu, N.~Adams,
  \href{http://linkinghub.elsevier.com/retrieve/pii/S0021999111003342}{Scale
  separation for implicit large eddy simulation}, Journal of Computational
  Physics 230 (2011) 7240--7249.
\newblock \href {https://doi.org/10.1016/j.jcp.2011.05.023}
  {\path{doi:10.1016/j.jcp.2011.05.023}}.
\newline\urlprefix\url{http://linkinghub.elsevier.com/retrieve/pii/S0021999111003342}

\bibitem{Balsara2000}
D.~S. Balsara, C.~W. Shu, Monotonicity preserving weighted essentially
  non-oscillatory schemes with increasingly high order of accuracy, Journal of
  Computational Physics 160 (2000) 405--452.
\newblock \href {https://doi.org/10.1006/jcph.2000.6443}
  {\path{doi:10.1006/jcph.2000.6443}}.

\bibitem{Fu2016}
L.~Fu, X.~Y. Hu, N.~A. Adams, A family of high-order targeted eno schemes for
  compressible-fluid simulations, Journal of Computational Physics 305 (2016)
  333--359.
\newblock \href {https://doi.org/10.1016/j.jcp.2015.10.037}
  {\path{doi:10.1016/j.jcp.2015.10.037}}.

\bibitem{vanLeer1979}
B.~van Leer,
  \href{https://linkinghub.elsevier.com/retrieve/pii/0021999179901451}{Towards
  the ultimate conservative difference scheme. v. a second-order sequel to
  godunov's method}, Journal of Computational Physics 32 (1979) 101--136.
\newblock \href {https://doi.org/10.1016/0021-9991(79)90145-1}
  {\path{doi:10.1016/0021-9991(79)90145-1}}.
\newline\urlprefix\url{https://linkinghub.elsevier.com/retrieve/pii/0021999179901451}

\bibitem{Russo2000}
G.~Russo, P.~Smereka, \href{http://www.idealibrary.com}{A remark on computing
  distance functions}, Journal of Computational Physics 163 (2000) 51--67.
\newblock \href {https://doi.org/10.1006/jcph.2000.6553}
  {\path{doi:10.1006/jcph.2000.6553}}.
\newline\urlprefix\url{http://www.idealibrary.com}

\bibitem{Hickel2014b}
S.~Hickel, C.~P. Egerer, J.~Larsson, {Subgrid-scale modeling for implicit large
  eddy simulation of compressible flows and shock-turbulence interaction},
  Physics of Fluids 26~(10) (2014).
\newblock \href {https://doi.org/10.1063/1.4898641}
  {\path{doi:10.1063/1.4898641}}.

\bibitem{Batten1997}
P.~Batten, N.~Clarke, C.~Lambert, D.~M. Causon,
  \href{http://epubs.siam.org/doi/10.1137/S1064827593260140}{On the choice of
  wavespeeds for the hllc riemann solver}, SIAM Journal on Scientific Computing
  18 (1997) 1553--1570.
\newblock \href {https://doi.org/10.1137/S1064827593260140}
  {\path{doi:10.1137/S1064827593260140}}.
\newline\urlprefix\url{http://epubs.siam.org/doi/10.1137/S1064827593260140}

\bibitem{Hu2004}
X.~Y. Hu, B.~C. Khoo,
  \href{http://linkinghub.elsevier.com/retrieve/pii/S0021999104000178}{An
  interface interaction method for compressible multifluids}, Journal of
  Computational Physics 198 (2004) 35--64.
\newblock \href {https://doi.org/10.1016/j.jcp.2003.12.018}
  {\path{doi:10.1016/j.jcp.2003.12.018}}.
\newline\urlprefix\url{http://linkinghub.elsevier.com/retrieve/pii/S0021999104000178}

\end{thebibliography}

\end{document}